\documentclass[fleqn,usenatbib]{mnras}

\usepackage{newtxtext,newtxmath}

\usepackage[T1]{fontenc}

\DeclareRobustCommand{\VAN}[3]{#2}
\let\VANthebibliography\thebibliography
\def\thebibliography{\DeclareRobustCommand{\VAN}[3]{##3}\VANthebibliography}

\usepackage{graphicx}	
\usepackage{amsmath}	
\usepackage{pdflscape}

\title[Ly\(\alpha\) Halos Around ERQs]{Compact and Quiescent Circumgalactic Medium and Ly\(\alpha\) Halos around Extremely Red Quasars (ERQs)}

\author[J. Gillette et al.]{
Jarred Gillette,$^{1}$\thanks{E-mail: jgill016@ucr.edu }
Marie Wingyee Lau,$^{1}$
Fred Hamann,$^{1}$
Serena Perrotta,$^{2}$
David S. N. Rupke,$^{3}$
\newauthor Dominika Wylezalek,$^{4}$ 
Nadia L. Zakamska,$^{5}$
and Andrey Vayner$^{5}$\\
$^{1}$Department of Physics \& Astronomy, University of California, Riverside, CA 92521, USA\\
$^{2}$Center for Astrophysics and Space Sciences, University of California, San Diego, CA 92093, 
USA\\
$^{3}$Department of Physics, Rhodes College, Memphis, TN 38112, USA\\
$^{4}$Astronomical Calculation Institute, University of Heidelberg, D-69120 Heidelberg, Germany\\
$^{5}$Department of Physics \& Astronomy, Johns Hopkins University, Baltimore, MD 21218, USA}

\date{Accepted XXX. Received YYY; in original form ZZZ}

\pubyear{2023}

\begin{document}
\label{firstpage}
\pagerange{\pageref{firstpage}--\pageref{lastpage}}
\maketitle

\begin{abstract}
Red quasars may represent a young stage of galaxy evolution that provide important feedback to their host galaxies. We are studying a population of extremely red quasars (ERQs) with exceptionally fast and powerful outflows, at median redshift $z$~=~2.6. We present Keck/KCWI integral field spectra of 11 ERQs, which have a median color $i-W3$~=~5.9~mag, median $\left\langle L_{\text{bol}} \right\rangle$~$\approx$~5~$\times$~$10^{47}$~erg~s$^{-1}$, Ly$\alpha$ halo luminosity $\left\langle L_{\text{halo}} \right\rangle$~$=$~5~$\times$~$10^{43}$~erg~s$^{-1}$, and maximum linear size $>128$~kpc. The ERQ halos are generally similar to those of blue quasars, following known trends with $L_{\text{bol}}$ in halo properties. ERQs have halo symmetries similar to Type-I blue quasars, suggesting Type-I spatial orientations. ERQ $\left\langle L_{\text{halo}} \right\rangle$ is $\sim$2~dex below blue quasars, which is marginal due to scatter, but consistent with obscuration lowering photon escape fractions. ERQ halos tend to have more compact and circularly symmetric inner regions than blue quasars, with median exponential scale lengths of $\sim$9~kpc, compared to $\sim$16~kpc for blue quasars. When we include the central regions not available in blue quasar studies (due to PSF problems), the true median ERQ halo scale length is just $\sim$6~kpc. ERQ halos are also kinematically quiet, with median velocity dispersion 293~km~s$^{-1}$, consistent with expected virial speeds. Overall we find no evidence for feedback on circumgalactic scales, and the current episode of quasar activity, perhaps due to long outflow travel times, has not been around long enough to affect the circumgalactic medium. We confirm the narrow Ly$\alpha$ emission spikes found in ERQ aperture spectra are halo features, and are useful for systemic redshifts and measuring outflow speeds in other features.

\end{abstract}

\begin{keywords}
galaxies: active -- quasars: emission lines -- galaxies: intergalactic medium -- galaxies: halos -- galaxies: evolution -- galaxies: high-redshift
\end{keywords}



\section{Introduction}
\label{sec:sec_intro}

Quasars are supermassive black holes which grow as they rapidly accrete infalling material at the center of their host galaxy. At high redshift, accretion can coincide with galactic assembly via galaxy mergers, or streams of infalling gas. Major merger activity could trigger both star formation and growth of the central black hole \citep{Hopkins+06,Hopkins+08,Somerville+08,Glikman+15}. The process of infalling gas, cold-mode accretion, may also be a dominant mechanism for supplying matter for the formation of galaxies, triggering starbursts, and fueling quasars \citep{Keres+09,Dekel+09,FaucherGiguereKeres11,Fumagalli+14}. Accretion by infall can coincide with outflows as the subsequent galaxy activity generates feedback, influencing the galaxy's formation \citep{CostaSijackiHaehnelt14,Nelson+15,Suresh+19}. A possible evolutionary scheme is where the central back hole grows in obscurity until feedback generates outflows, which clear the obscuring interstellar medium, to reveal a luminous quasar \citep{Sanders+88,DiMatteo+05, Hopkins+06,Hopkins+08,Hopkins+16,RupkeVeilleux11,RupkeVeilleux13,Liu+13,Stacey+22}. 

Red quasars are important for testing the hypothesis that obscured quasars may be in a young evolution phase. Young quasars may be reddened by dust created in a major starburst inside the galaxies, triggered by a merger or cold-mode accretion. They may show signs of early evolution such as rapid accretion or more powerful outflows from the quasar, or more infall from the intergalactic medium, as they transition in to blue quasars. Studies of red/obscured quasars have found many to be in mergers or high-accretion phases \citep{Glikman+15,Wu+18,Zakamska+19}. Extremely Red Quasars (ERQs) display powerful outflow signatures, and are interesting candidates for studying a young quasar phase in the galaxy/quasar evolution scheme \citep{Hamann+17}. 

 ERQs were first discovered among the Baryon Oscillation Spectroscopic Survey \citep[BOSS,][]{Paris+17} in the Sloan Digital Sky Survey-III \citep[SDSS,][]{Eisenstein+11} and the ALLWISE data release \citep{Cutri+11,Cutri+13} of the Wide-field Infrared Survey Explorer \citep[WISE,][]{Wright+10}. They were initially described in \citet{Ross+15}, and were further refined in \citet{Hamann+17} to $\sim$200 objects, characterized by their extremely red colors \citep[\(i-W3\) \(>4.6\) mag;][]{Hamann+17}. Currently known ERQs are at comic noon redshifts $z$~$\sim$~2$-$4, and are on the high end of quasar bolometric luminosities $L_{bol}$~>~$10^{47}$~erg~s$^{-1}$.

ERQs exhibit other extreme spectral properties that make them unique compared to other red quasar samples, beyond being extremely red. Many have unusually strong and/or blueshifted broad emission lines, with C~IV rest equivalent widths >~100~\AA, and peculiar wingless profiles with high kurtosis \citep{MonadiBird22}. They also frequently have unusual emission line flux ratios (e.g. high N~V~$\lambda$1240/Ly$\alpha$ or  N~V~$\lambda$1240/C~IV~$\lambda$1549). These broad line features indicate outflow properties controlled by accretion, and reside on scales of tens of parsecs \citep{Zhang+17,Alexandroff+18}. ERQs also have the most blueshifted [O~III] $\lambda$5007 emission lines ever reported, with speeds >~6000~km~s$^{-1}$ \citep{Zakamska+16,Perrotta+19}. [O~III] is significant because it traces gases on galactic scales, at tens of kpc, and are low density forbidden transitions \citep{Hamann+11,Vayner+21}. Outflows at these galactic scales carry energy that could generate feedback in the host galaxy. This suite of properties found in ERQ spectra indicate more extreme physical conditions than what are found in typical quasar populations, and beyond orientation effects. 

Understanding ERQs and their evolutionary nature has launched a multi-faceted study of their physical environments. ERQ host galaxies have been directly imaged with \textit{Hubble Space Telescope} to look for signs of merger activity \citep{Zakamska+19}. Major merger activity was not found in most of their objects, but observations of high redshift quasars have added difficulty subtracting point-spread function in two dimensional imaging. Another approach is to study the kinematics and physical extent of the outflowing gas, specifically, the [O~III] emission (\citet{Perrotta+19}; \citet{Vayner+21}; Lau~et~al. in preparation). Past studies have found [O~III] outflows to be spatially compact, not extended to circumgalactic regions, and at kpc scales of the nuclear regions of host galaxies \citep{Vayner+21}. A third method of investigating ERQ environments is through their halo emission. The circumgalactic medium (CGM) has an important role in understanding the evolution of galaxies \citep{TumlinsonPeeplesWerk17}, and is the site of interaction for outflows from galaxies and the inflows from the intergalactic medium. 

High redshift quasars have been found to have large gas reservoirs in the CGM, observed in fluorescent Ly$\alpha$ emission, and have been interpreted as filaments or inflowing gas in cold-mode accretion \citep{Borisova+16b,ArrigoniBattaia+19,Cai+19}. ERQs are a well defined and uniform sample from BOSS, and make good targets for studying their Ly$\alpha$ halos for quasar evolution.

In this paper, we aim to investigate whether ERQs represent an earlier evolutionary stage compared to regular luminous blue quasars, which are characterized by strong rest-UV emission lines and blue color across the rest-frame mid-UV to Near-IR spectral range. With the embedded quasar evolution scheme, we investigate ERQs for evidence they are different or unusual from normal blue quasars at the same redshifts and luminosities. This could be evidence for feedback, but also signatures of more intense infall or chaotic mergers, more/less asymmetries, and/or suppressed Ly$\alpha$ halo emission due to the quasar obscuration. Detailed analysis of the Ly\(\alpha\) emission from ERQs can provide key insights into this question. 

Secondarily we want to investigate the spatial origin of a peculiar, and narrow, Ly\(\alpha\) emission component frequently seen in spatially unresolved ERQ spectra. It often appears as a ``spike'' on top of the broad Ly$\alpha$ emission, and it's narrow width (FWHM~<~1,000~km~s$^{-1}$) indicates the emission is far from the quasar's broad line region (BLR). The spike sometimes coincides with the rest frame of other narrow emission lines (e.g. He~II and narrow [O~III], \citet{Hamann+17} and \citet{Perrotta+19}, respectively). We want to determine whether this sharp feature originates in the extended inner-halo region, as hypothesized in \citet{Hamann+17}, and discussed or used in \citet{Perrotta+19} and \citet{Lau+22}. This investigation is motivated from observations that many broad emission lines in ERQs are often blueshifted, and involved in outflow, that would typically be at the systemic in other quasar populations \citep{Hamann+17,Perrotta+19}. This blueshifting would systemically bias automated redshift estimates generated in BOSS, making outflows appear weaker. If the source of the Ly$\alpha$ spike is far from the galactic center, where outflows would take place, and is extended in the galactic halo, then it would not be involved in strong outflows and would provide a better systemic redshift estimate. Confirming the narrow Ly\(\alpha\) emission can be used as a better systemic redshift of ERQs also confirms outflows hypothesized in the discussions aforementioned, and helps constrain velocities for future work on ERQ outflows (J.~Gillette~et~al.~2023b~in~preparation).

Our team already analyzed one Ly$\alpha$ halo from our sample that is the reddest known ERQ (J0006+1215, in \citet{Lau+22}), which displays an extremely fast [O~III]~$\lambda$5007 outflow at $\sim$6000~km~s$^{-1}$ \citep{Perrotta+19}. The Ly$\alpha$ halo spans $\sim$100~kpc, and the narrow Ly$\alpha$ emission spike in the quasar spectrum originates from the inner halo. It is kinematically quiet, with velocity dispersion of $\sim$300~km~s$^{-1}$ and no broadening above the dark matter circular velocity down to the spatially resolved limit $\sim$6~kpc from the quasar. \citet{Lau+22}, hereafter L22, proposed that the He~II~$\lambda$1640/Ly$\alpha$ ratio of the inner halo and the asymmetry level of the overall halo in J0006+1215 are dissimilar to Type-II quasars, suggesting unique physical conditions for J0006+1215 that are beyond orientation differences from other quasar populations. We note the Type-I/II quasar classification dichotomy, because other studies find that Type-IIs tend to show more asymmetric halos, consistent with more edge-on views of the quasar. It also correlates strongly with the level of obscuration toward the nucleus \cite{Greene+14}, and we want to distinguish ERQs from obscured quasar populations explained by orientation effects.

In this paper we analyze Ly$\alpha$ halos of a sample of ERQs, and compare them to those around other quasars as a unique population. It is organized as follows. Section 2 describes the selection method for ERQ observations, their data reduction, calibration, and post-processing. Section 3 describes the measured halo properties of ERQ sample, such as the extended line emission surrounding the quasar, and analysis results of size, morphology, surface brightness (SB), and kinematics of the extended emission. In Section 4 we discuss implications for quasar youth, feedback, and further for quasar studies. Section 5 concludes the paper. Throughout this paper we adopt a \(\Lambda\)-CDM cosmology with \(H_0\)~=~69.6~km~s\(^{-1}\)~Mpc\(^{-1}\), \(\Omega_\text{M}\)~=~0.286 and \(\Omega_\Lambda\)~=~0.714, as adopted by the online cosmology calculator developed by \citet{Wright06}. All magnitudes are on the AB system. Reported wavelengths are in vacuum and in the heliocentric frame.



\section{Target Selection, Observations, and Data Reduction}

Here we summarize the KCWI observations, data reduction, and post-processing for the full sample of 12 ERQs, following the procedures described in detail by L22. Our aim is to test whether the most extreme properties of ERQs correlate in any way with particular or unusual Ly\(\alpha\) halo properties. Our second goal is to test the suggestion in \citet{Hamann+17} that the narrow Ly\(\alpha\) spikes in the BOSS spectra of some ERQs are halo emission.

\subsection{Sample Selection}
\label{sec:selection_label} 

We select targets for our study from the sample of 205 ERQs in \citet{Hamann+17}, and in the redshift range \(2.0 \leq z \leq 3.6\), all have coverage of the Ly\(\alpha\) feature at rest-frame 1215.7\AA  . ERQs showing a narrow Ly\(\alpha\) emission in the profile of their BOSS spectra, by visual inspection, were prioritized during observing to determine if the emission originates from an extended halo. We also prioritize ERQs in \citet{Perrotta+19}, which have [O~III] measurements, which were shown to have color-correlated powerful outflows at tens of kpc scales, and may be more likely to show evidence of feedback in the halo. We also prefer the reddest available ERQs. We chose the best available targets subject to weather and scheduling constraints, and prioritize targets fitting several of the criteria above.

Finally, after applying our selection criteria to our four nights  we present a total sample of 12 ERQs, with redshifts $2.31 \le z \le 3.14$. Half of them have the narrow component in their Ly\(\alpha\) emission. The median reddening of the sample is $i-W3 \approx 5.9$, Table~\ref{tab:table_catalogue} contains basic sample properties.

\subsection{Observations}

We observed our final sample of 11 ERQs using the Keck Cosmic Web Imager \citep[KCWI,][]{Morrissey+18}, on the Keck II telescope. KCWI is a wide field, integral field spectrograph optimized for observing low-surface brightness targets. It provides both spatial and spectral information for resolved targets, allowing us to make pseudo-narrow-band images and obtaining spectra from individual spatial-pixels, or ``spaxels.''

Observations were conducted over four nights from 2018 to 2020, with identical instrument configuration across the sample. Observation details are noted in Table~\ref{tab:table_catalogue}.  Conditions during observation often varied through any night, but our work had typical seeing FWHM of about (0.8-1.4) arcseconds. KCWI currently has only a blue filter, and we used the BL grating, which has the best efficiency and widest wavelength bandpass ($\Delta\lambda \approx$ 2,000\AA). KCWI uses a ``slicer'' to slice the field of view into rows or columns before going to the grating. Slicers come in three different sizes to optimize field of view, spectral resolution, and spatial sampling. We used the medium slicer to have considerable field of view, spectral resolution comparable to SDSS, and sufficient spatial sampling. This configuration yields a field of view of 15.7 arcsec \(\times\) 18.9 arcsec, corresponding to a physical scale of approximately 128~kpc \(\times\) 154~kpc at our sample's median redshift (\(z_{em} \approx 2.61\)). With 24 slices, the instrument configuration provides a spatial sampling of 0.68 arcsec, and seeing limited, along the slices. Each exposure was dithered 0.35 arcsec across slices to sub-sample the long spatial dimension of the output spaxels, which have sizes of 0.68 arcsec \(\times\) 0.29 arcsec. The spectral resolution is \(R = 1800\). The full spectral range is approximately 3500 to 5625\AA, which is ample for coverage of the Ly\(\alpha\) emission profile across a wide range of quasar redshifts, \(2.0 \leq z \leq 3.6\). We used exposures of 20 minutes, optimally integrating for 2 to 3 hours total, and varied depending on target availability, observing conditions, and priority of the object. We calibrated using arclamps and spectroscopic standard stars at the beginning and end of each night.


\subsection{Data Reduction \& Post Processing}
\label{sec:reduction_label} 

We adopted the data reduction and post-processing approaches described in detail in L22. Here we only provide a summary of the major steps. We used the KCWI Data Extraction and Reduction Pipeline (KDERP, https://github.com/Keck-DataReductionPipelines/KcwiDRP) written in the Interactive Data Language (IDL) for flat-fielding, cosmic-ray removal, and for the first-pass instrument noise removal and background subtraction. We made sure that no prominent skyline residuals are present near Ly\(\alpha\) in each spectrum. We then used CWITools \citep{OSullivan+20}, and the IDL library IFSFIT \citep{RupkeTo21}, for the second-pass background subtraction and removing internally scattered light. Specifically, IFSFIT uses a spatially-unresolved quasar emission template, generated from a one arcsecond aperture around the object, and is scaled to subtract all emission from the quasar at every spaxel position. A key step that was customized for half of the sample was the quasar template which is subtracted to leave only halo emission in the residual, described in Section \ref{sec:sec_narrow_lya}. We also used CWITools' simplified algorithm to subtract any foreground continuum sources convolved with the point spread function.

\subsection{Narrow Halo Emission}
\label{sec:sec_narrow_lya}

L22 confirmed that the narrow Ly\(\alpha\) ``spike'' emission in ERQ spectra originates in the halo. In Section \ref{sec:SB_maps} we further confirm this in the rest of our ERQ sample. This insight into the emission's origin allows us to treat the emission differently in processing the spectra and images. For objects with no spike the PSF spectral template removes all emission within the seeing disk resulting in a hollow at the quasar position in the SB map, like in blue quasar studies. 

\begin{figure*}
    \includegraphics[width=0.33\textwidth]{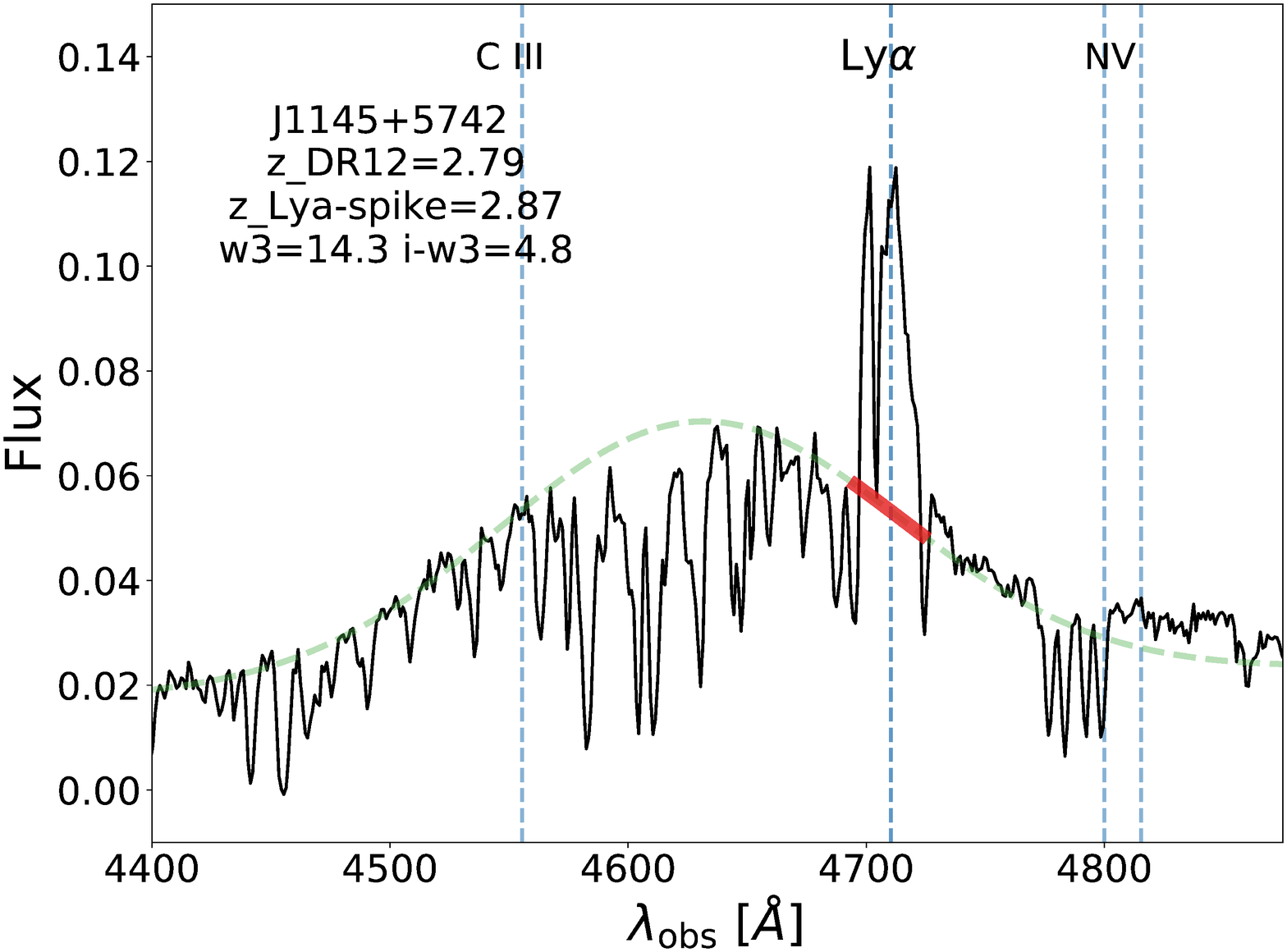}
    \includegraphics[width=0.33\textwidth]{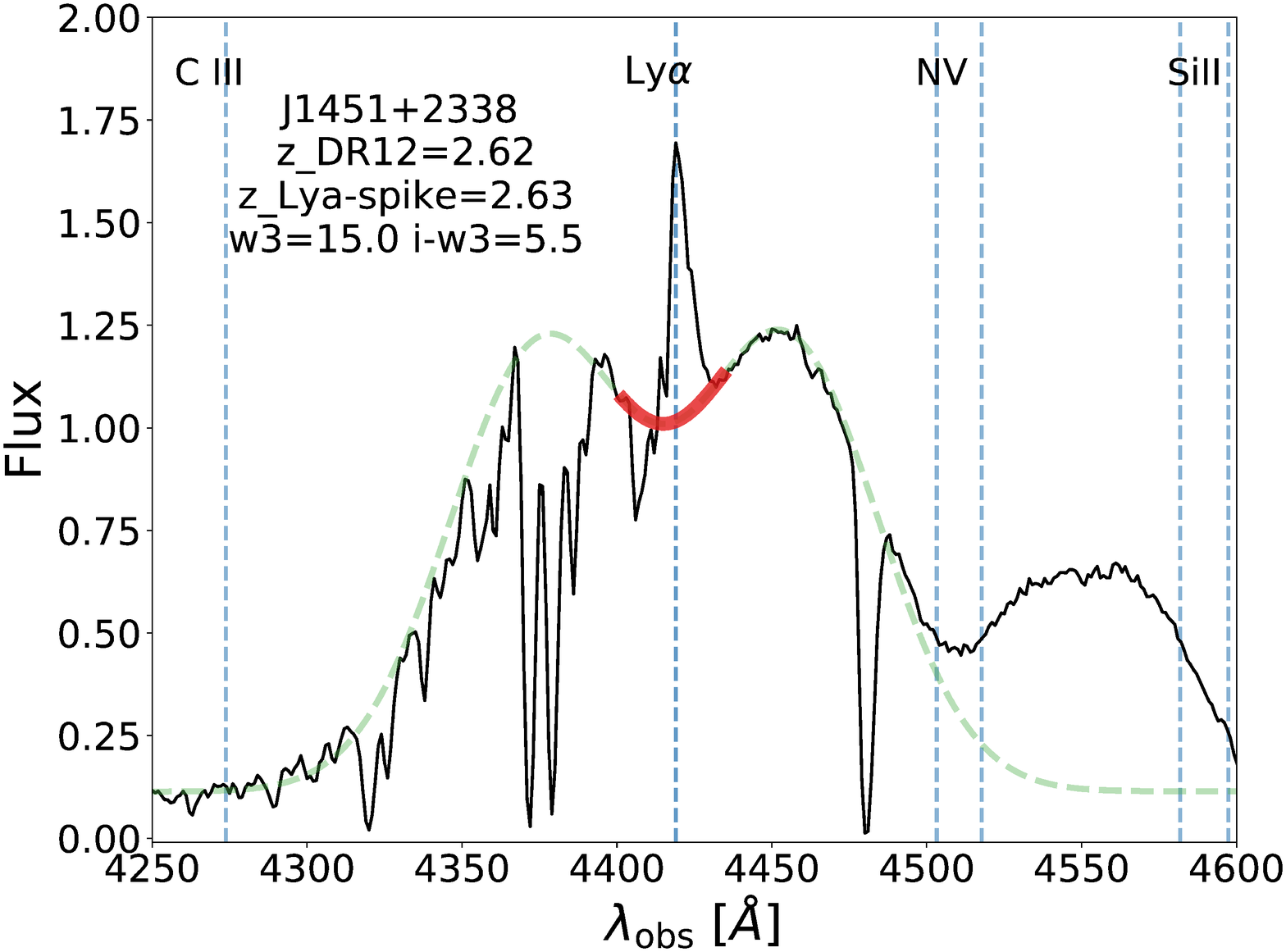}
    \includegraphics[width=0.33\textwidth]{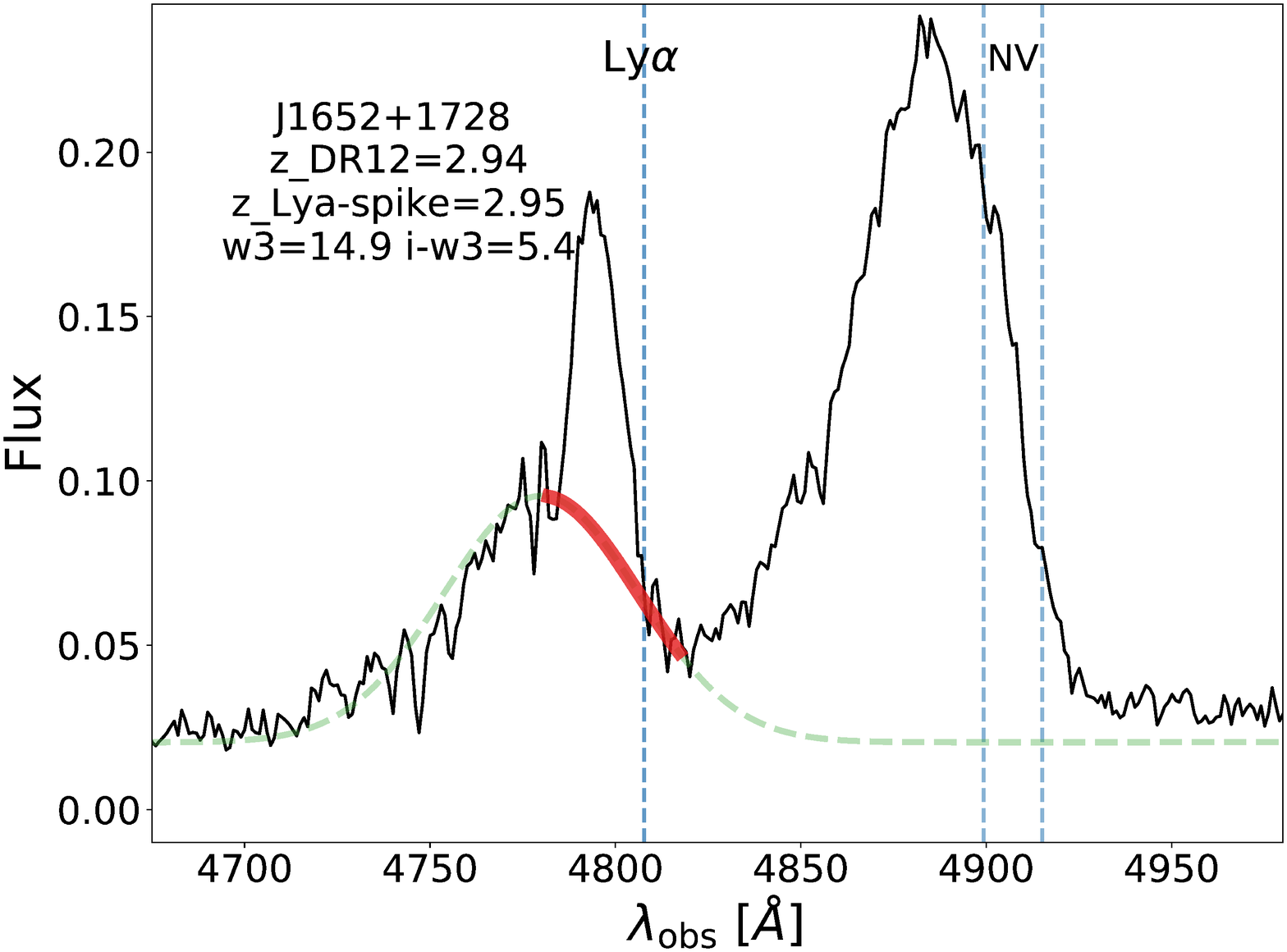}
    \caption{Example spectral templates extracted from an inner 1 arcsecond aperture spectra near Ly\(\alpha\) of, from left to right, J1145+5742, J1451+2338, and J1652+1728. The spectrum is shown in black, the interpolation function is in dashed green, and the ``clipped'' narrow feature in the spectrum is replaced by the thick red line segment. All emission lines are labelled from the rest frame of the Ly\(\alpha\) extended halo emission. Function fitting procedures would become convoluted by absorption in the Ly\(\alpha\) forest, and so the interpolation function was hand-fit to account for smooth emission features and continuum on the red side of Ly\(\alpha\) and deep absorption on the blue side. In the first panel, J1145+5742 has a straightforward and distinct separation between the broad and narrow emission components. The second panel shows J1451+2338, shown to have blended broad emission lines from Ly\(\alpha\) and N~V~$\lambda 1240$. The third panel of J1652+1728 also has unusual emission lines, and the blueshifted central emission of Ly\(\alpha\) caused uncertainties in clipping the narrow emission. We do not assume this interpolation function describes any characteristics of the broad emission, but simply isolates the emission from the narrow component of Ly\(\alpha\).}
    \label{fig:fig_psf_template}
\end{figure*}

We followed the procedure for generating quasar spectral templates as described in L22, and made customisations to the quasar template for each ERQ identified to have narrow Ly\(\alpha\) emission. We isolated the feature within the one arcsecond template by interpolating underneath the narrow line, effectively ``clipping'' it, and left the extended emission in the residual. Examples of this template interpolation is shown in Figure~\ref{fig:fig_psf_template} (see also L22). In cases where the spectral location and profile shape of the spike was more ambiguous, we compared extracted spectra from gradually larger aperture sizes. We confirmed the spike was the only emission that increases with larger aperture size, and interpolated underneath its narrow profile.

\subsection{Line Variability}
We inspected ERQ spectra for variability between our KCWI observations and the BOSS catalog. The median time between observations is two years at rest frame of the the median ERQ. In general, characteristics of line profiles do not significantly change between SDSS and our observations, in agreement with comments in L22. Similarities between these observations indicate the unusual profile features that distinguish ERQs from typical quasars persist beyond two years in the quasar frame, and are consistent with other observations of ERQs from \cite{Hamann+17}.

\begin{landscape}
    \begin{table}
	    \centering
	    \caption{ERQ properties in our program. \(
	    z_{em}\) is from the SDSS DR12Q BOSS catalog emission-line measurements, \(
	    z_{halo}\) is computed from the Ly\(\alpha\) line emission from the extended halo. C IV FWHM is from emission line fitting done in \citet{Hamann+17}. [O~III] emission blueshift measurements, when available, are from \citet{Perrotta+19}, and blueshifted from CO emission measurements by F. Hamann et al. (2023 in prep.), and will be later discussed in Gillette et al. (2023b in prep.). Bolometric luminosity is estimated from the W3 magnitude. ERQ J0834+0159 was observed under cloudy conditions, and is omitted from analysis of Ly$\alpha$ halos. \\ \textit{Notes.} \(^a\) Quasars with their narrow Ly\(\alpha\) component clipped from the quasar emission template. \(^b\) Redshift approximated by Ly\(\alpha\) line peak from an inner 1-arcsec aperture spectrum. \(^c\) Blueshift relative to a redshift approximated by Ly\(\alpha\) line peak from an inner 1-arcsec aperture spectrum.}
	    \begin{tabular}{lccccccccccr} 
		    \hline
            ERQ Name & RA & Dec & \textit{z}\(_{em}\) & \textit{z}\(_{halo}\) & C IV FWHM & [O III] v\(_{98}\) & W3 & i-W3 & Bolometric Luminosity & Observation Date & Exposure Time \\
            & (J2000) & (J2000) & & & (km s\(^{-1}\)) & (km s\(^{-1}\)) & (mag) & (mag) & (erg/s) & & (s) \\
            \hline
            J0006+1215\(^a\) & 00:06:10.67 & 12:15:01.2 & 2.31 & 2.3184 & 4540$\pm$200 & $-$6224.99 & 14.1 & 8.0 & 7.58e+47 & 2019-10-02 & 4,200.0 \\
            J0220+0137\(^a\) & 02:20:52.11 & 01:37:11.1 & 3.14 & 3.1375 & 2613$\pm$161 & ... & 15.8 & 6.2 & 3.68e+47 & 2019-10-02 & 8,400.0 \\
            J0834+0159 & 08:34:48.48 & 01:59:21.2 & 2.59 & 2.5882\(^b\) & 2863$\pm$65 & $-$4426.08 & 14.9 & 6.0 & 4.98e+47 & 2019-03-02 & 6,000.0 \\
            J1145+5742\(^a\) & 11:45:08.00 & 57:42:58.6 & 2.79 & 2.8747 & 9103$\pm$446 & ... & 14.3 & 4.8 & 1.14e+48 & 2019-03-02 & 10,800.0 \\
            J1232+0912 & 12:32:41.75 & 09:12:09.3 & 2.38 & 2.4034 & 4787$\pm$52 & $-$7026.20 & 14.3 & 6.8 & 6.33e+47 & 2020-05-26 & 4,800.0 \\
            J1451+0132\(^a\) & 14:51:13.61 & 01:32:34.1 & 2.77 & 2.8130 & 6231$\pm$156 & ... & 14.7 & 5.7 & 6.78e+47 & 2019-03-02 & 10,800.0 \\
            J1451+2338\(^a\) & 14:51:48.01 & 23:38:45.4 & 2.62 & 2.6348 & 4166$\pm$124 & ... & 15.0 & 5.5 & 4.56e+47 & 2020-05-26 & 8,400.0 \\
            J1652+1728\(^a\) & 16:52:02.61 & 17:28:52.3 & 2.94 & 2.9548 & 2403$\pm$45 & $-$2534.00\(^c\) & 14.9 & 5.4 & 6.80e+47 & 2020-05-26 & 4,800.0 \\
            J1705+2736 & 17:05:58.56 & 27:36:24.7 & 2.45 & 2.4461 & 1301$\pm$22 & ... & 15.5 & 5.1 & 2.43e+47 & 2020-05-26 & 3,600.0 \\
            J2215\(-\)0056 & 22:15:24.03 & $-$00:56:43.8 & 2.51 & 2.5074 & 4280$\pm$112 & $-$3876.95 & 16.0 & 6.2 & 1.51e+47 & 2019-10-02 & 8,400.0 \\
            J2254+2327 & 22:54:38.33 & 23:27:14.5 & 3.09 & 3.0825 & 4412$\pm$146 & ... & 16.5 & 5.5 & 1.68e+47 & 2018-10-05 & 10,800.0 \\
            J2323\(-\)0100 & 23:23:26.17 & $-$01:00:33.1 & 2.36 & 2.3831 & 3989$\pm$62 & $-$6458.03 & 15.2 & 7.2 & 2.88e+47 & 2019-10-02 & 7,800.0 \\
            Median & ... & ... & 2.61 & 2.6348 & 4223 & ... & 15.0 & 5.9 & 4.56e+47 & ... & ... \\
            \hline
            \end{tabular}
            \label{tab:table_catalogue}
    \end{table}

    \begin{table}
	    \centering
	    \caption{ERQ measurements of the Ly\(\alpha\) halo luminosity and morphology. Area covered and size are shown as lower limits when extended halo is detected to the edge of the FOV. Ly\(\alpha\) halo luminosities are measured from modified PSF subtraction, as referenced in Section \ref{sec:sec_narrow_lya}. }
	    \begin{tabular}{lccccccr}
	        \hline
            ERQ Name & Area Covered & Maximum Linear Size & Halo Luminosity & Peak to Quasar Distance & Centroid to Quasar Distance & \(e_\text{weight}\) & \(e_\text{unweight}\) \\
            & (kpc\(^2\)) & (kpc) & (erg/s) & (kpc) & (kpc) & & \\
            \hline
            J0006+1215\(^a\) & 6851 & 140 & 5.00e+43 & 3.44 & 5.56 & 0.44 & 0.69 \\
            J0220+0137\(^a\) & >5693 & >104 & 6.26e+43 & 5.04 & 9.59 & 0.35 & 0.66 \\
            J1145+5742\(^a\) & >10529 & >148 & 9.97e+43 & 2.31 & 17.09 & 0.74 & 0.71 \\
            J1232+0912 & 6016 & 119 & 5.40e+43 & 6.83 & 6.17 & 0.72 & 0.64 \\
            J1451+0132\(^a\) & >9970 & >148 & 2.29e+44 & 0.00 & 4.00 & 0.43 & 0.59 \\
            J1451+2338\(^a\) & >11645 & >157 & 3.80e+44 & 5.29 & 2.87 & 0.57 & 0.59 \\
            J1652+1728\(^a\) & >8596 & >128 & 2.14e+44 & 2.30 & 3.72 & 0.46 & 0.49 \\
            J1705+2736 & >6703 & >122 & 2.37e+43 & 4.81 & 23.15 & 0.91 & 0.78 \\
            J2215\(-\)0056 & 3812 & 114 & 9.78e+42 & 28.91 & 21.50 & 0.87 & 0.85 \\
            J2254+2327 & 2181 & 80 & 9.36e+42 & 8.18 & 8.41 & 0.65 & 0.67 \\
            J2323\(-\)0100 & >3640 & >137 & 3.08e+42 & 9.97 & 19.95 & 0.86 & 0.65 \\
            Median & >6702 & >128 & 5.40e+43 & 5.04 & 6.50 & 0.65 & 0.66 \\
            \hline
        \end{tabular}
        \label{tab:table_SB_maps}
    \end{table}
    
\end{landscape}

\subsection{Sample Properties}
\label{sec:sec_sample_properties}
Table \ref{tab:table_catalogue} lists basic properties of the sample, measured redshift of the Ly\(\alpha\) emission, and details of the observations. We include the SDSS BOSS catalog emission-line redshift measurements $z_{em}$, and $z_{halo}$ computed from the extended Ly$\alpha$ halo emission centroid, without the quasar emission. We include C IV FWHM, measured from C IV~$\lambda$1549 emission-line profile fitting done in \citet{Hamann+17}, and when available [O III]~$\lambda$5007 emission blueshift from \citet{Perrotta+19}. We include catalog magnitudes and color ($W3$ and $i-W3$), and our computed bolometric luminosities ($L_{bol}$). We apply galactic extinction corrections to all luminosity and SB measurements.

ERQs are heavily extincted in the visible and UV, but the amount of extinction is difficult to determine. In \cite{Hamann+17} it is estimated that the median SED is suppressed in ERQs by three magnitudes in the rest-frame UV, in comparison to normal blue quasars. Uncertain extinction in ERQs also makes bolometric luminosities difficult to determine. Therefore, we assume ERQs have intrinsic SEDs like typical blue quasars, and use their measured WISE \(W\)3 fluxes (assumed to be unaffected by extinction) to estimate the bolometric luminosities, similar to the ERQ luminosities computed by \citet{Perrotta+19}. 

We compare the medians quantities from our sample to medians from other surveys of blue quasars, discussion of these comparisons are in Section \ref{sec:sec_compare}. One quasar, J0834+0159, was observed under cloudy conditions, and was omitted from analysis of Ly\(\alpha\) halos. We do not have detection of other extended emission lines such as C IV~$\lambda$1549 or He~II~$\lambda$1640 except for J0006+1215, which is discussed in L22.

\subsection{Halo Detection}
We used an optimal extraction algorithm for detecting diffuse emission in Ly\(\alpha\) instead of pseudo narrowband images to obtain morphology and kinematics of the halo. This algorithm takes into account pixels of the three dimensional data cube, or ``voxels,'' that are of good signal to noise (S/N > 2) and connected at sides, edges, or corners to each other. Our algorithm is described with more detail in L22, which followed algorithms in \citet{Borisova+16a, ArrigoniBattaia+19, Cai+19, Farina+19}. Our surface brightness's $1\sigma$ limit, and root-mean-square value, are measured in a 1-arcsec aperture for a single 1-angstrom channel. Surface brightness maps probe the halo Ly$\alpha$ emission down to a median $1\sigma$ SB limit of 2.16~$\times$~$10^{-19}$ erg~s$^{-1}$~cm$^{-2}$~arcsec$^{-2}$, across spatial scales from 40~kpc to $>80$~kpc from the quasars. Surface brightness root-mean-square has median of 2.26~$\times$~$10^{-19}$~erg~s$^{-1}$~cm$^{-2}$~arcsec$^{-2}$, for a channel in rest-frame wavelengths between 1255 and 1275 \AA, and where there are no prominent quasar emission lines.

\section{Measurements \& Results} 
\label{sec:measurements}

To investigate the environment of ERQs, we measure basic properties of the Ly$\alpha$ halo emission, such as surface brightness, linear size, and velocity dispersion. Peculiarities in halo properties compared to blue quasars may be evidence ERQ inhabit a different quasar evolutionary phase.

Panels in Figure~\ref{fig:fig_maps} show, from left to right, halo surface brightness, 1st velocity moment of the line flux distribution (i.e. velocity shift), velocity dispersion, circularly-averaged surface brightness (SB) radial profile, and spatially-integrated spectrum, for 11 of the 12 observed ERQs. J0834+0159 was observed under cloudy conditions, and is not included in the Figure~\ref{fig:fig_maps} and subsequent tables. Tables 2 - 4 list a variety of properties measured from these maps.

\subsection{Morphology \& Brightness Maps}
\label{sec:SB_maps}

Morphology measurements such as extent and asymmetry may give insights into the gaseous environment of the quasar, and could be influenced by large scale structure, the presence of merger activity, and quasar luminosity.

Morphology of ERQ Ly$\alpha$ halos varies across the sample, with some halos extending to the edge of the FOV (>~60-70 kpc), and others compact near the central quasar ($\sim$40 kpc). These Ly$\alpha$ halos also show varying asymmetry. We present Ly\(\alpha\) SB maps in the first column of Figure~\ref{fig:fig_maps}. One unique aspect of this project is showing continuous halo emission down to zero projected distance from the quasar, and thus the six ERQs with Ly$\alpha$ spikes do not show a hollow due to our quasar PSF subtraction (Section \ref{sec:sec_narrow_lya}).

Table \ref{tab:table_SB_maps} presents the measured projected distance from the quasar to the Ly\(\alpha\) halo emission peak, and to the halo centroid. The median distance from quasar to peak in our sample is $r_\text{peak} = 5.0$~kpc, and the median distance from quasar to halo centroid is $r_\text{centroid} = 6.5$~kpc. Uncertainty in these projected distances is about half a pixel in size, or $\sim$1~kpc. The position of the Ly\(\alpha\) halo centroid is used in analysis of spatial asymmetry in this section, and the position of the halo peak is used in analysis of aperture kinematics in Section \ref{sec:sec_vel_maps}.

Six of the halos extend beyond the FOV, and $> 55$~kpc from the central quasar. Our sample has a median maximum linear size $> 128$~kpc, and median halo luminosity 5.40~$\times$~$10^{43}$~erg~s$^{-1}$. 

We quantify the asymmetry level of the morphology of the Ly\(\alpha\) halo with the elliptical eccentricity parameters $e_{\text{weight}}$ and $e_{\text{unweight}}$ (identical to L22). We define $e_{\text{weight}}$ using flux-weighted second-order spatial moments with respect to the Ly\(\alpha\) halo centroid, and follow the formula in \citet{OSullivanChen20}. $e_{\text{unweight}}$ uses flux-unweighted spatial moments with respect to the quasar position. Values of $e \approx 0$ correspond to circular morphologies, and values near 1 correspond to being more elliptical. Parameter $e_{\text{weight}}$ is defined as $e_\text{weight} = \sqrt{1 - \alpha_\text{weight}^{2}}$, where the flux-weighted eccentricity parameter $\alpha_\text{weight}$ is defined in \citet{ArrigoniBattaia+19} or \citet{Cai+19}.  $e_{\text{weight}}$ tends to characterize central regions of the halo with high SB. Conversely, $e_{\text{unweight}}$ is defined as $e_\text{unweight} = \sqrt{1 - \alpha_\text{unweight}^{2}}$, where the flux unweighted eccentricity parameter $\alpha_\text{unweight}$ is defined in \citet{denBrok+20} or \citet{Mackenzie+21}.  $e_{\text{unweight}}$  better describes the large-scale shape of the diffuse halo surrounding the quasar position. 

Large differences between $e_{\text{weight}}$ and $e_{\text{unweight}}$ could indicate significant changes from inner halo to extended regions that could be affected by filamentary structures. For example, J1145+5742 has a luminous Ly$\alpha$ halo that extends East from the quasar position, and is visibly asymmetric about the halo centroid, with a measured $e_{\text{weight}} \approx 0.74$. Asymmetric extending of the halo away from the quasar position is also reflected in the large $e_{\text{unweight}} \approx 0.71$. An unusual case is J2323$-$0100, which displays an asymmetric halo with a patch of emission to the North-West that causes the Ly$\alpha$ halo centroid to be offset. This offset away from the luminous bulk of halo emission results in a large $e_{\text{weight}} \approx 0.86$. Considering the more diffuse mission around the quasar position, the asymmetry is more moderate, and thus has lower $e_{\text{unweight}} \approx 0.65$. Median eccentricities for the optimally extracted halos (11 objects) are $e_{\text{weight}} = 0.65$ and $e_{\text{unweight}} = 0.66$. $e_{\text{unweight}}$ values larger than $e_{\text{weight}}$ could reflect the sample generally having a more circularly symmetric inner halo, where the flux is strongest, versus the slightly more asymmetric outer halo. These median eccentricity values are comparable to each other, but are given further context by comparing values to other quasar samples in Section \ref{sec:sec_compare}.

To characterize the radial extent of the Ly$\alpha$ halo, we calculate the circularly averaged SB radial profiles for the ERQ sample (Figure~\ref{fig:fig_maps}, 4th column). While optimally extracted line maps yield measurements of morphology and kinematics, they do not allow direct comparison with other samples because other surveys used pseudo-narrowband imaging to compute their radial profiles. Therefore, we choose to generate a pseudo-narrowband image with fixed width to recover all possible fluxes in extended regions. Using a similar method to other studies, we adopt a fixed wavelength width of $\pm 1,000$~km~s$^{-1}$ centered at the Ly\(\alpha\) rest wavelength. For the full sample of ERQs, the median SB in the innermost annulus (2$-$4 kpc) is 1.53~$\times 10^{-14}$~erg~s$^{-1}$~cm$^{-2}$ arcsec$^{-2}$, and 1.61~$\times 10^{-16}$~erg~s$^{-1}$~cm$^{-2}$ arcsec$^{-2}$ in the most distant annulus (32$-$63 kpc). Six of the 11 ERQs which have modified PSF subtraction, due to the presence of a Ly$\alpha$ spike in their spectra, have an inner annulus (2$-$4 kpc) median SB 11.4~$\times 10^{-14}$ erg~s$^{-1}$~cm$^{-2}$ arcsec$^{-2}$, and most distant annulus (32$-$63 kpc) median SB 4.69 $\times 10^{-16}$ erg s$^{-1}$~cm$^{-2}$ arcsec$^{-2}$. We calculate the averaged SB at each radial distance measured by annuli centered on the quasar position for each ERQ, as well as the full sample median (Table \ref{tab:table_rad_profile}). We compute a full sample SB median omitting the inner region data in order to compare these ERQ radial profiles with the spike to other samples, which cannot perform modified PSF subtraction. To demonstrate the potential difference between our quasars and other samples, we compute the median radial profile of the sub-sample of six ERQs, where a Ly$\alpha$ spike is present in the spectra and modified PSF subtraction was performed. SB of ERQs with detected Ly\(\alpha\) emission increases monotonically as the radial distance decreases, indicating that the innermost regions are the most luminous part of these halo. ERQs with the Ly$\alpha$ spike emission demonstrate the most centrally concentrated emission.

We use an exponential fit to the binned radial profiles, SB$_{\text{Ly}\alpha} (r) = C_{e} \text{exp} (-r/r_{h})$, where $C_{e}$ is the normalization, $r$ is the projected distance from the quasar, and $r_h$ the scale length of the profile. Figure~\ref{fig:fig_maps} shows the fits of this model in the fourth column, and appear to reasonably describe the the radial SB profile. For individual ERQs, a steep decline in SB near their inner regions (<4kpc) appear only for ERQs with quasar PSF subtraction that did not involve clipping the Ly\(\alpha\) spike. Offset between the profile fit and data in the outer regions (>32kpc, eg. J1451+0132 and J1652+1728) are generally from ERQs with an asymmetric shape in the Ly\(\alpha\) halo at large scales. These profiles are better fit by exponential functions than power laws, and show that SB steeply declines at large distances from the quasar. Table~\ref{tab:table_rad_profile} presents averaged SB at each annular bin and exponential fit parameters. Our full sample median exponential scale length is $r_h = 9.4$~kpc, and the median of the six ERQ with a Ly\(\alpha\) spike have $r_h = 8.7$~kpc. Generally, ERQ Ly$\alpha$ halos are more centrally concentrated than blue quasar samples, and ERQs that have the Ly$\alpha$ spike are the most compact of our sample. Section \ref{sec:sec_compare} will have further discussion of quasar population comparisons.

\begin{landscape}
    \begin{table}
	    \centering
	    \caption{Our sample's Ly\(\alpha\) halo circularly averaged surface brightness radial profiles. The first five columns make up the data points for radial profiles in Figure~\ref{fig:fig_maps}. For modeling the radial profiles, the inner radius (2-4)kpc is not included when there is no inner halo clipped from the quasar PSF spectral-template, shown in parentheses. The last two columns are the parameters from exponential-profile fit to the radial profile data. Median-errors shown are the standard deviation of the values that compose the median.}
	    \begin{tabular}{lccccccc} 
	        \hline
	        ERQ Name & Radii Brightness & Radii Brightness & Radii Brightness & Radii Brightness & Radii Brightness & Exponential Amplitude & Exponential Scale Length \\
	        & (2-4) kpc & (4-8) kpc & (8-16) kpc & (16-32) kpc & (32-63) kpc & & \\
	        & (erg/s/cm\(^{^2}\)/arcsec\(^{^2}\)) & (erg/s/cm\(^{^2}\)/arcsec\(^{^2}\)) & (erg/s/cm\(^{^2}\)/arcsec\(^{^2}\)) & (erg/s/cm\(^{^2}\)/arcsec\(^{^2}\)) & (erg/s/cm\(^{^2}\)/arcsec\(^{^2}\)) & (erg/s/cm\(^{^2}\)/arcsec\(^{^2}\)) & (kpc) \\
	        \hline
	        J0006+1215\(^a\) & 1.63e-14 & 1.22e-14 & 5.88e-15 & 1.48e-15 & 1.78e-16 & (2.36 \(\pm\) 0.04)e-14 & 8.6 \(\pm\) 0.1 \\
	        J0220+0137\(^a\) & 1.43e-14 & 1.28e-14 & 7.68e-15 & 1.93e-15 & $-$4.76e-17 & (2.50 \(\pm\) 0.05)e-14 & 9.4 \(\pm\) 0.1 \\
	        J1145+5742\(^a\) & 2.16e-14 & 1.60e-14 & 7.09e-15 & 2.49e-15 & 7.98e-16 & (2.43 \(\pm\) 0.02)e-14 & 10.8 \(\pm\) 0.1 \\
	        J1232+0912 & (1.03e-15) & 9.25e-15 & 6.75e-15 & 2.60e-15 & 1.61e-16 & (1.77 \(\pm\) 0.11)e-14 & 11.9 \(\pm\) 0.5 \\
	        J1451+0132\(^a\) & 2.10e-13 & 8.23e-14 & 1.53e-14 & 3.85e-15 & 3.13e-16 & (4.68 \(\pm\) 0.02)e-13 & 3.51 \(\pm\) 0.01 \\
	        J1451+2338\(^a\) & 8.46e-14 & 8.59e-14 & 4.01e-14 & 9.42e-15 & 1.22e-15 & (1.57 \(\pm\) 0.02)e-13 & 8.7 \(\pm\) 0.1 \\
	        J1652+1728\(^a\) & 1.46e-13 & 7.26e-14 & 1.92e-14 & 4.96e-15 & 6.25e-16 & (2.41 \(\pm\) 0.03)e-13 & 5.01 \(\pm\) 0.04 \\
	        J1705+2736 & (6.69e-15) & 2.39e-15 & 1.50e-15 & 7.85e-16 & 7.91e-17 & (3.82 \(\pm\) 0.52)e-15 & 13.9 \(\pm\) 1.3 \\
	        J2215\(-\)0056 & (1.23e-16) & 1.33e-15 & 6.67e-16 & 4.69e-16 & 8.05e-17 & (1.64 \(\pm\) 0.17)e-15 & 16.9 \(\pm\) 1.4 \\
	        J2254+2327 & (3.15e-15) & 7.91e-15 & 2.24e-15 & 5.39e-16 & $-$2.10e-16 & (2.07 \(\pm\) 0.22)e-14 & 5.7 \(\pm\) 0.3 \\
	        J2323\(-\)0100 & ($-$1.59e-16) & 4.14e-16 & 3.77e-16 & 1.28e-16 & $-$2.48e-17 & (1.02 \(\pm\) 0.19)e-15 & 10.5 \(\pm\) 1.4 \\
	        Median & 1.53e-14 & 1.22e-14 & 6.75e-15 & 1.93e-15 & 1.61e-16 & (2.36 \(\pm\) 0.14)e-14 & 9.4 \(\pm\) 3.7 \\
        \hline
        \end{tabular}
        \label{tab:table_rad_profile}
    \end{table}
    
    \begin{table}
        \centering
        \caption{Halo kinematics measured from several methods from different columns in Figure~\ref{fig:fig_maps}. Measurements in the first column refers to the standard deviation of the 1st velocity-moments, computed from the emission line flux distribution at each spaxel. The second and third column refer to dispersion as computed by a single Gaussian fit to each Voronoi-bin in the velocity dispersion map. 2nd velocity-moments are not computed because there is not enough signal to reliably compute higher velocity-moments. The final column shows results from fitting a single Gaussian to the total integrated halo.}
        \begin{tabular}{lcccc}
            \hline
            ERQ Name & Spatial Standard Deviation of & Spatial Median of & Spatial Standard Deviation of & Spatially-Integrated \\
            & Velocity Centroid & Velocity Dispersion & Velocity Dispersion & Velocity Dispersion \\
            & (km~s\(^{-1}\)) & (km~s\(^{-1}\)) & (km~s\(^{-1}\)) & (km~s\(^{-1}\))  \\
            \hline
            J0006+1215\(^a\) & 238 & 275 & 73 & 293 \\
            J0220+0137\(^a\) & 456 & 259 & 44 & 264 \\
            J1145+5742\(^a\) & 288 & 491 & 124 & 526 \\
            J1232+0912 & 231 & 424 & 114 & 393 \\
            J1451+0132\(^a\) & 264 & 420 & 82 & 467 \\
            J1451+2338\(^a\) & 331 & 401 & 112 & 354 \\
            J1652+1728\(^a\) & 393 & 542 & 194 & 526 \\
            J1705+2736 & 424 & 251 & 108 & 235 \\
            J2215\(-\)0056 & 301 & 220 & 122 & 292 \\
            J2254+2327 & 195 & 374 & 216 & 261 \\
            J2323\(-\)0100 & 280 & 266 & 201 & 225 \\
            Median & 288 & 374 & 114 & 293 \\
            \hline
        \end{tabular}
        \label{tab:tab_dispersion}
    \end{table}

\end{landscape}

\begin{figure*}
\includegraphics[height=1.5in]{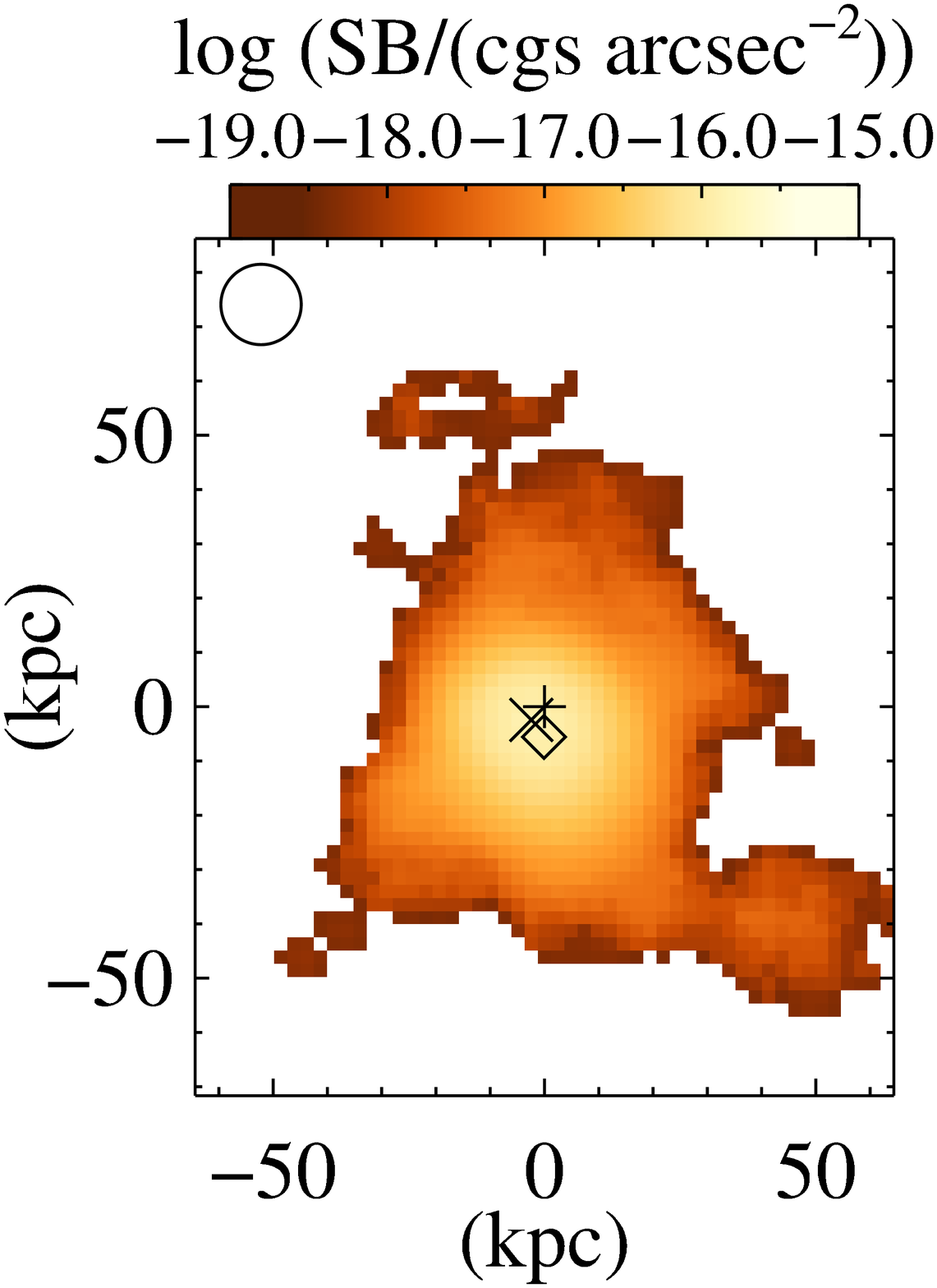}
\hspace{-0.01in}
\includegraphics[height=1.5in]{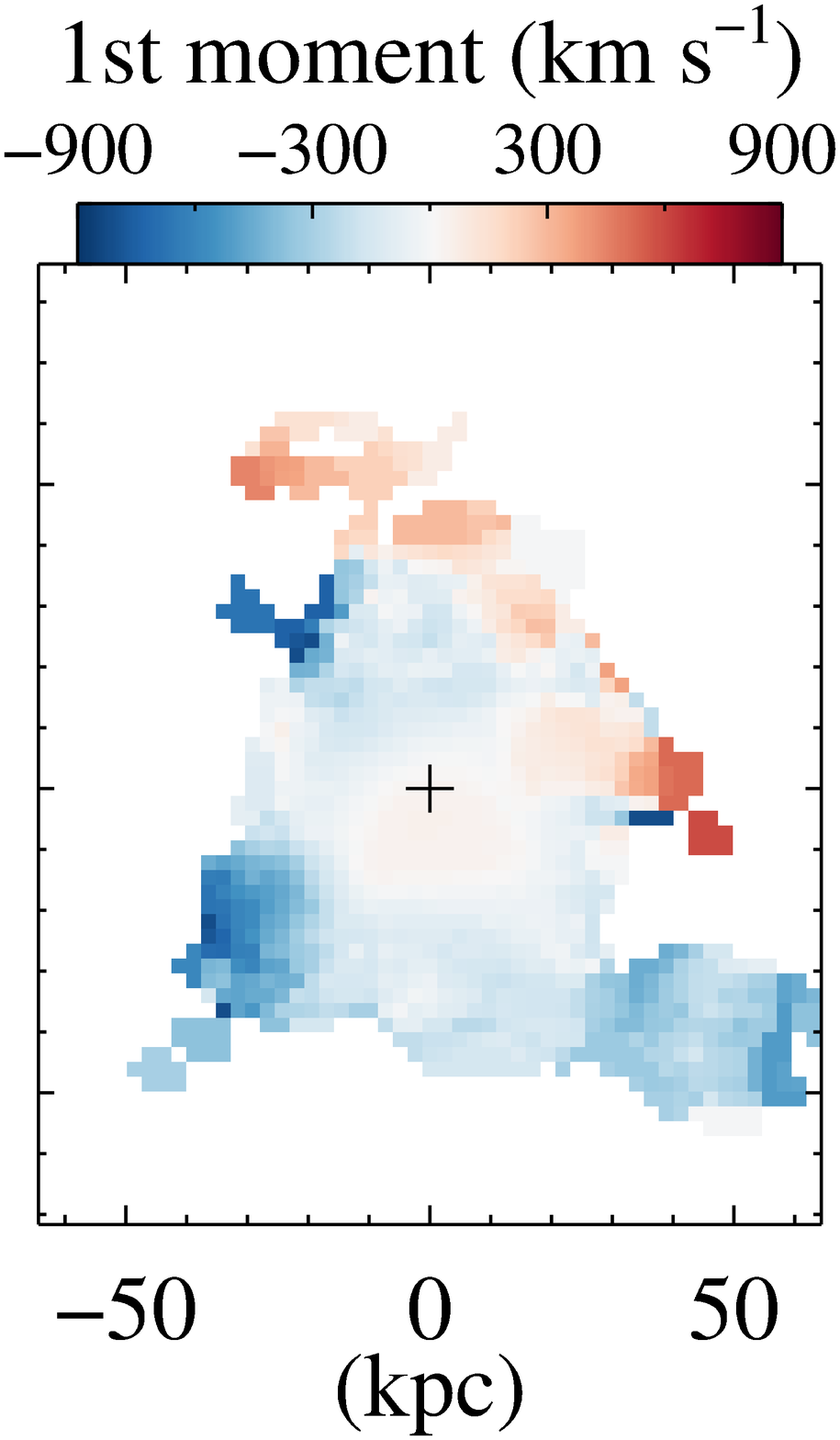}
\hspace{+0.01in}
\includegraphics[height=1.5in]{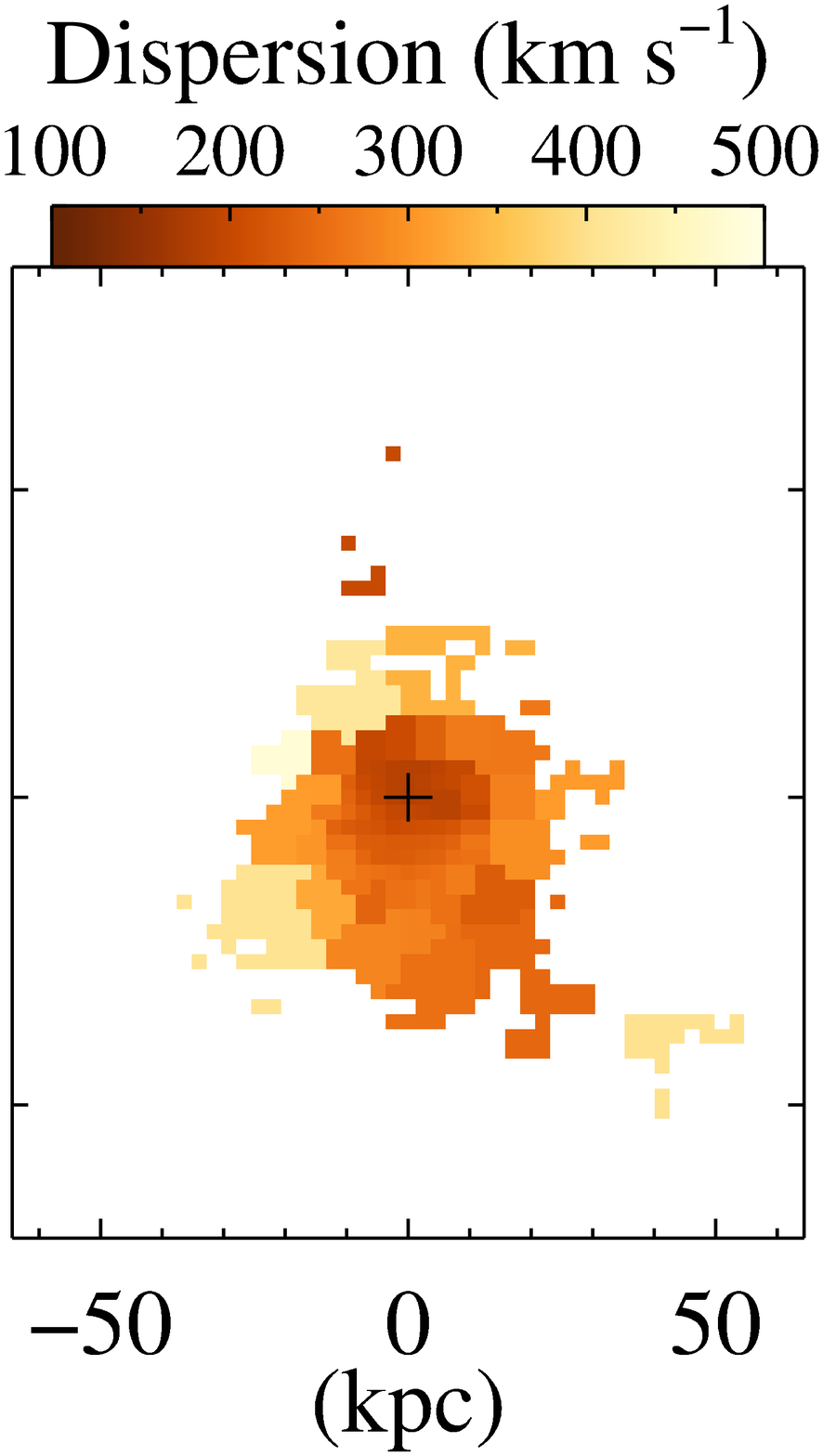}
\hspace{0.0in}
\includegraphics[width=0.27\textwidth]{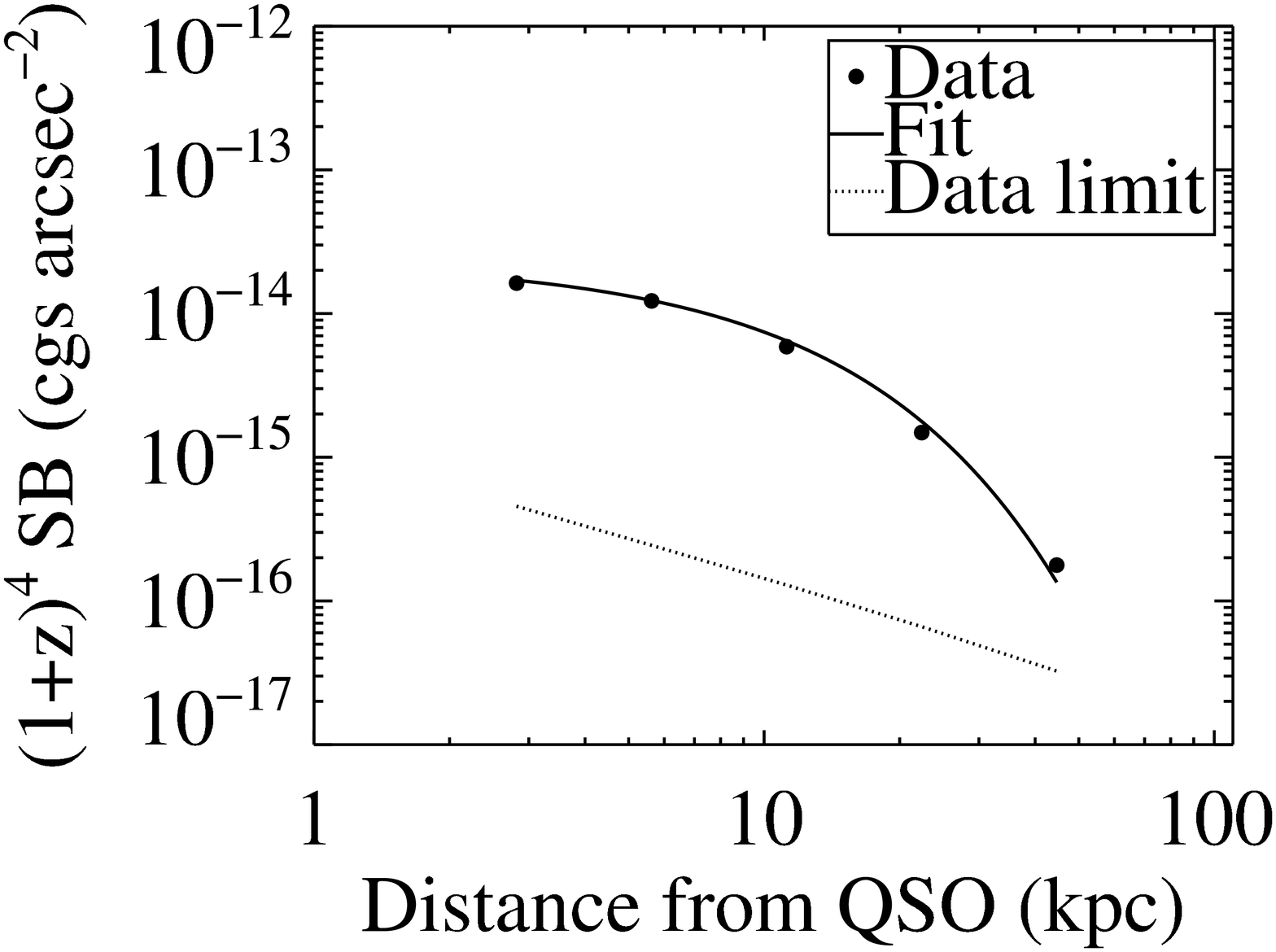}
\hspace{0.0in}
\includegraphics[width=0.27\textwidth]{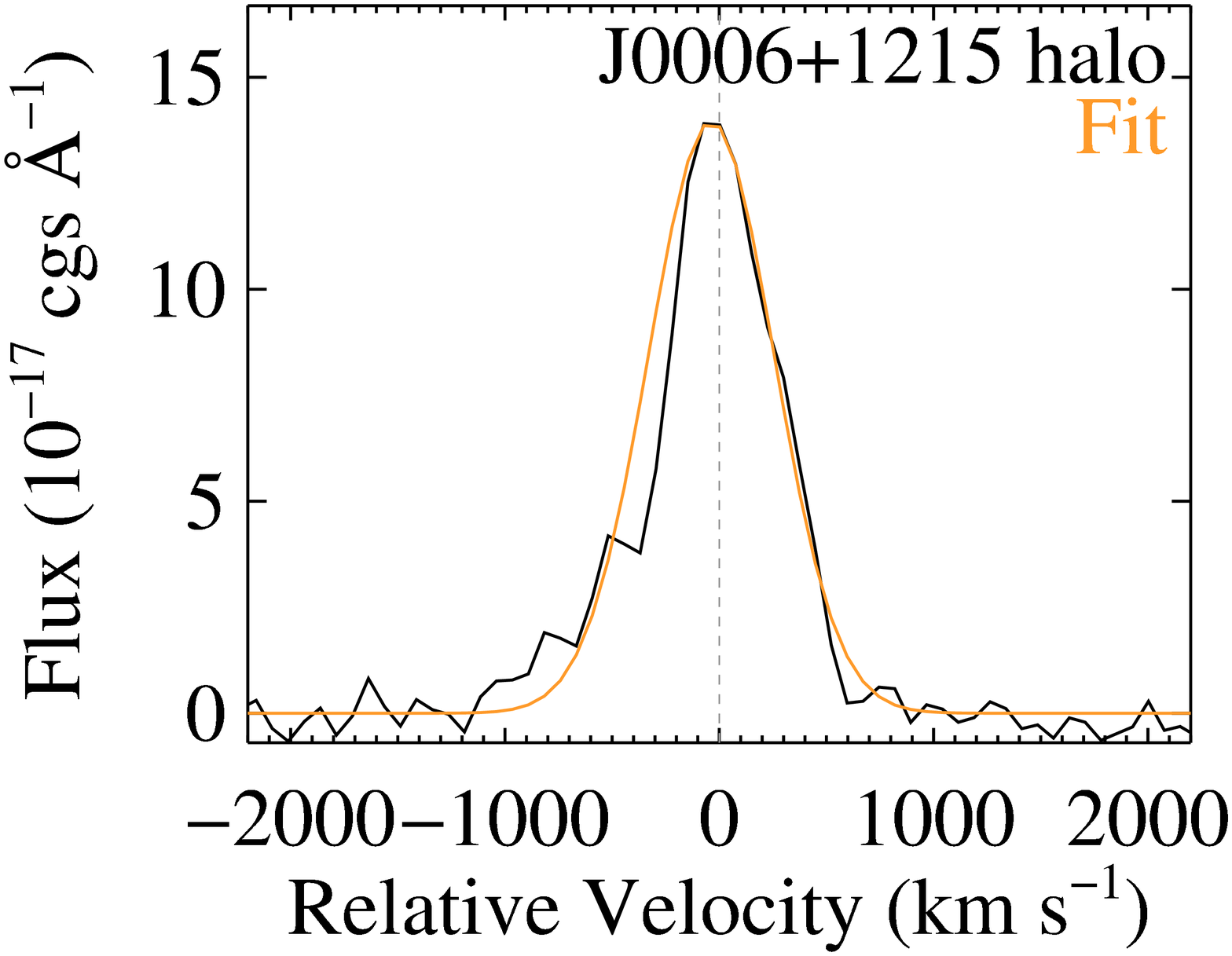}\\

\includegraphics[height=1.5in]{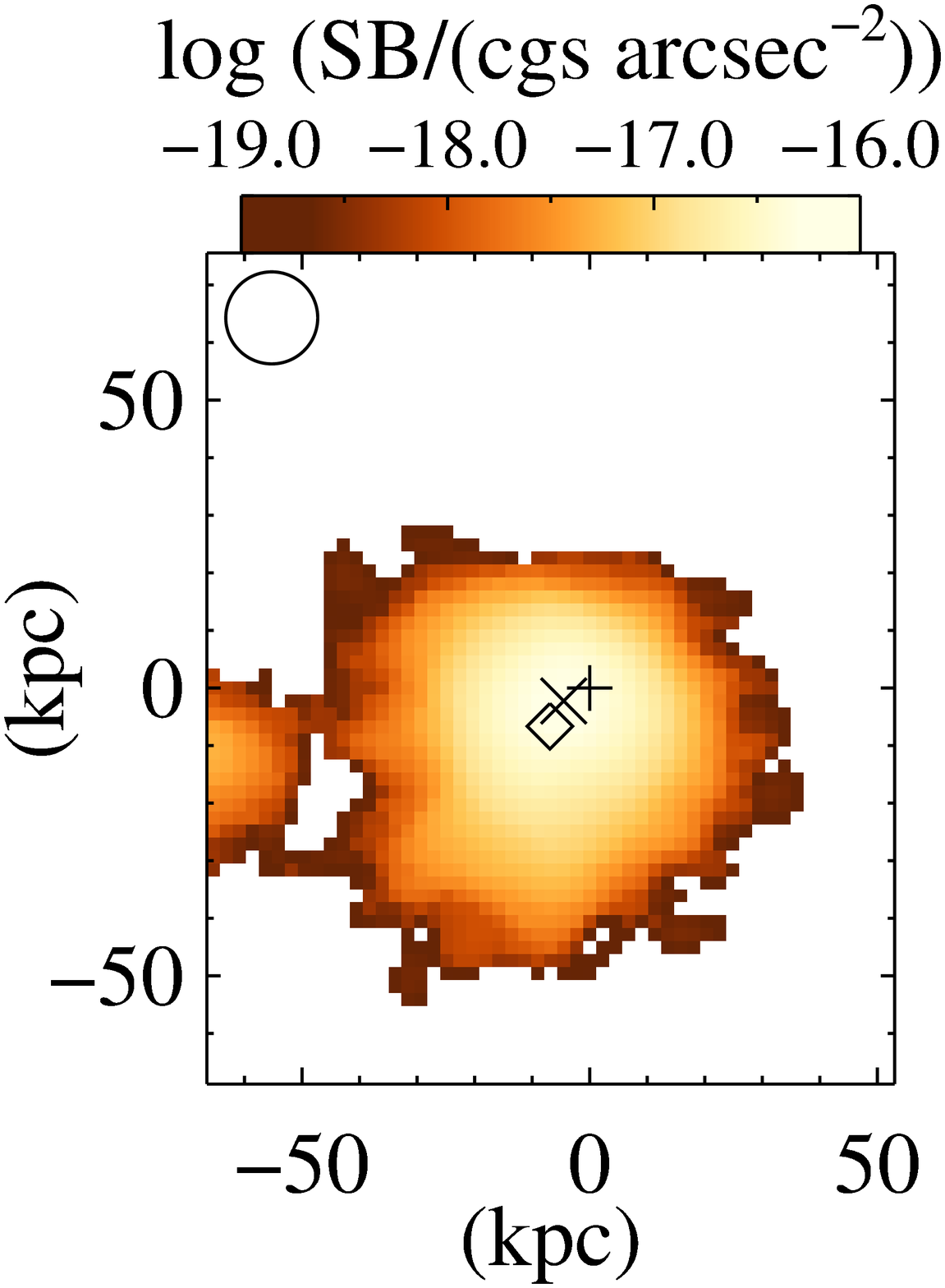}
\hspace{-0.01in}
\includegraphics[height=1.5in]{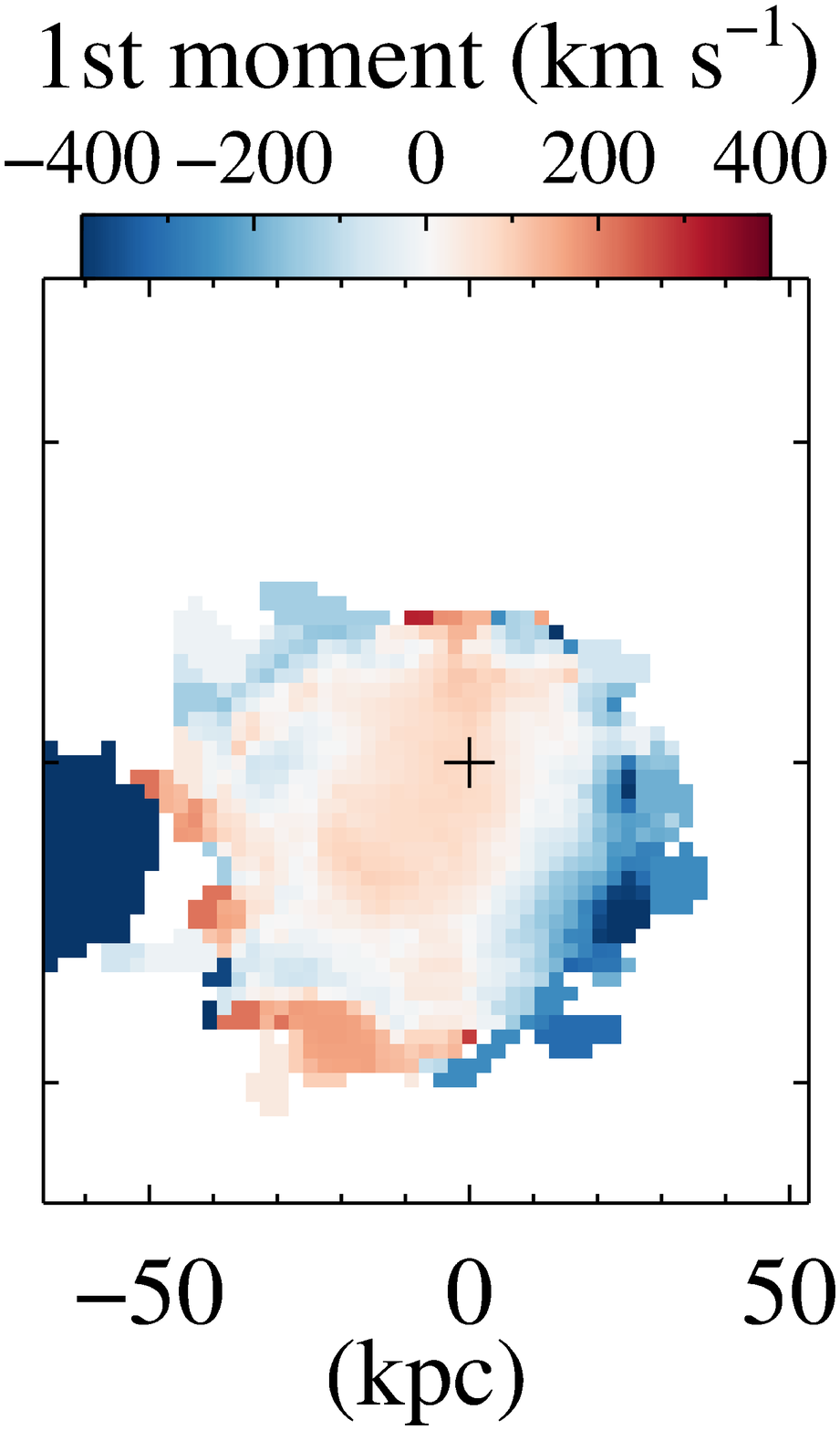}
\hspace{+0.01in}
\includegraphics[height=1.5in]{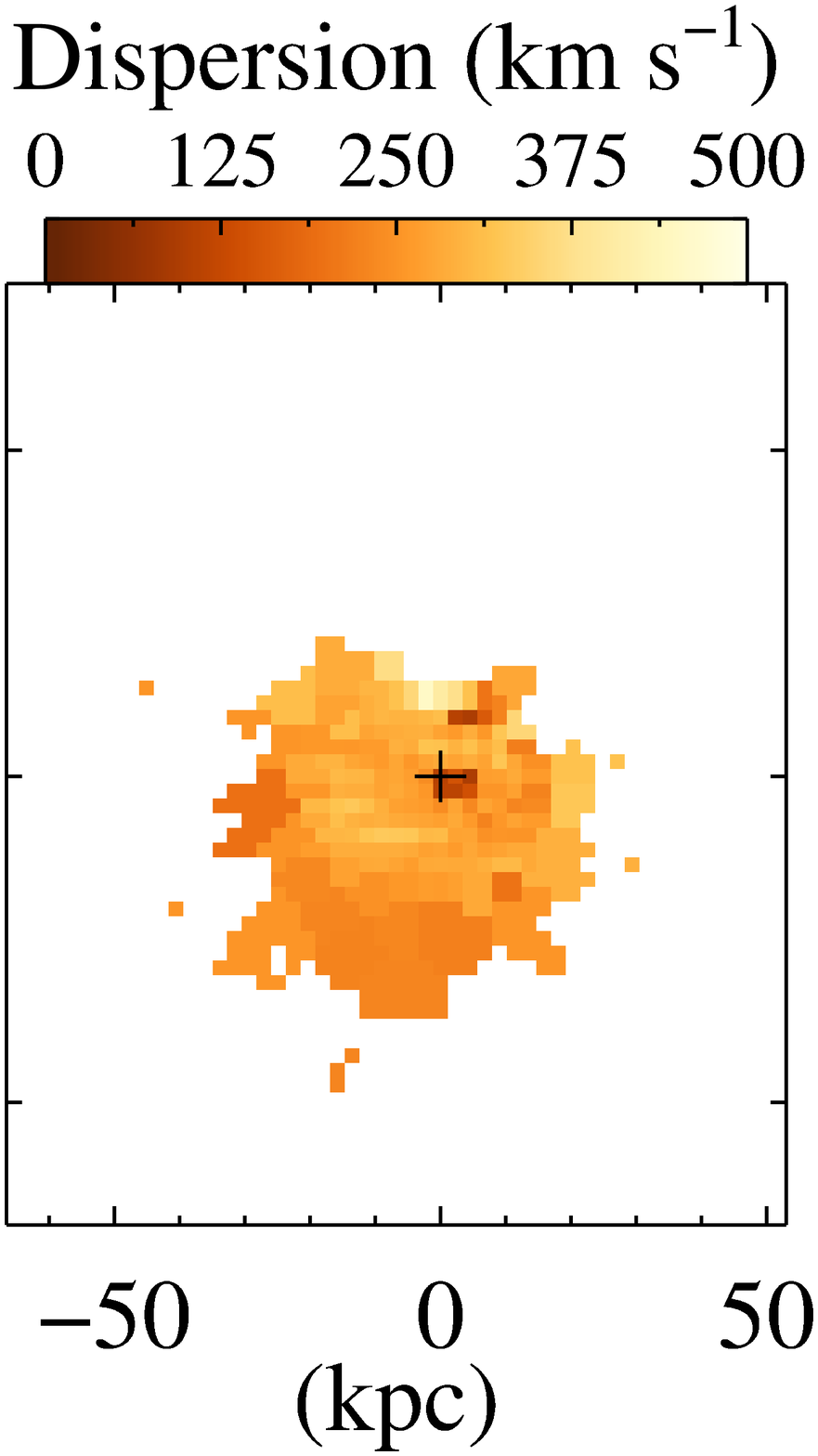}
\hspace{0.0in}
\includegraphics[width=0.27\textwidth]{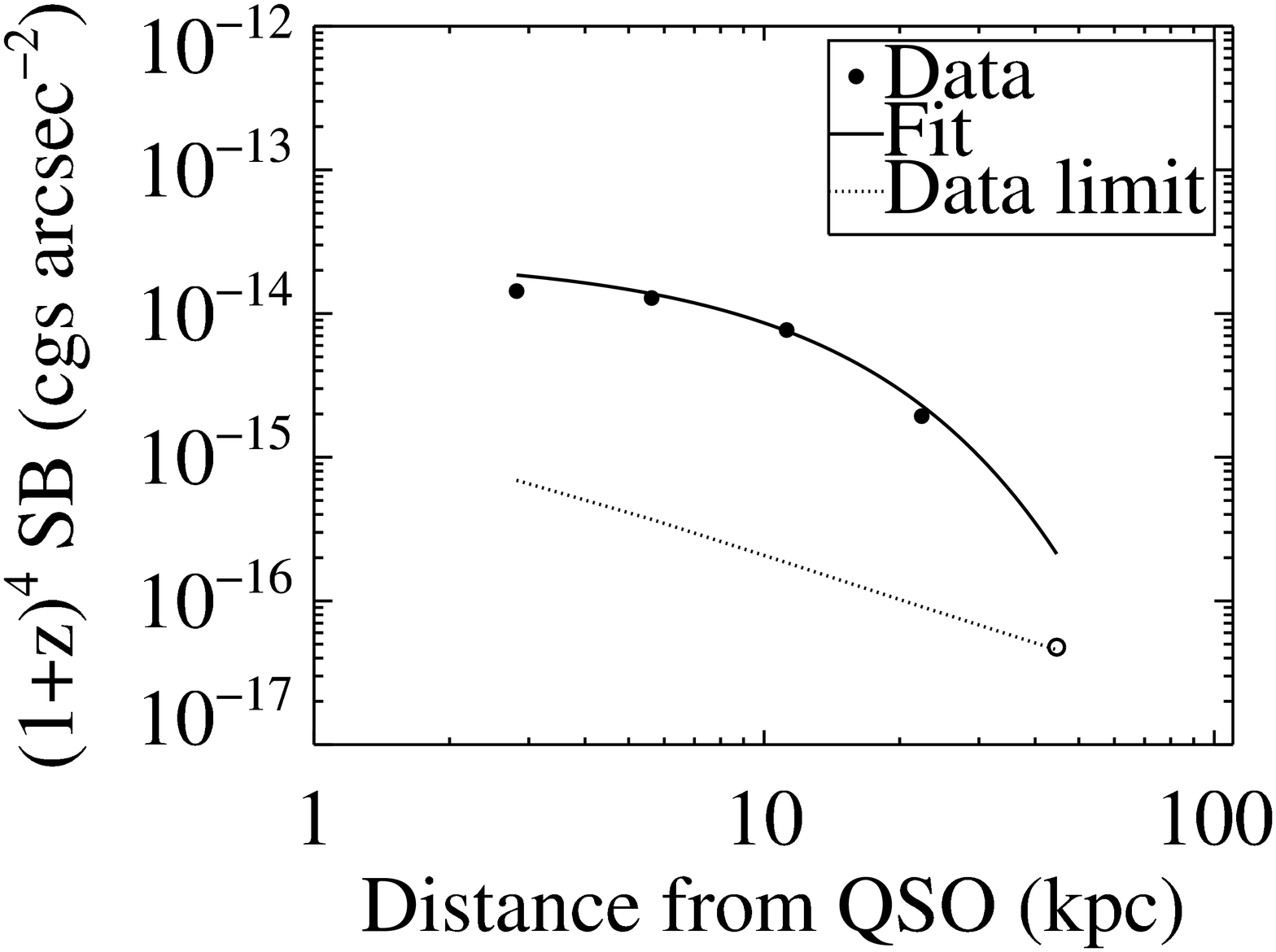}
\hspace{0.0in}
\includegraphics[width=0.27\textwidth]{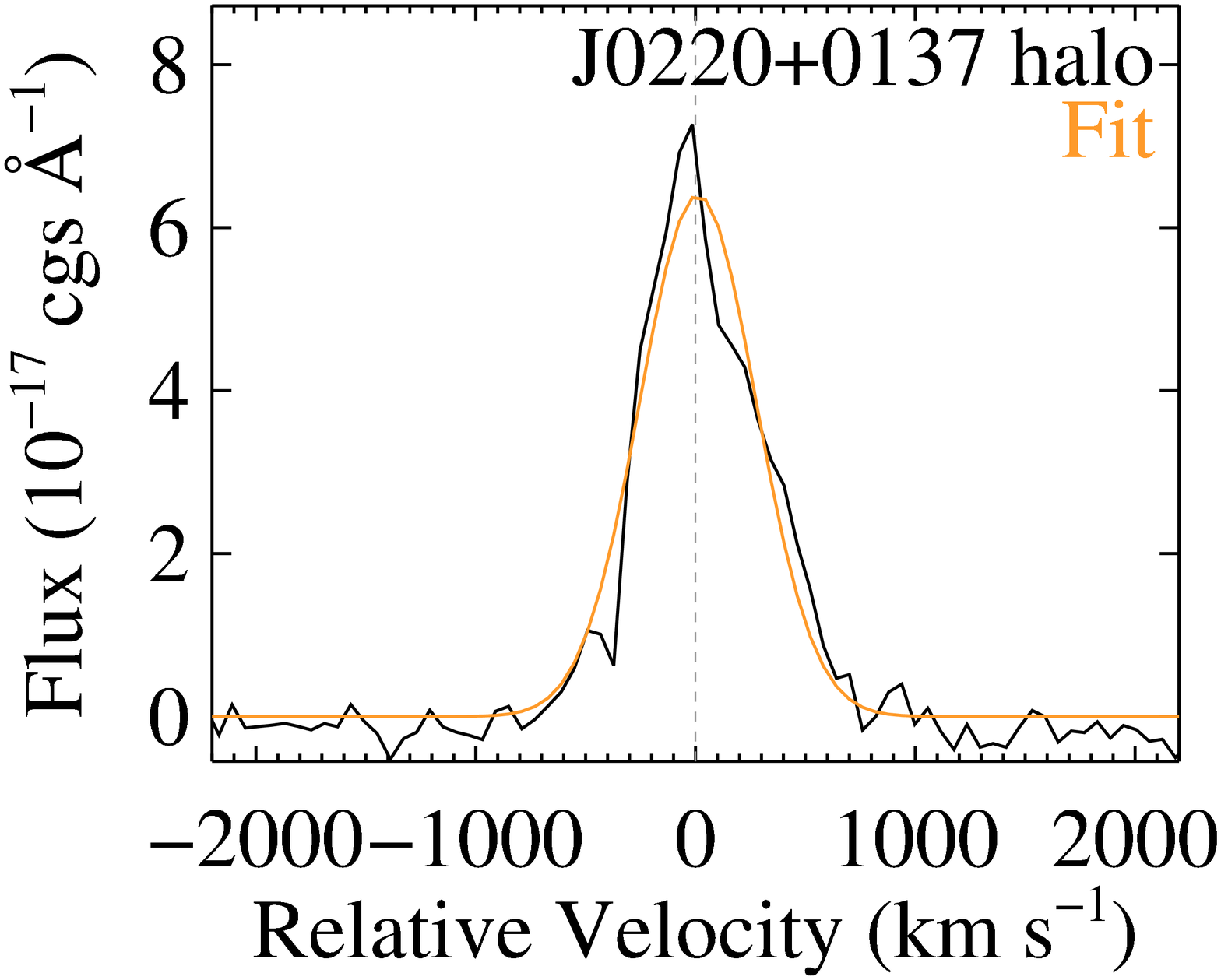}\\

\includegraphics[height=1.5in]{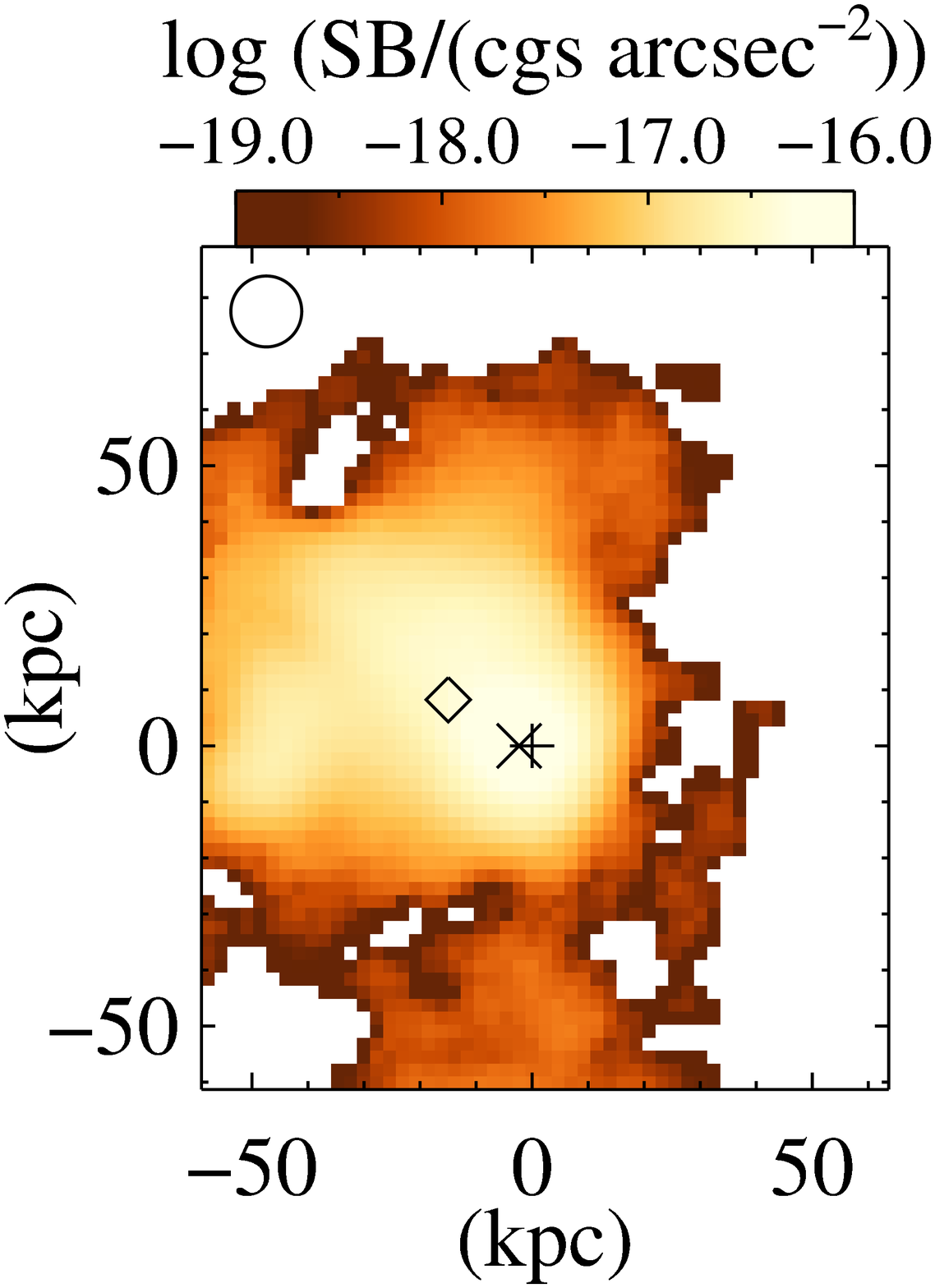}
\hspace{-0.01in}
\includegraphics[height=1.5in]{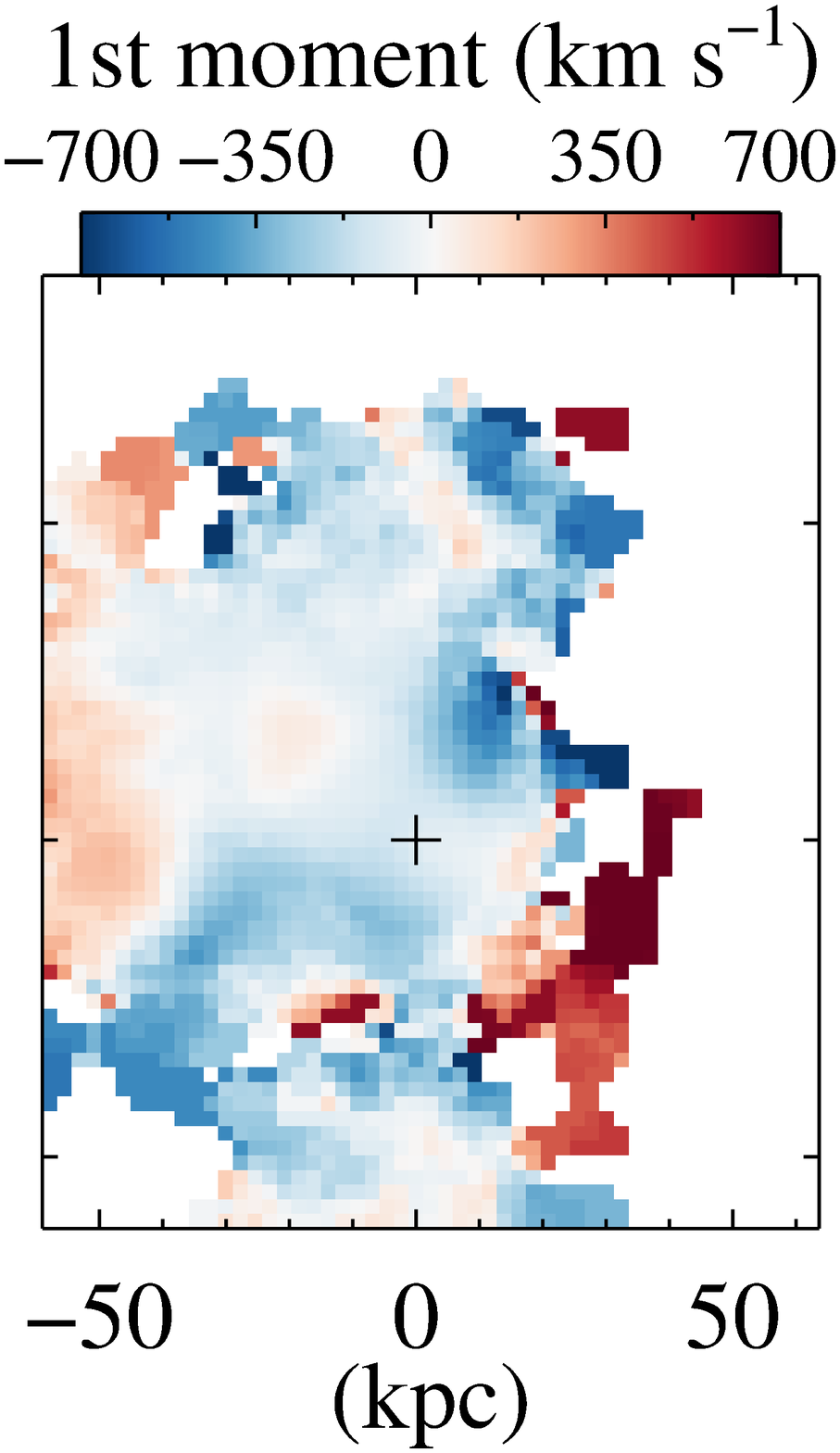}
\hspace{+0.01in}
\includegraphics[height=1.5in]{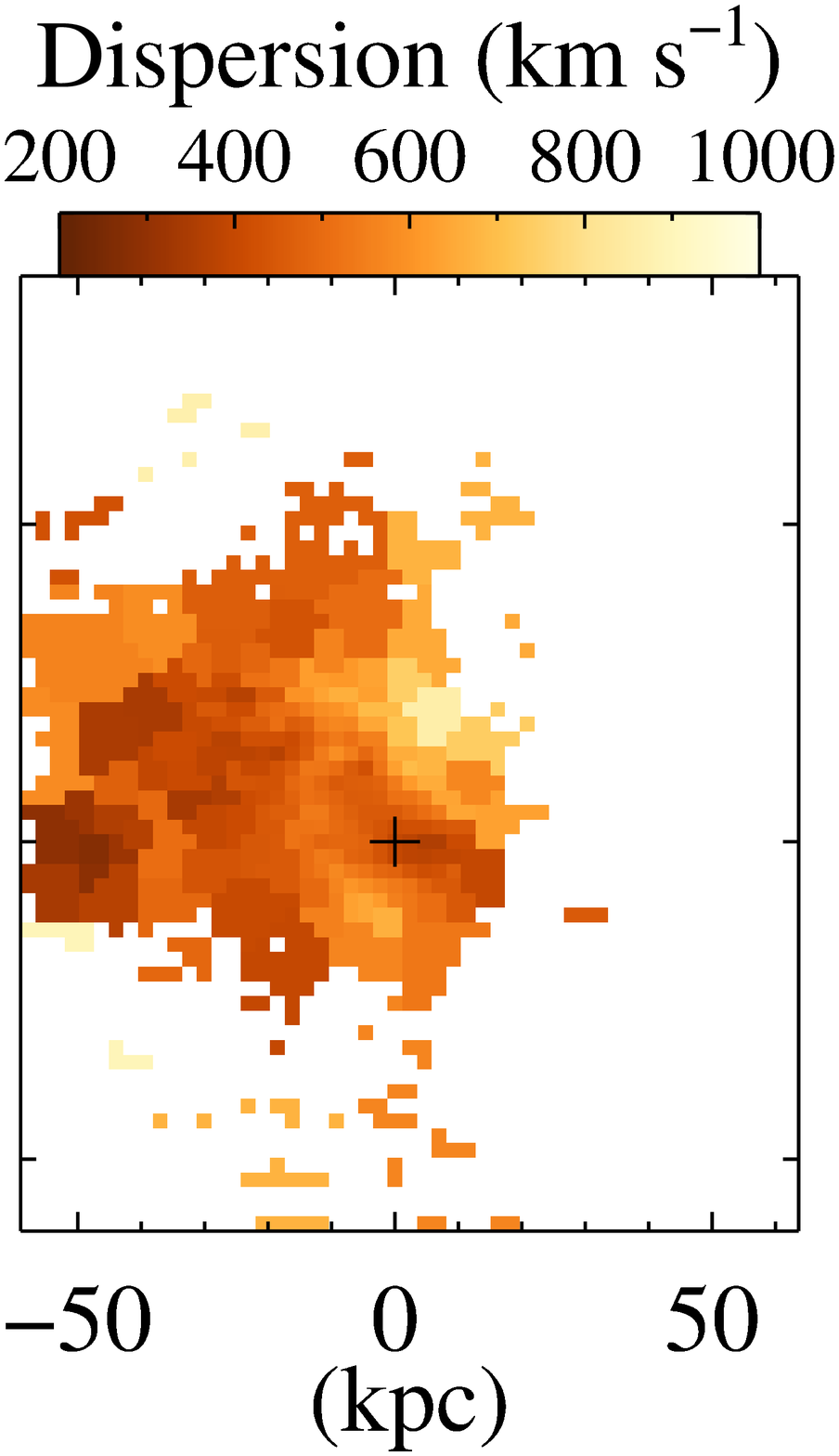}
\hspace{0.0in}
\includegraphics[width=0.27\textwidth]{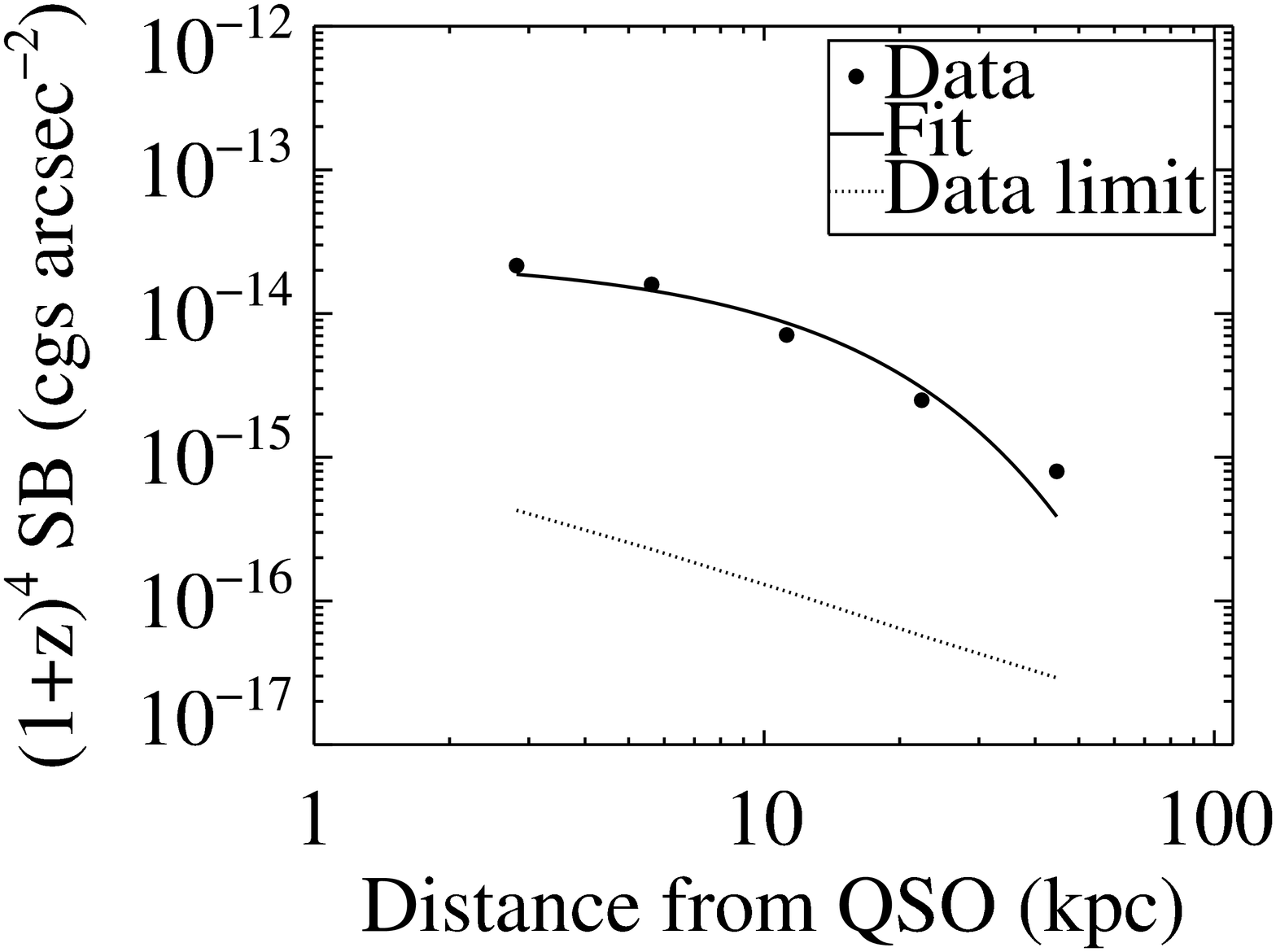}
\hspace{0.0in}
\includegraphics[width=0.27\textwidth]{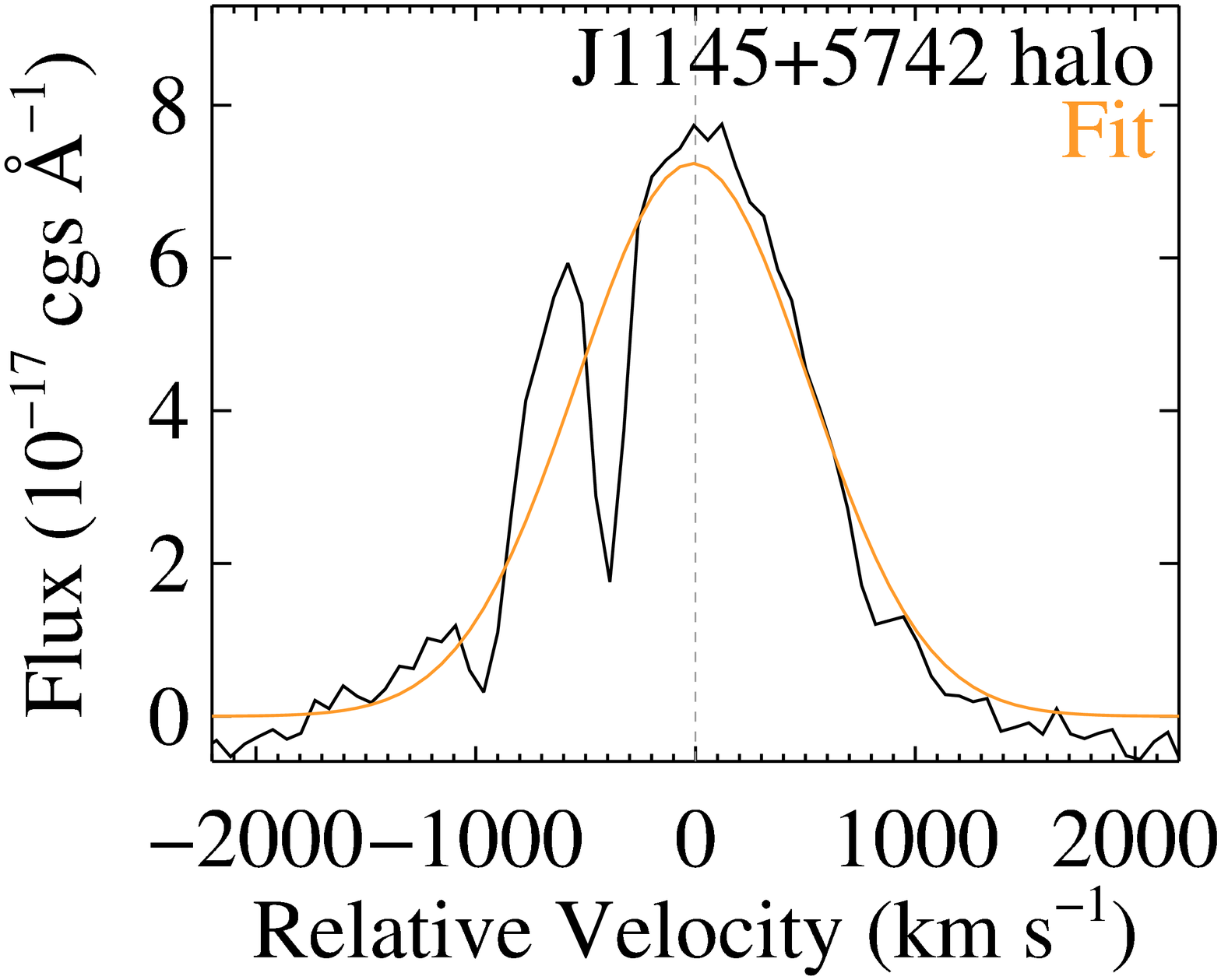}\\

\includegraphics[height=1.5in]{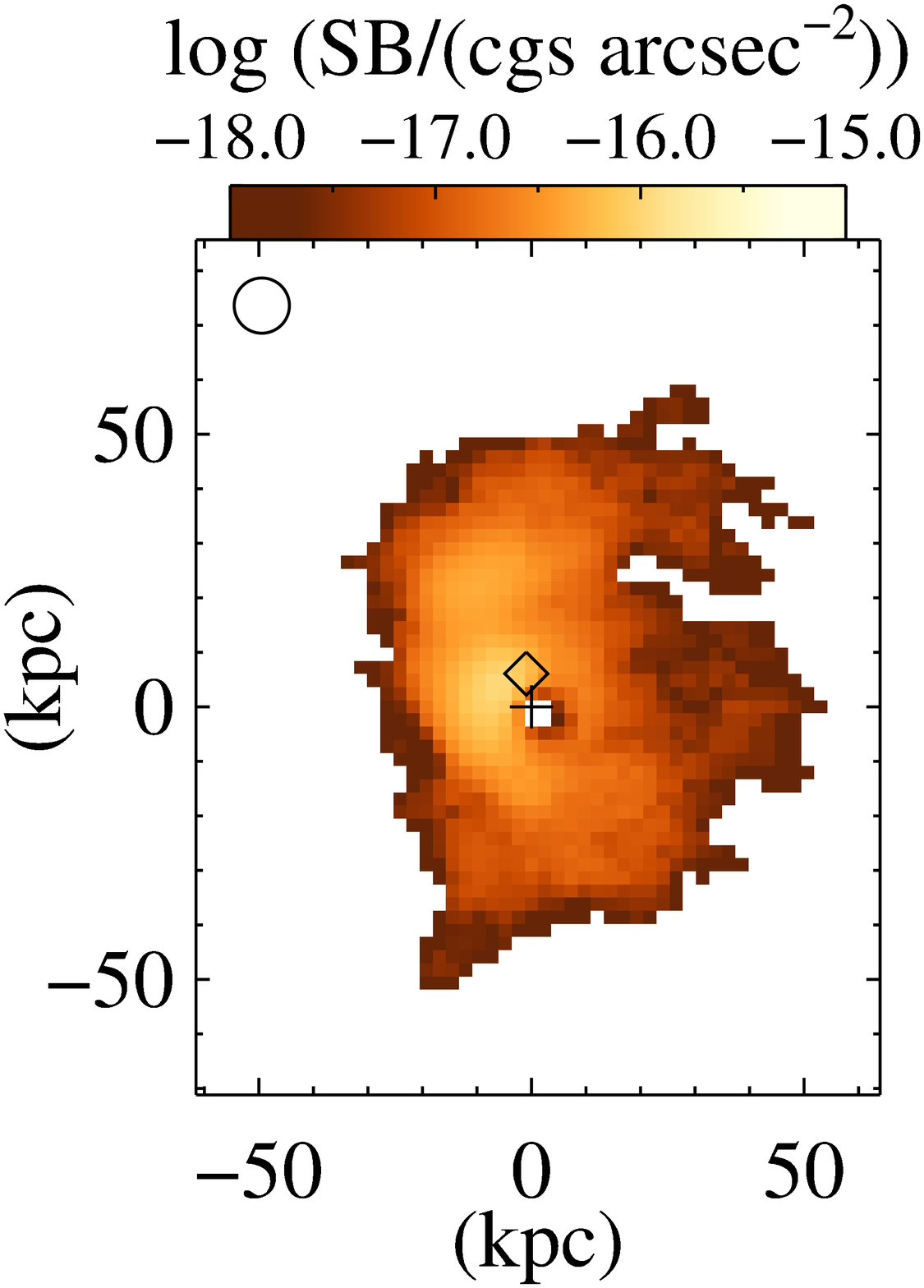}
\hspace{-0.01in}
\includegraphics[height=1.5in]{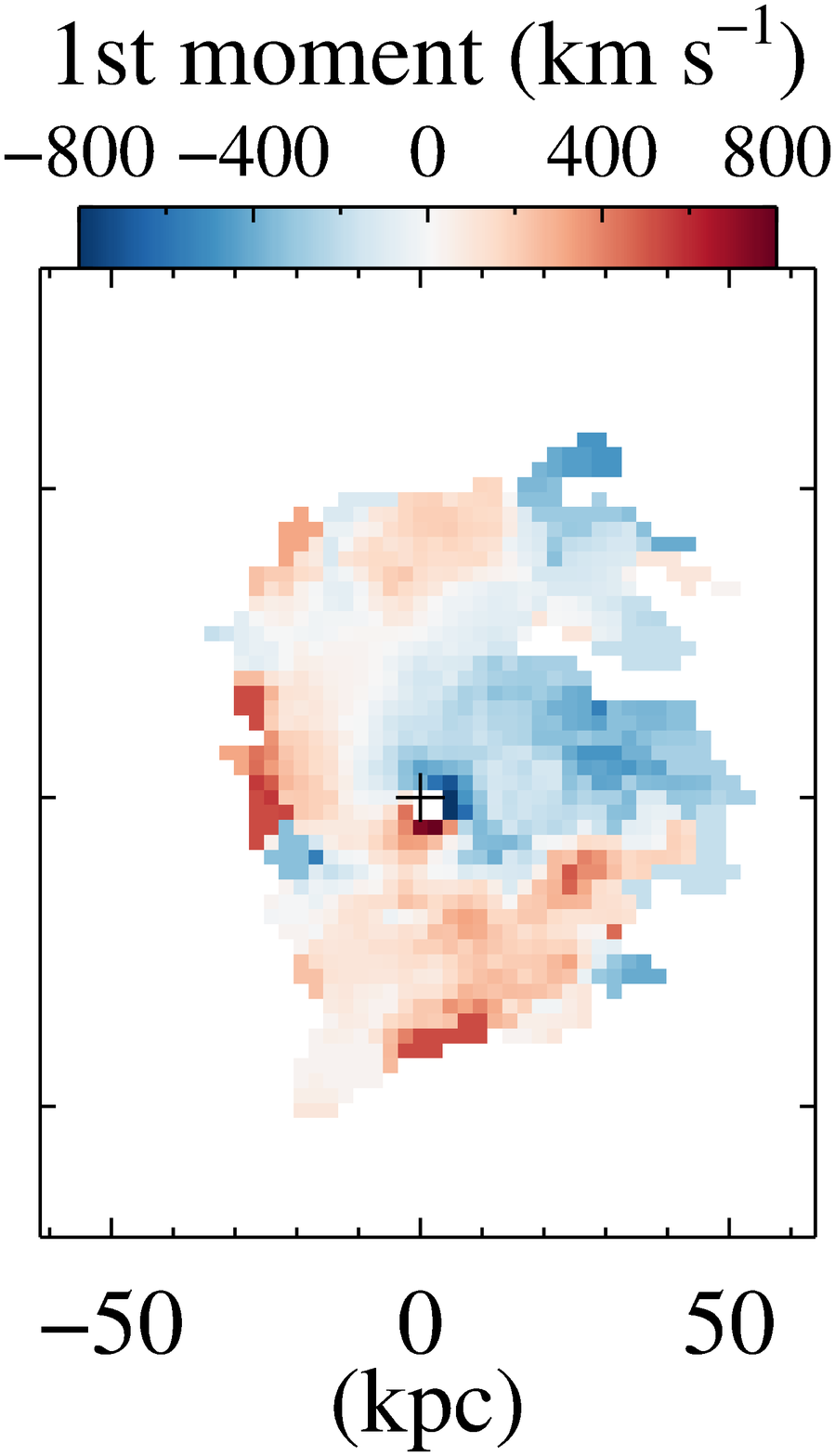}
\hspace{+0.01in}
\includegraphics[height=1.5in]{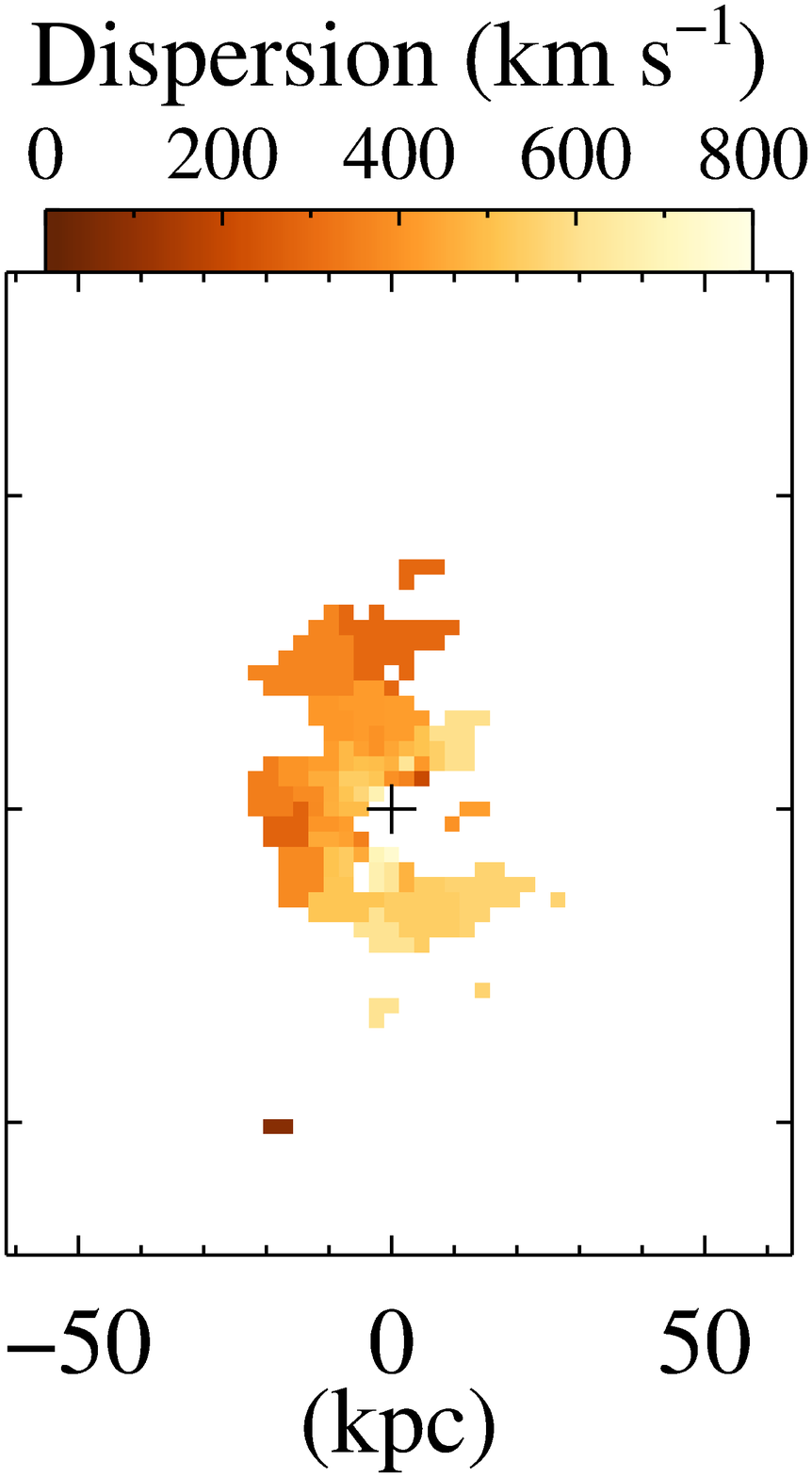}
\hspace{0.0in}
\includegraphics[width=0.27\textwidth]{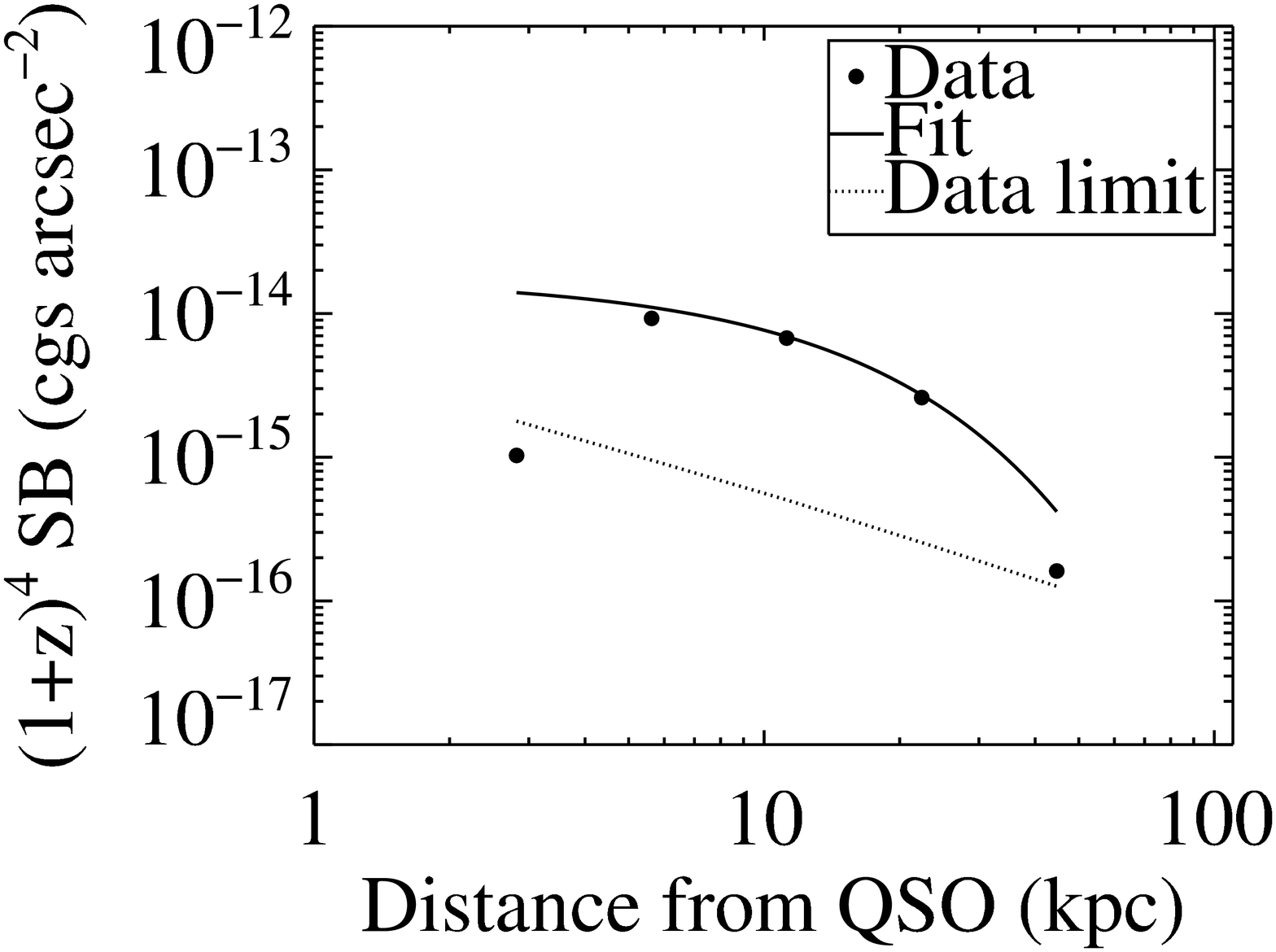}
\hspace{0.0in}
\includegraphics[width=0.27\textwidth]{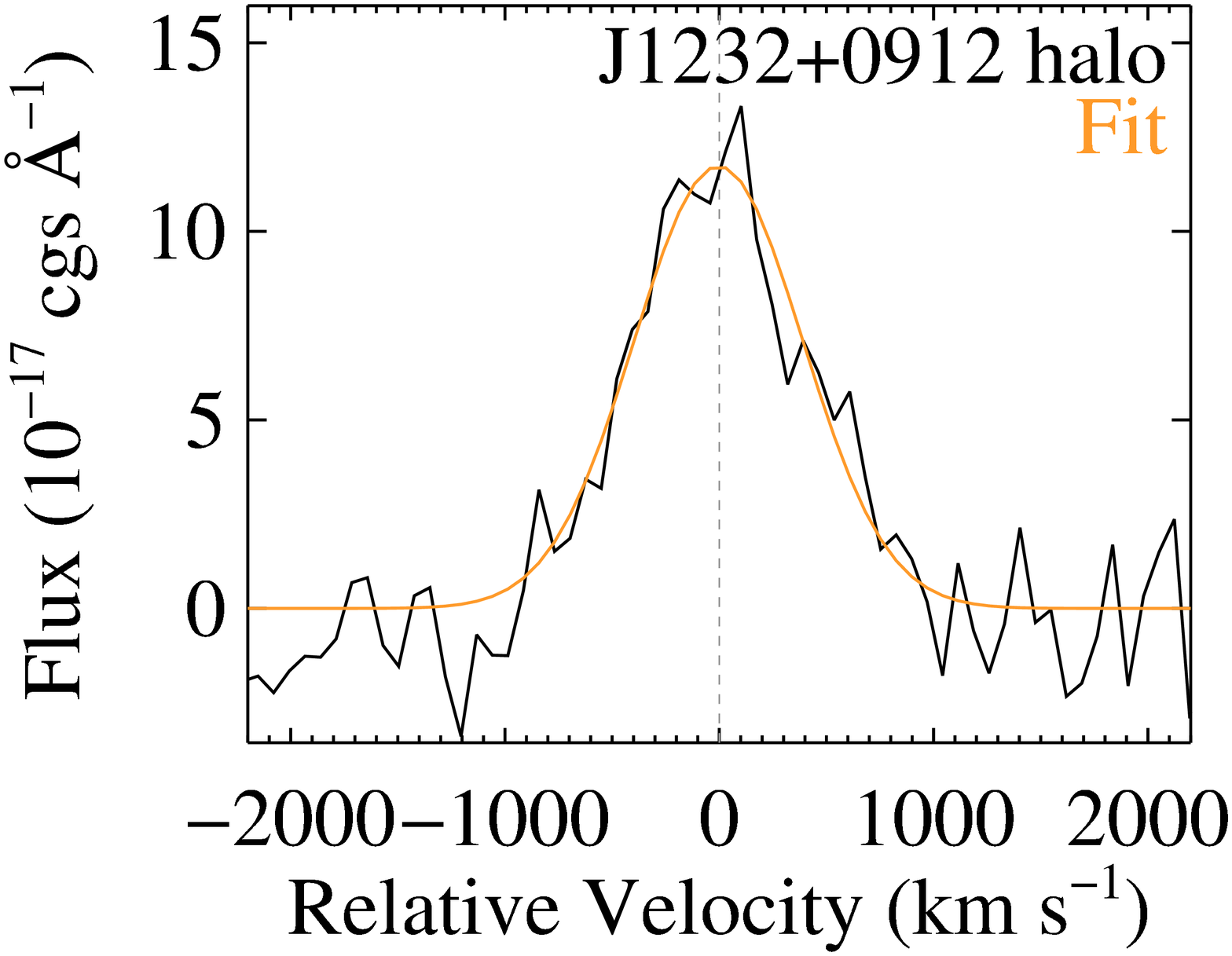}
\caption{Full sample of ERQ Ly\(\alpha\) halos after subtracting the quasar, and the best available measurements, where the Ly\(\alpha\) spike was isolated in six quasar spectra before subtraction. First column shows the optimally extracted halo, second is the 1st moment halo velocity map, third is Voronoi binned velocity dispersion, fourth is the circularly averaged surface brightness radial profile, and the fifth is the total integrated Ly\(\alpha\) halo spectrum. For all 2D maps, the plus is the location of the quasar. For the surface brightness panels the PSF size is indicated as a circle in the upper left corner, the diamond is the centroid of halo emission, the cross symbol is the location of peak Ly\(\alpha\) halo emission when the Ly\(\alpha\) spike is left in the residual map. For the surface brightness radial profiles hollow circles represent negative values that were included for profile fitting, and plotted as absolute values. They have also been corrected for cosmological dimming for comparison to other samples. The integrated halo spectrum shows a Gaussian that was fit to determine the overall halo velocity dispersion. Zero velocity is defined as the centroid of integrated halo emission, without removing the Ly\(\alpha\) spike in quasar subtraction. J0220+0137 has a cloud in the eastern edge of the FOV which is likely a foreground source not physically related to the quasar, and is omitted in measuring Ly$\alpha$ halo emission.}
\label{fig:fig_maps}
\end{figure*}

\newpage

\begin{figure*}
\includegraphics[height=1.5in]{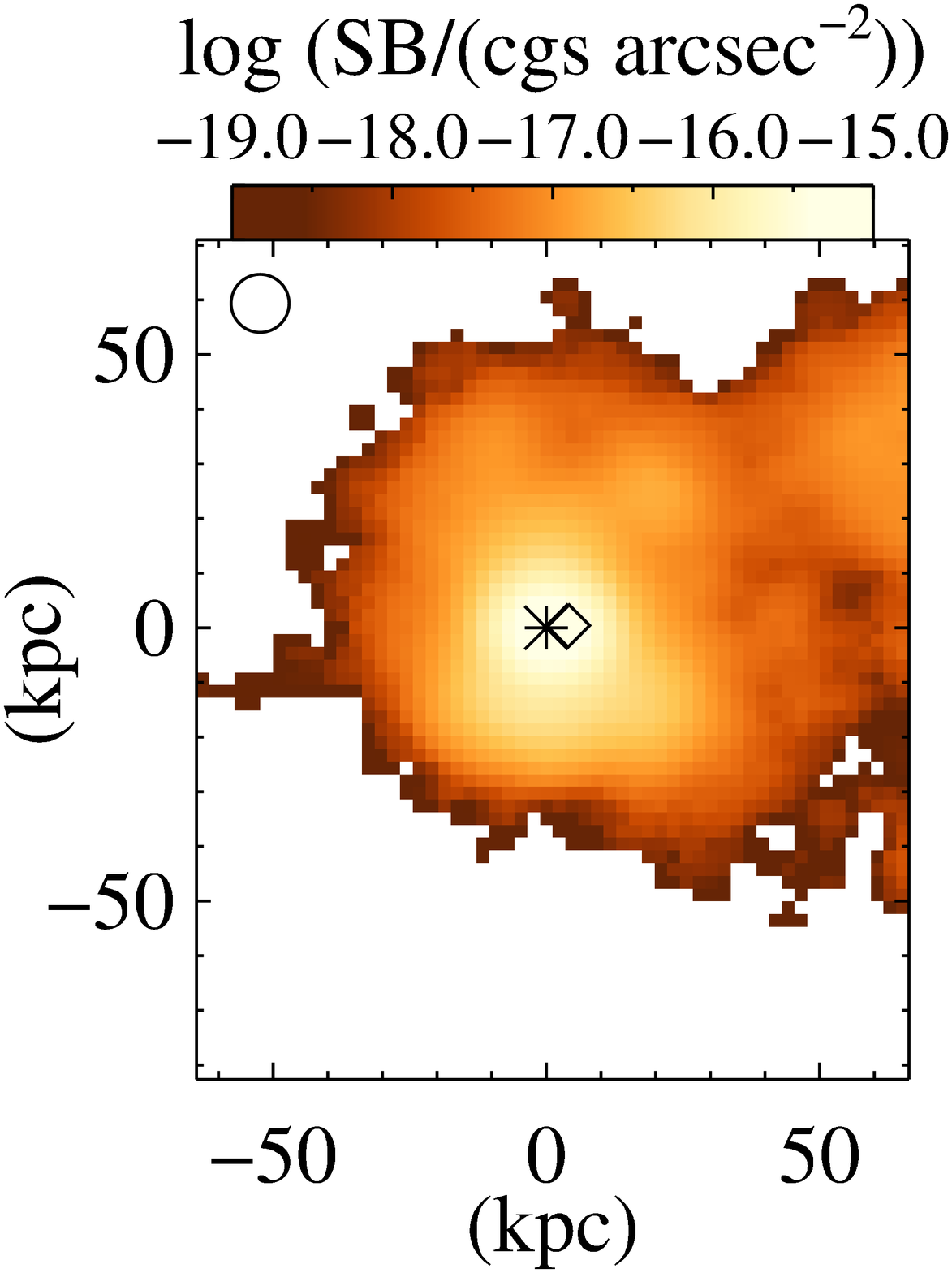}
\hspace{-0.01in}
\includegraphics[height=1.5in]{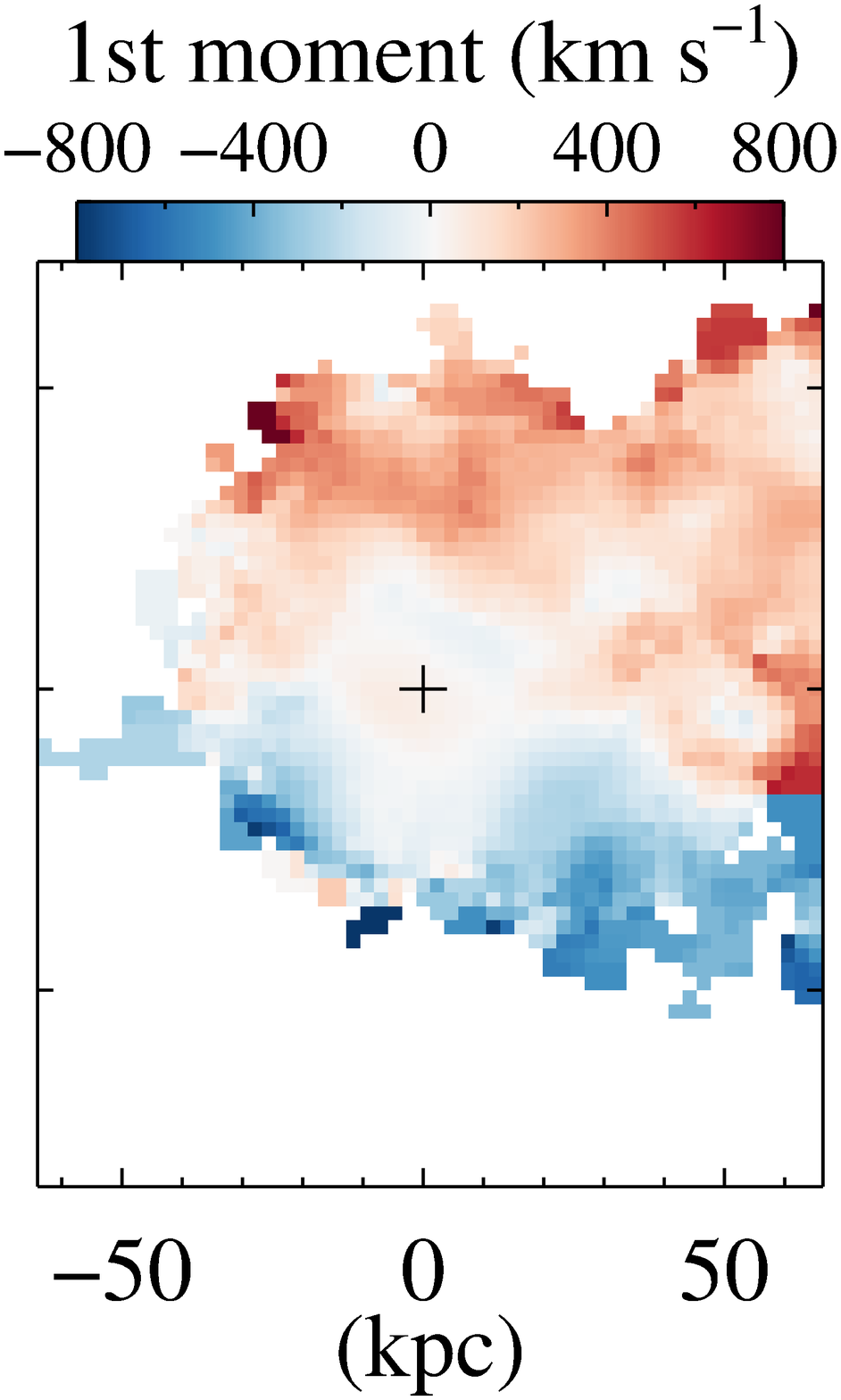}
\hspace{+0.01in}
\includegraphics[height=1.5in]{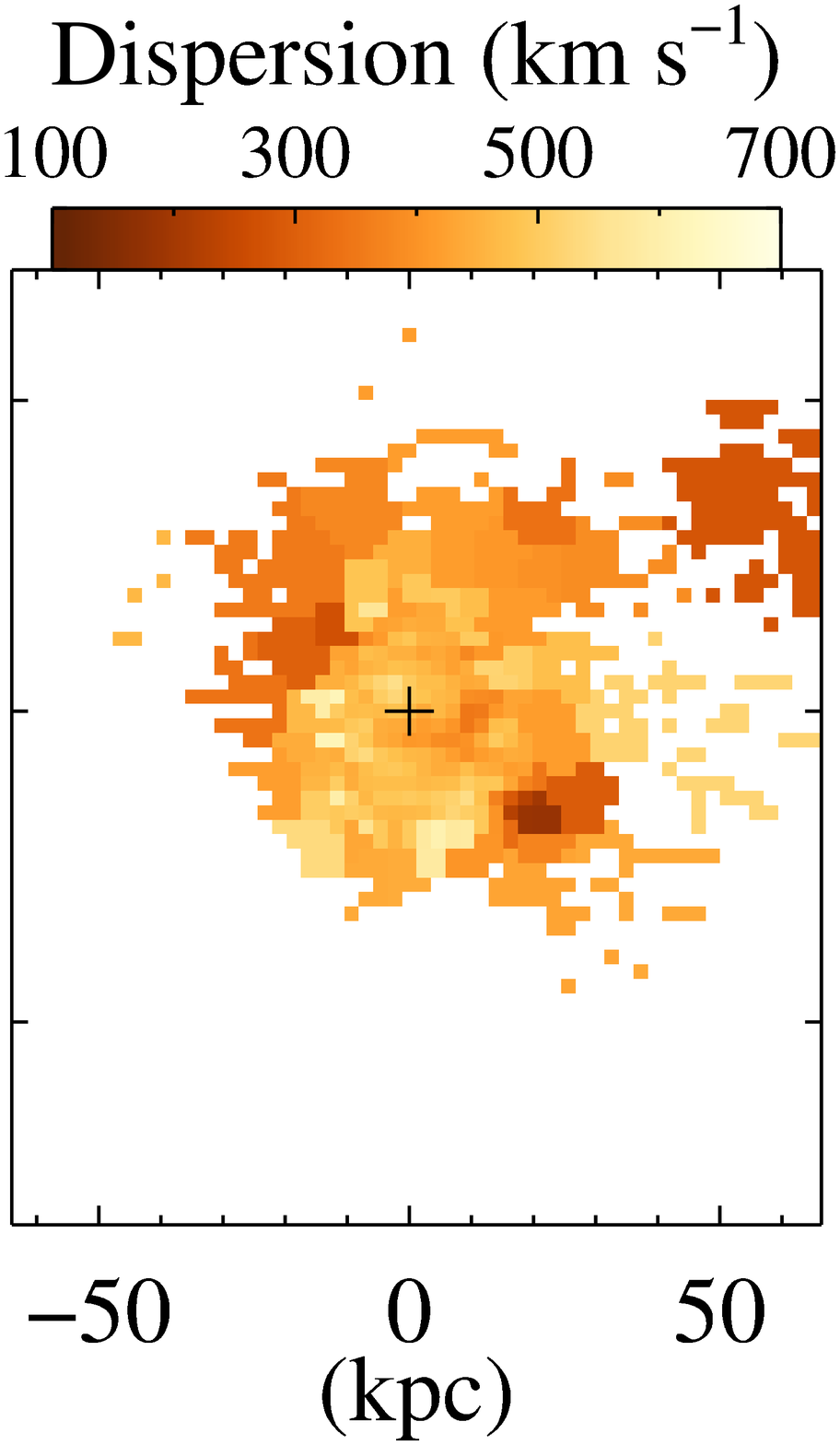}
\hspace{0.0in}
\includegraphics[width=0.27\textwidth]{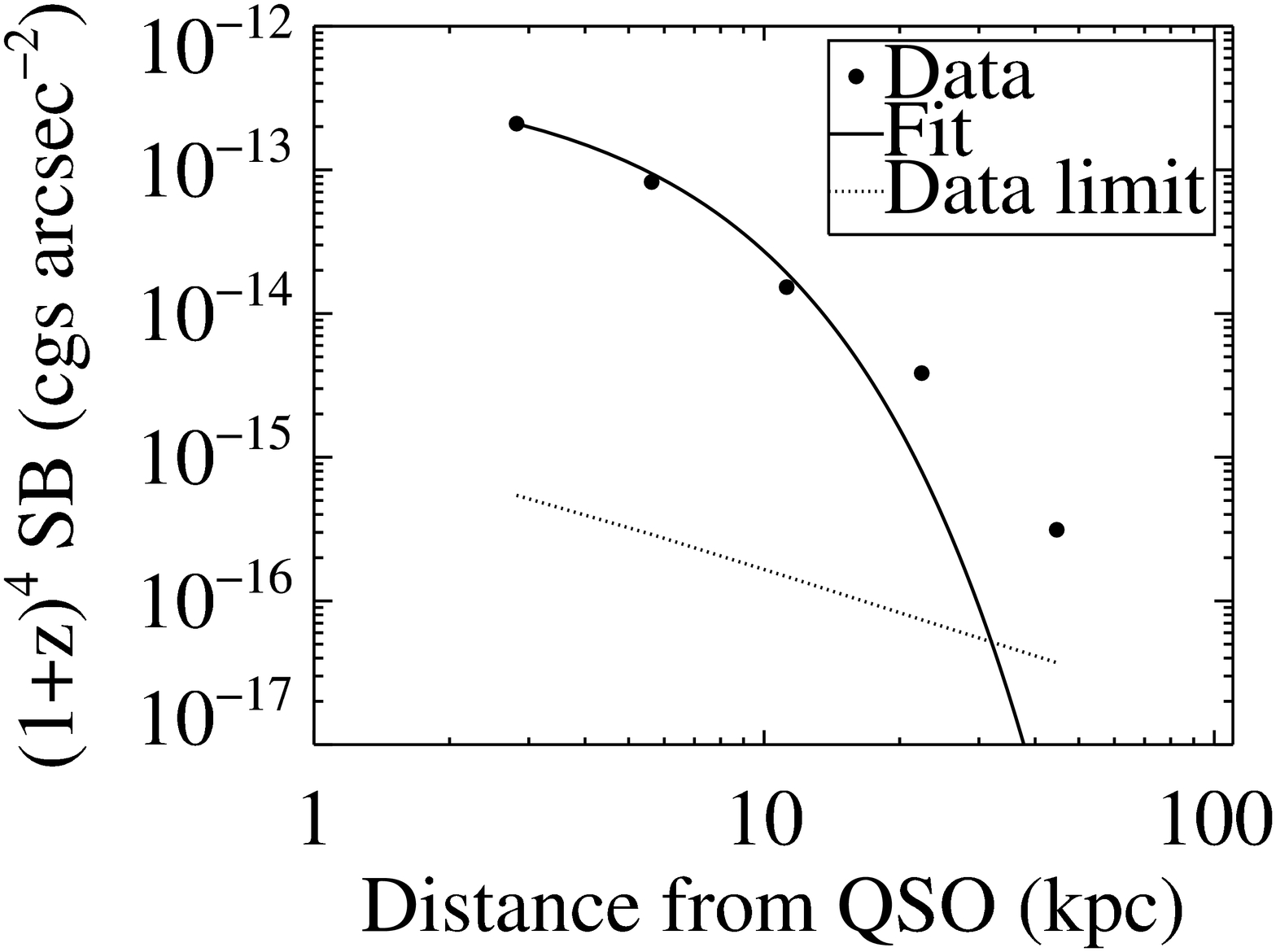}
\hspace{0.0in}
\includegraphics[width=0.27\textwidth]{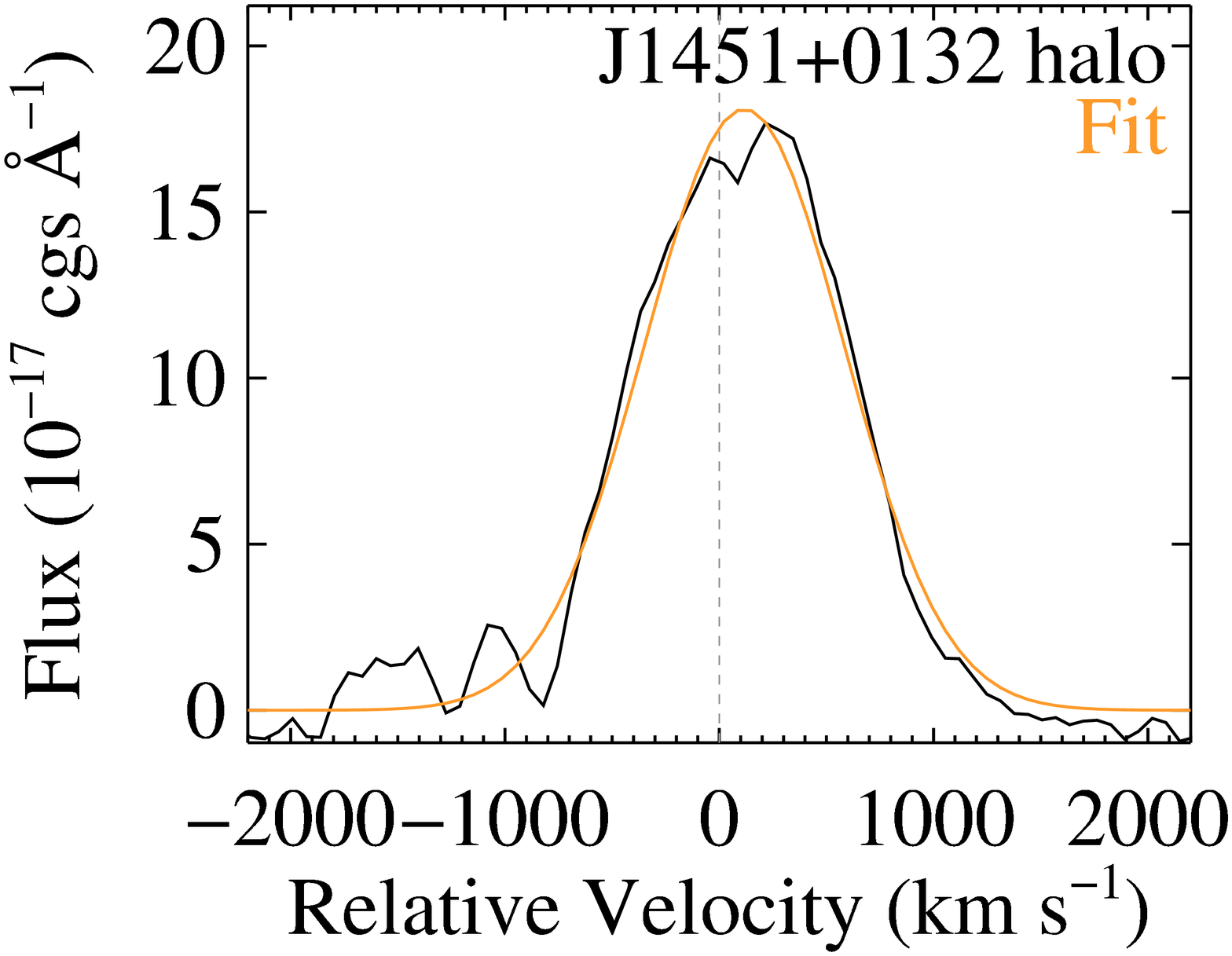} \\

\includegraphics[height=1.5in]{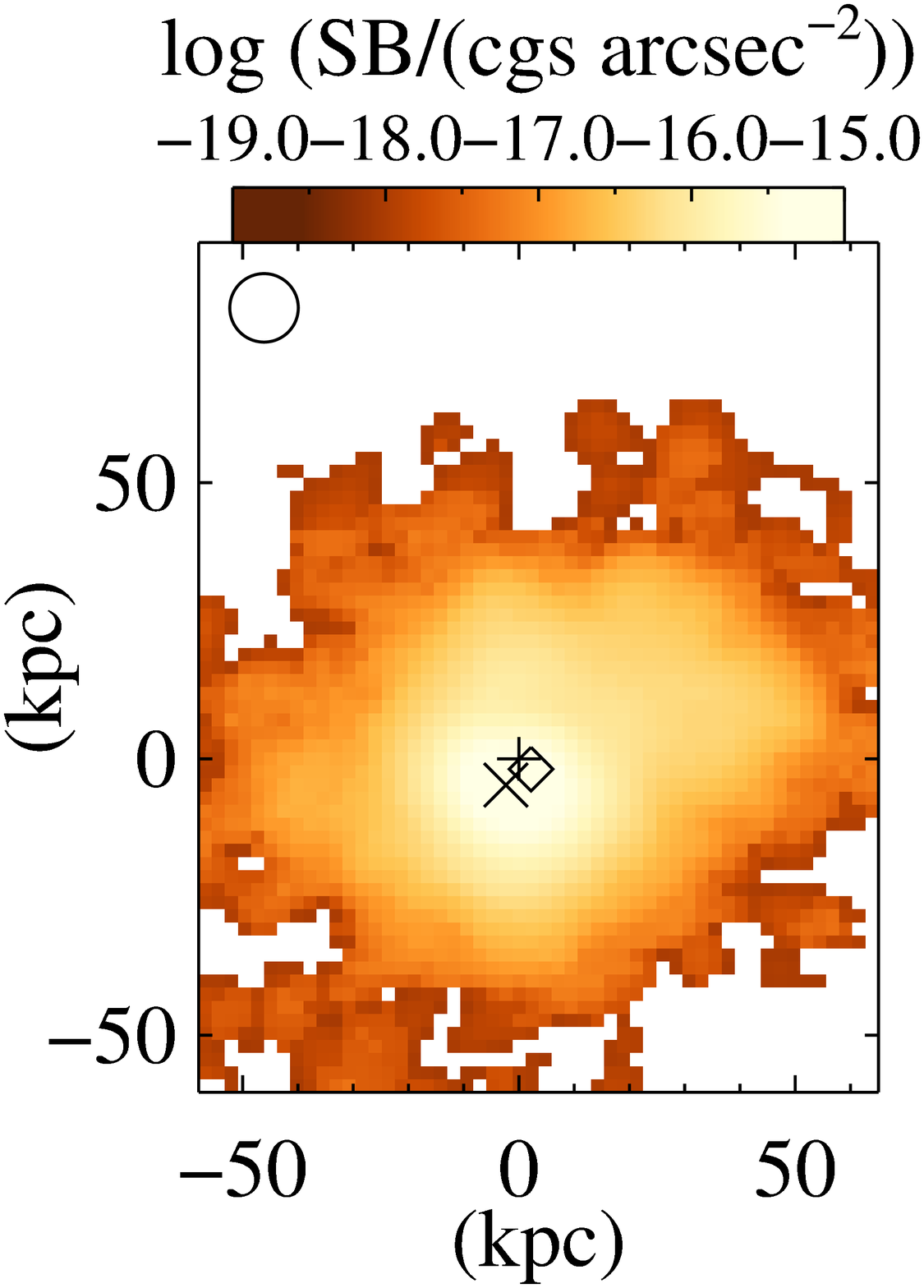}
\hspace{-0.01in}
\includegraphics[height=1.5in]{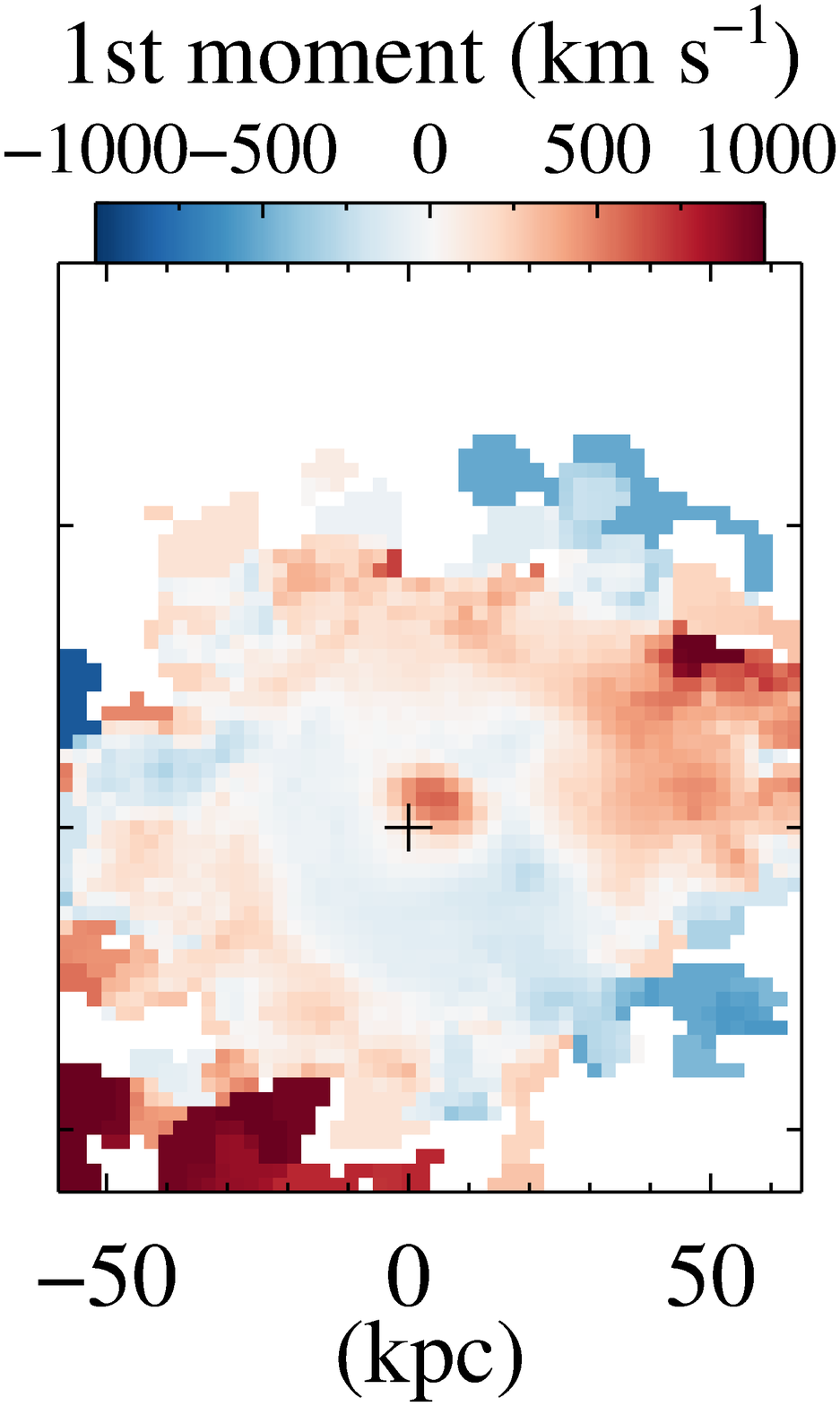}
\hspace{+0.01in}
\includegraphics[height=1.5in]{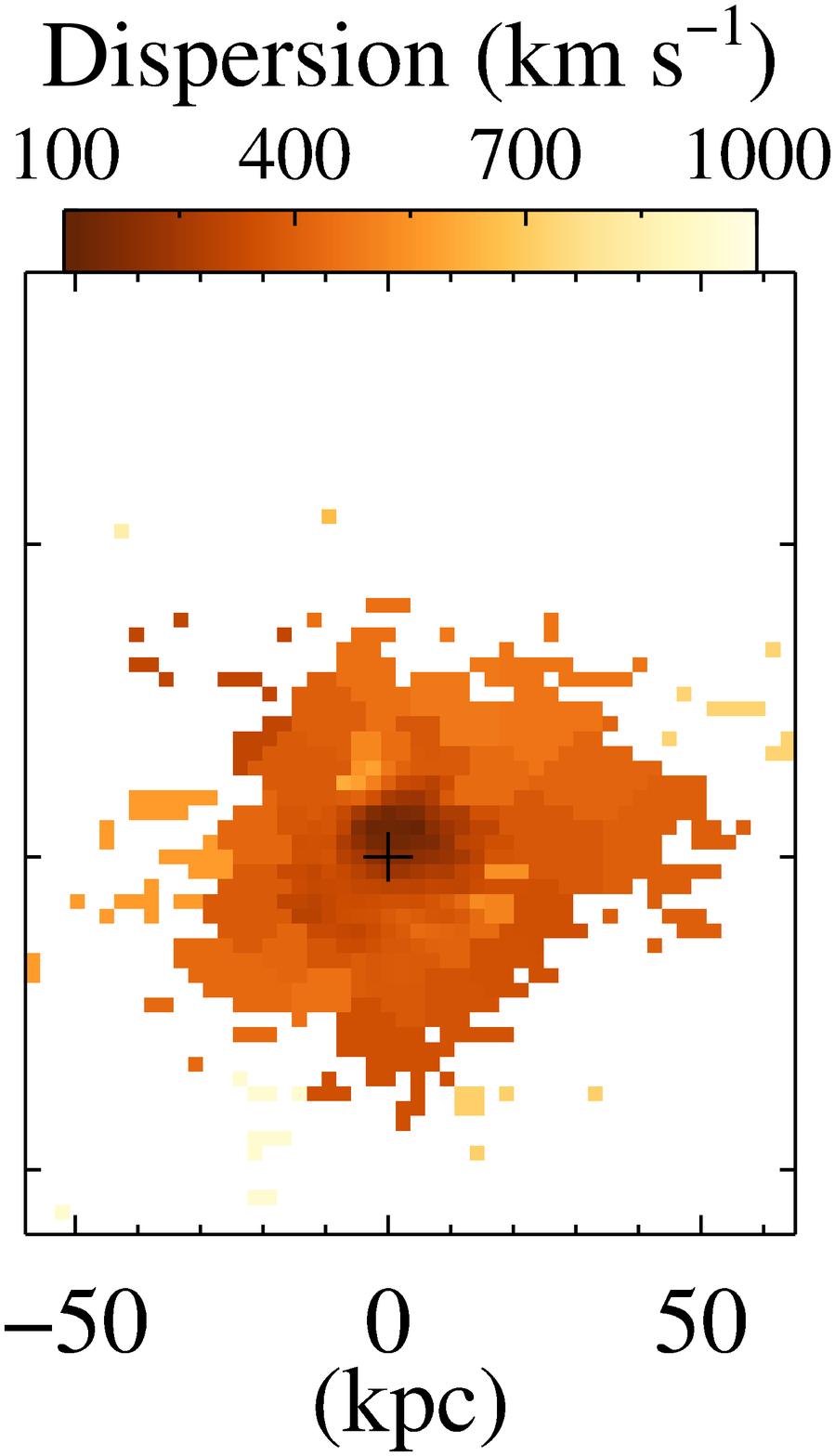}
\hspace{0.0in}
\includegraphics[width=0.27\textwidth]{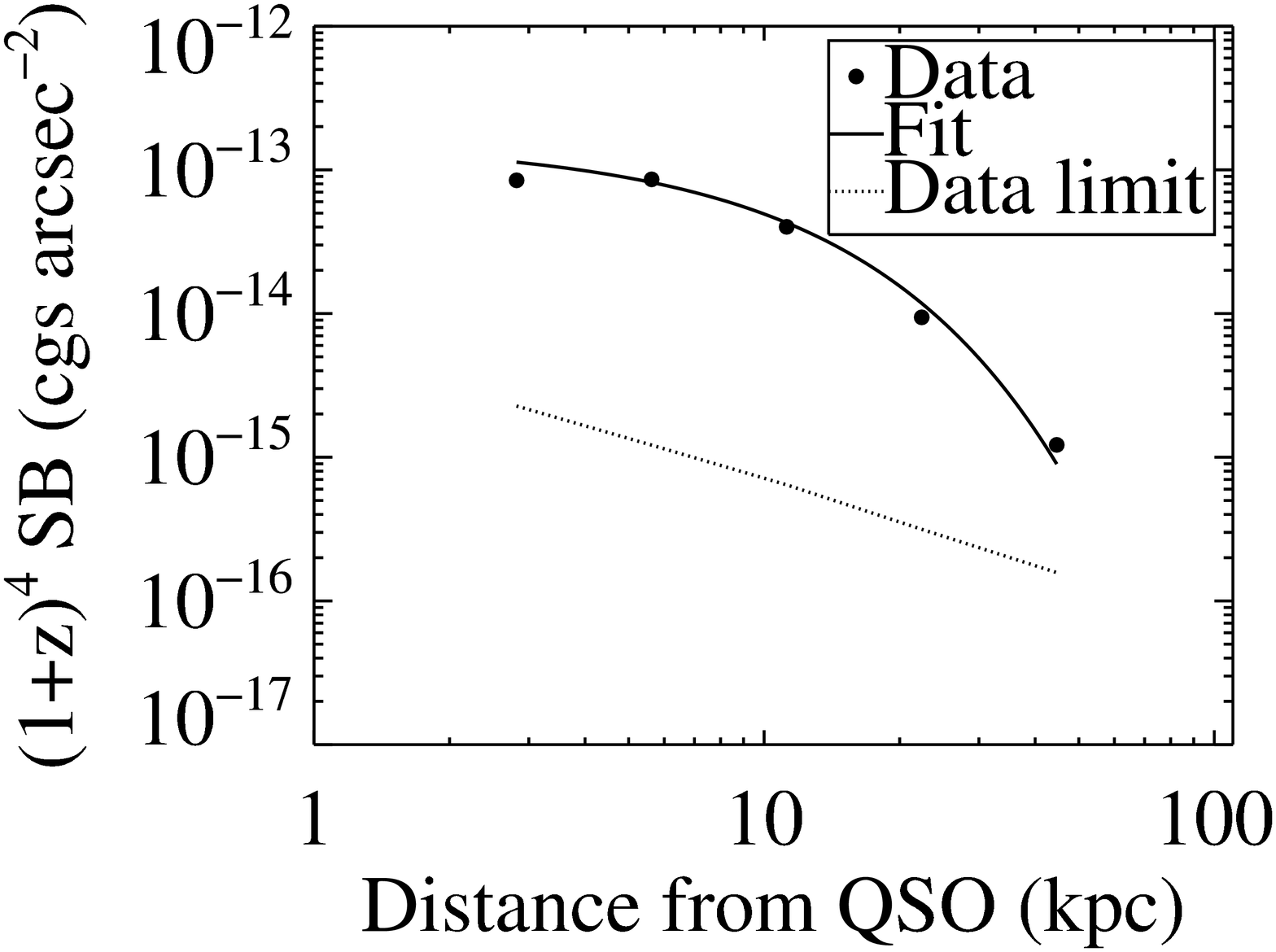}
\hspace{0.0in}
\includegraphics[width=0.27\textwidth]{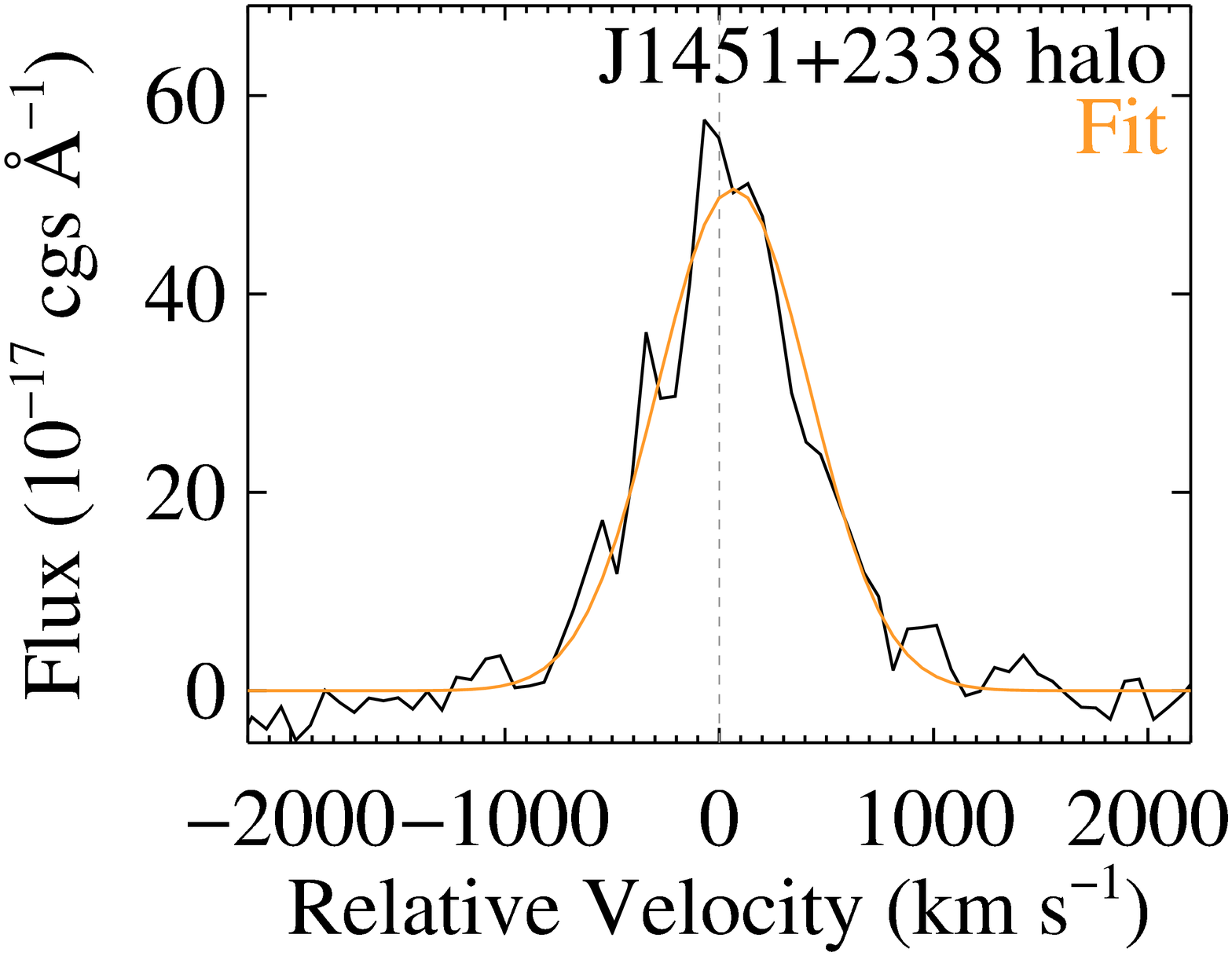}\\

\includegraphics[height=1.5in]{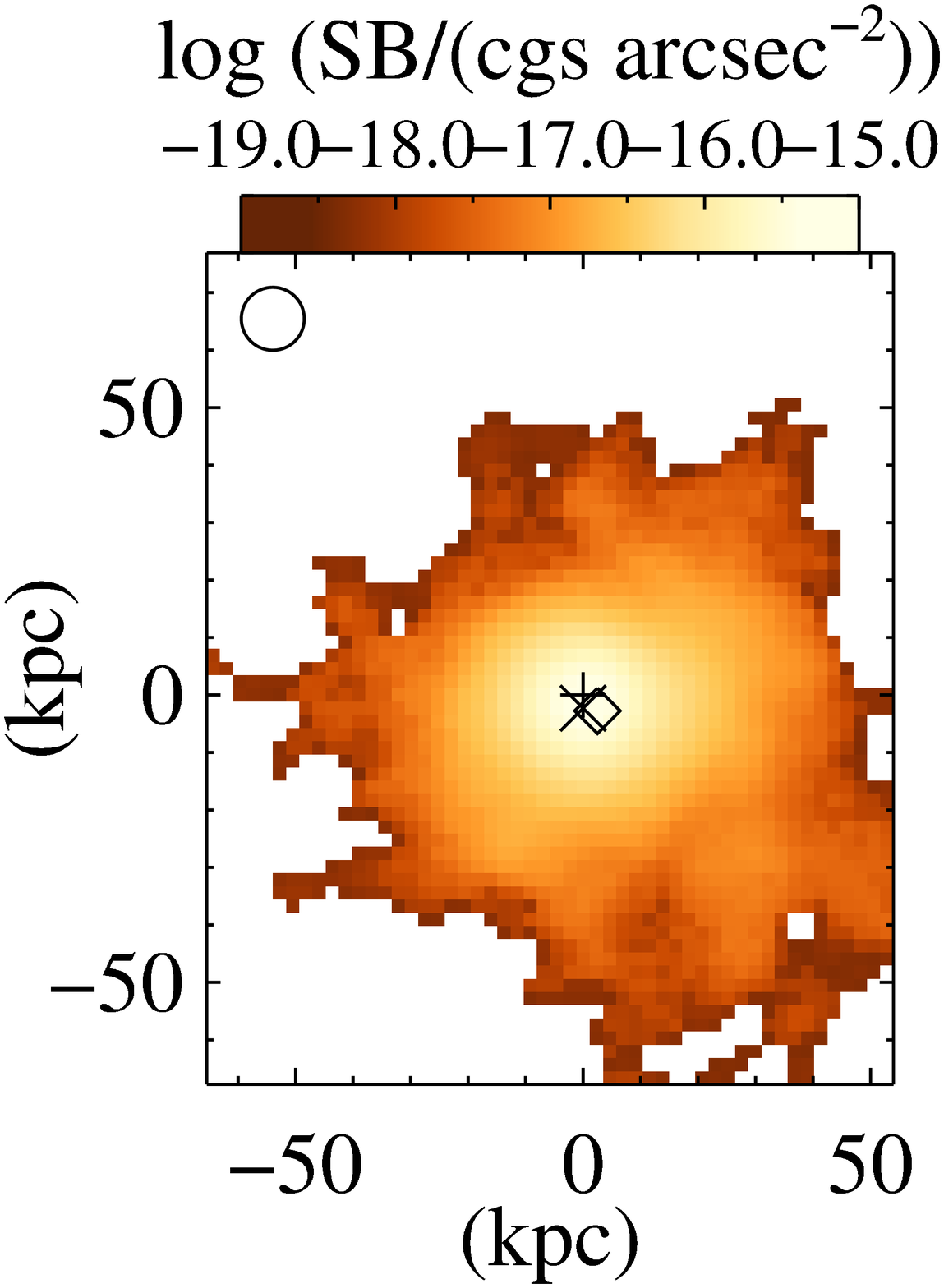}
\hspace{-0.01in}
\includegraphics[height=1.5in]{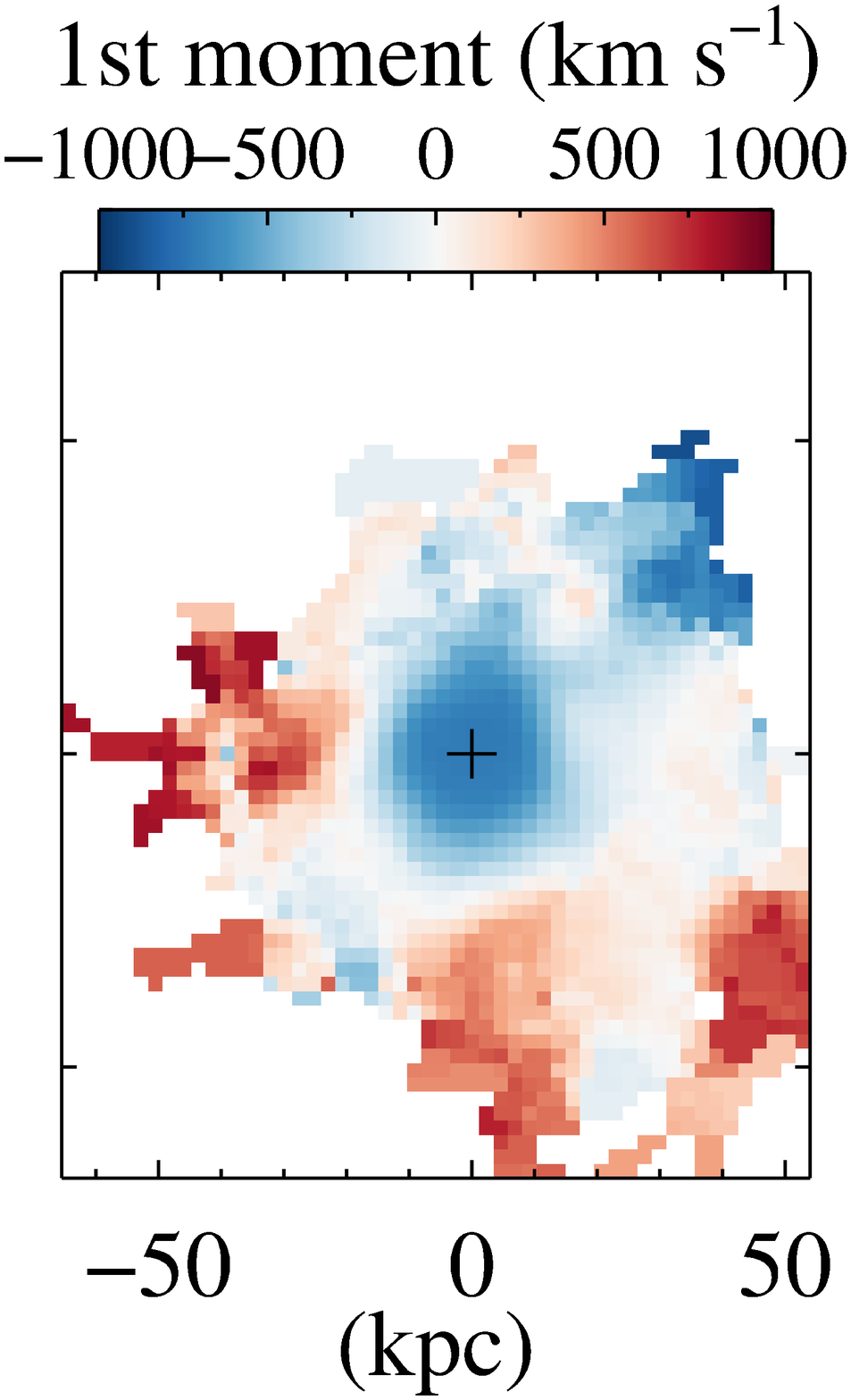}
\hspace{+0.01in}
\includegraphics[height=1.5in]{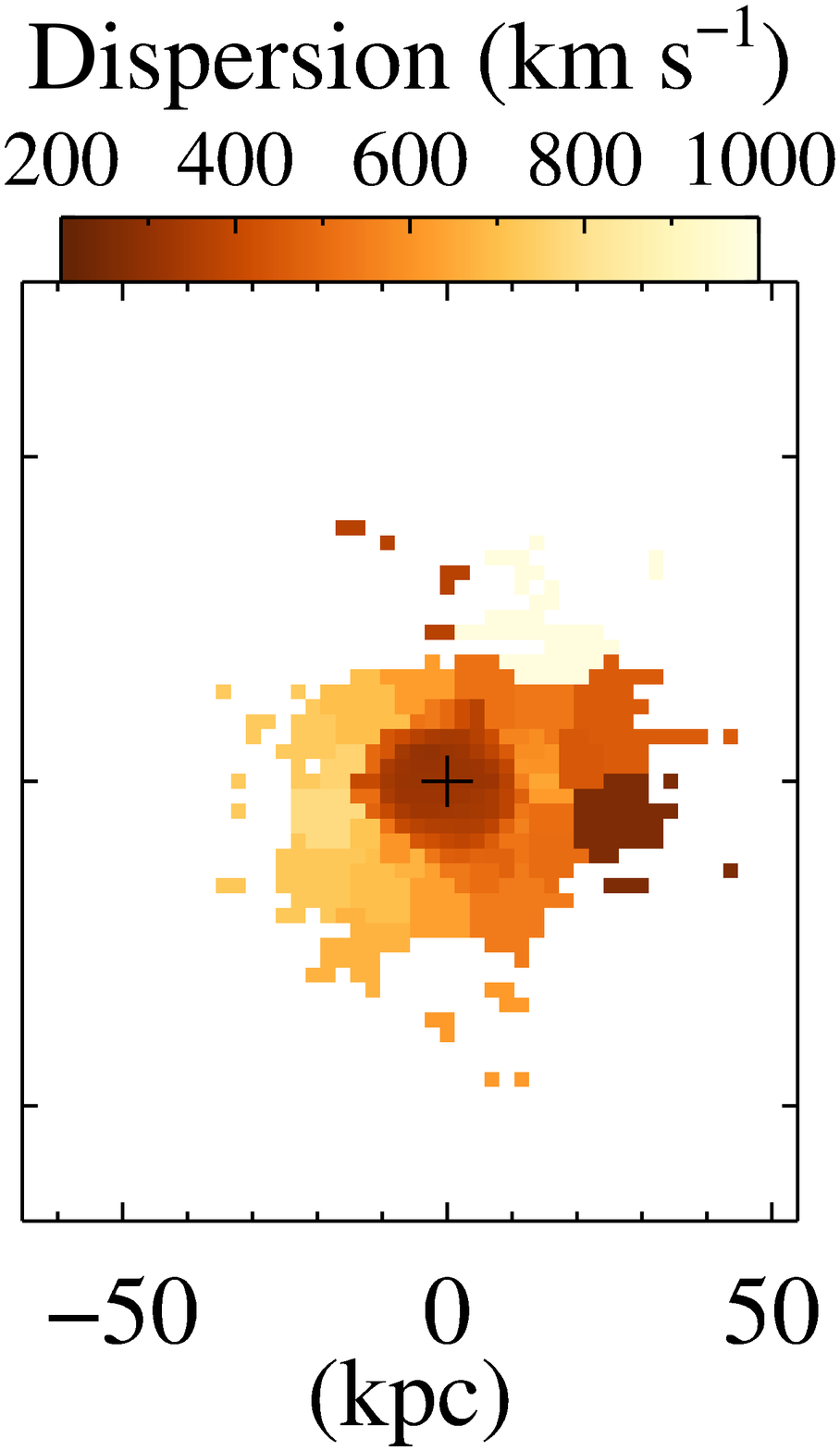}
\hspace{0.0in}
\includegraphics[width=0.27\textwidth]{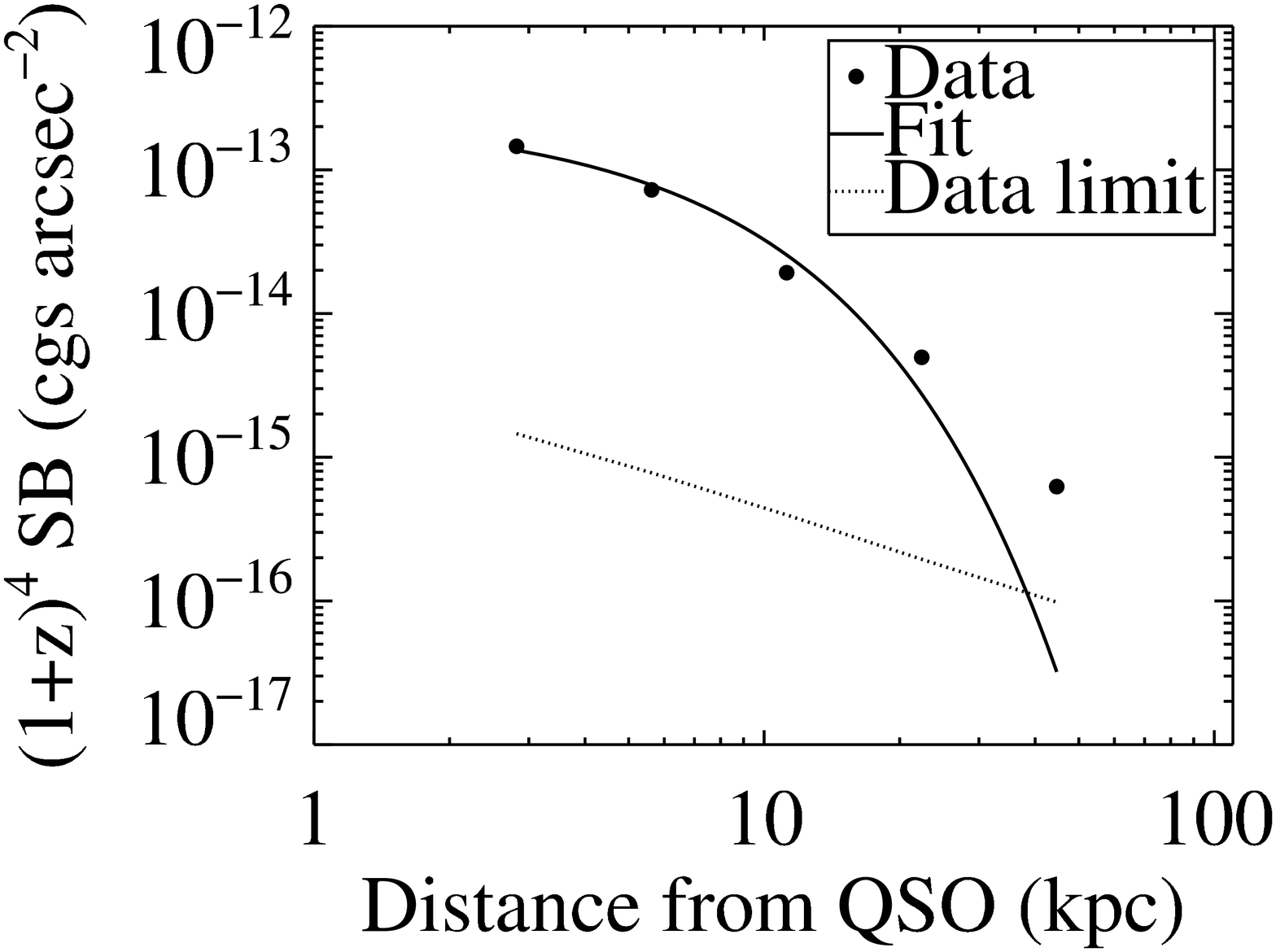}
\hspace{0.0in}
\includegraphics[width=0.27\textwidth]{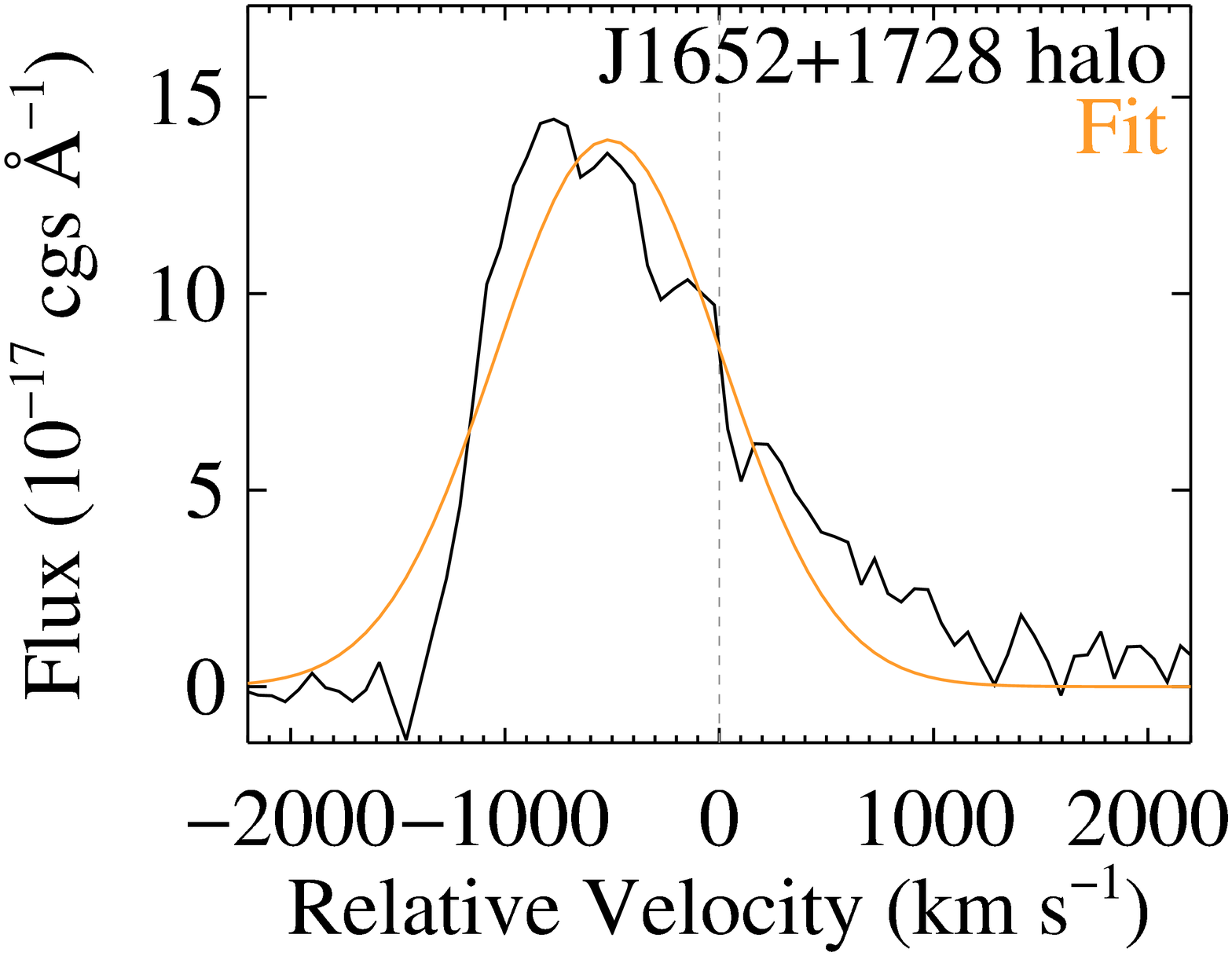}\\

\includegraphics[height=1.5in]{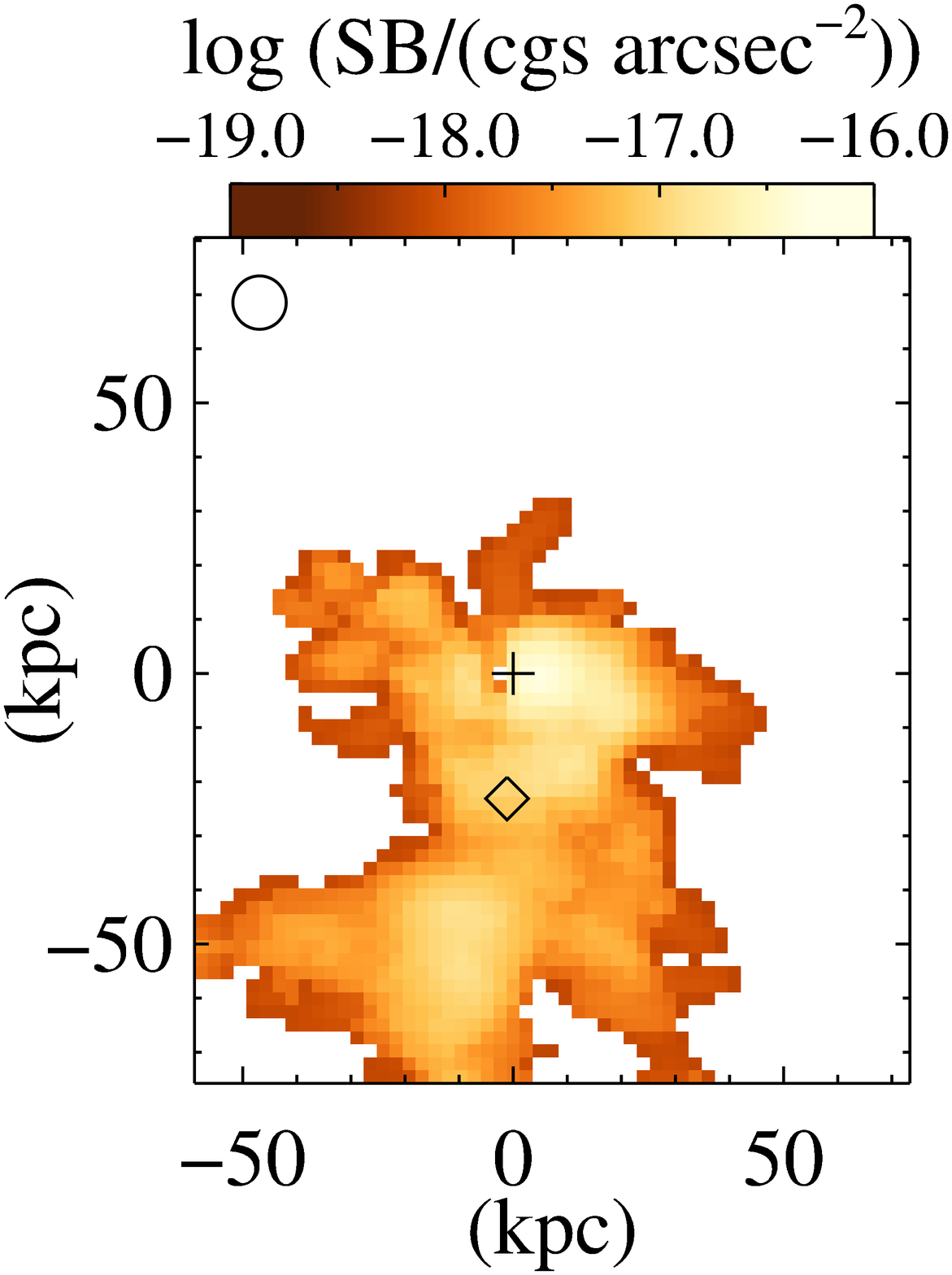}
\hspace{-0.01in}
\includegraphics[height=1.5in]{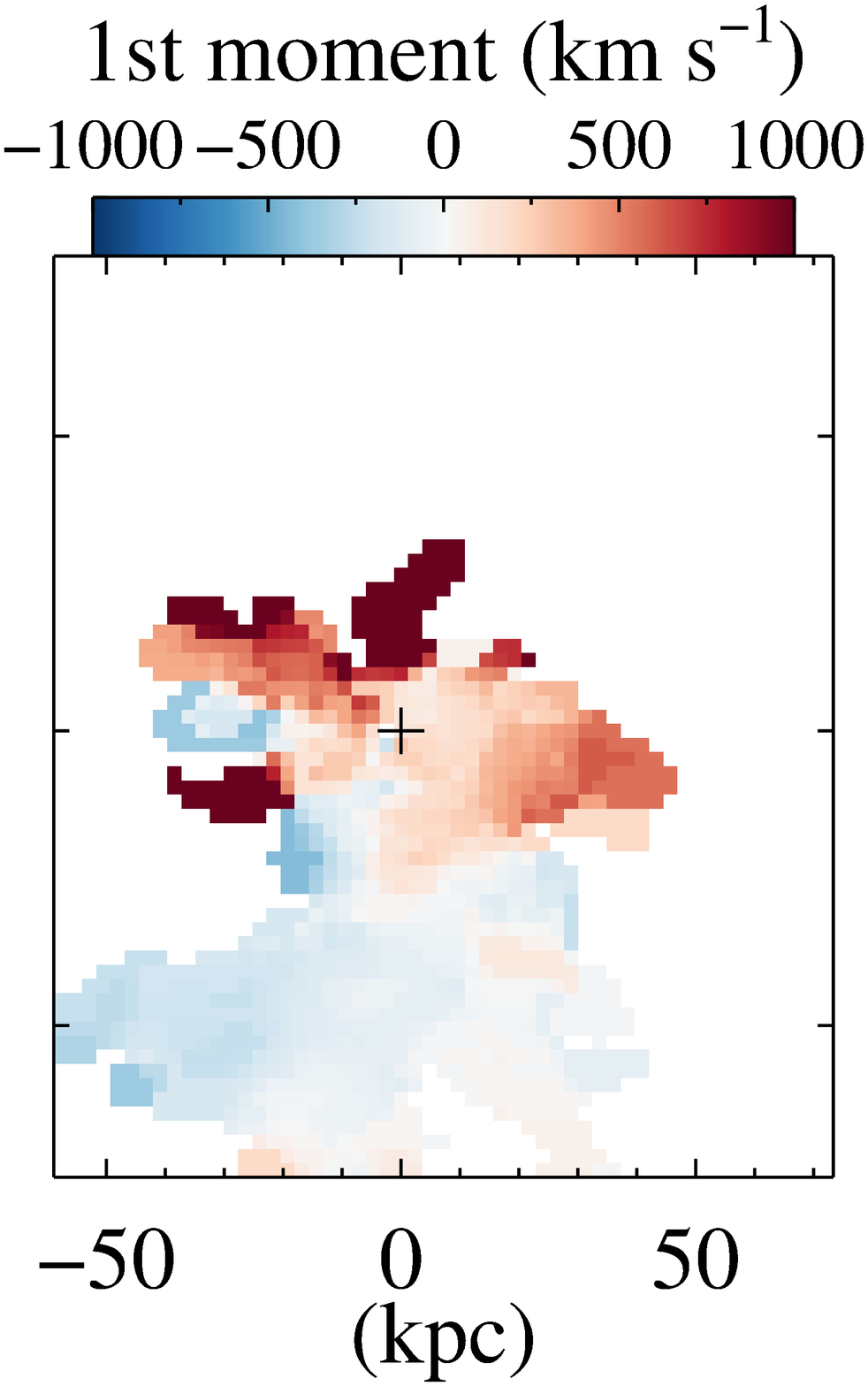}
\hspace{+0.01in}
\includegraphics[height=1.5in]{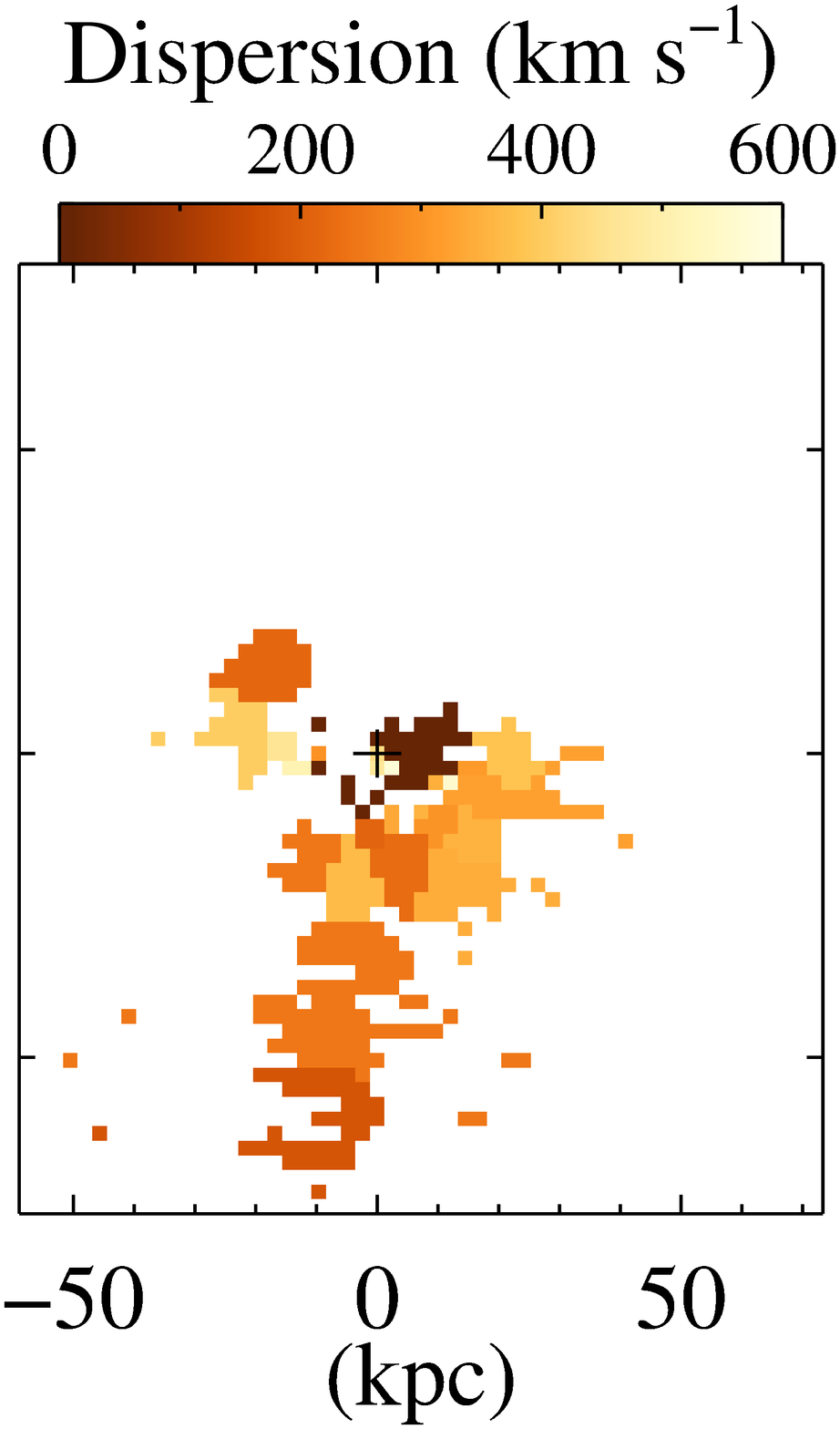}
\hspace{0.0in}
\includegraphics[width=0.27\textwidth]{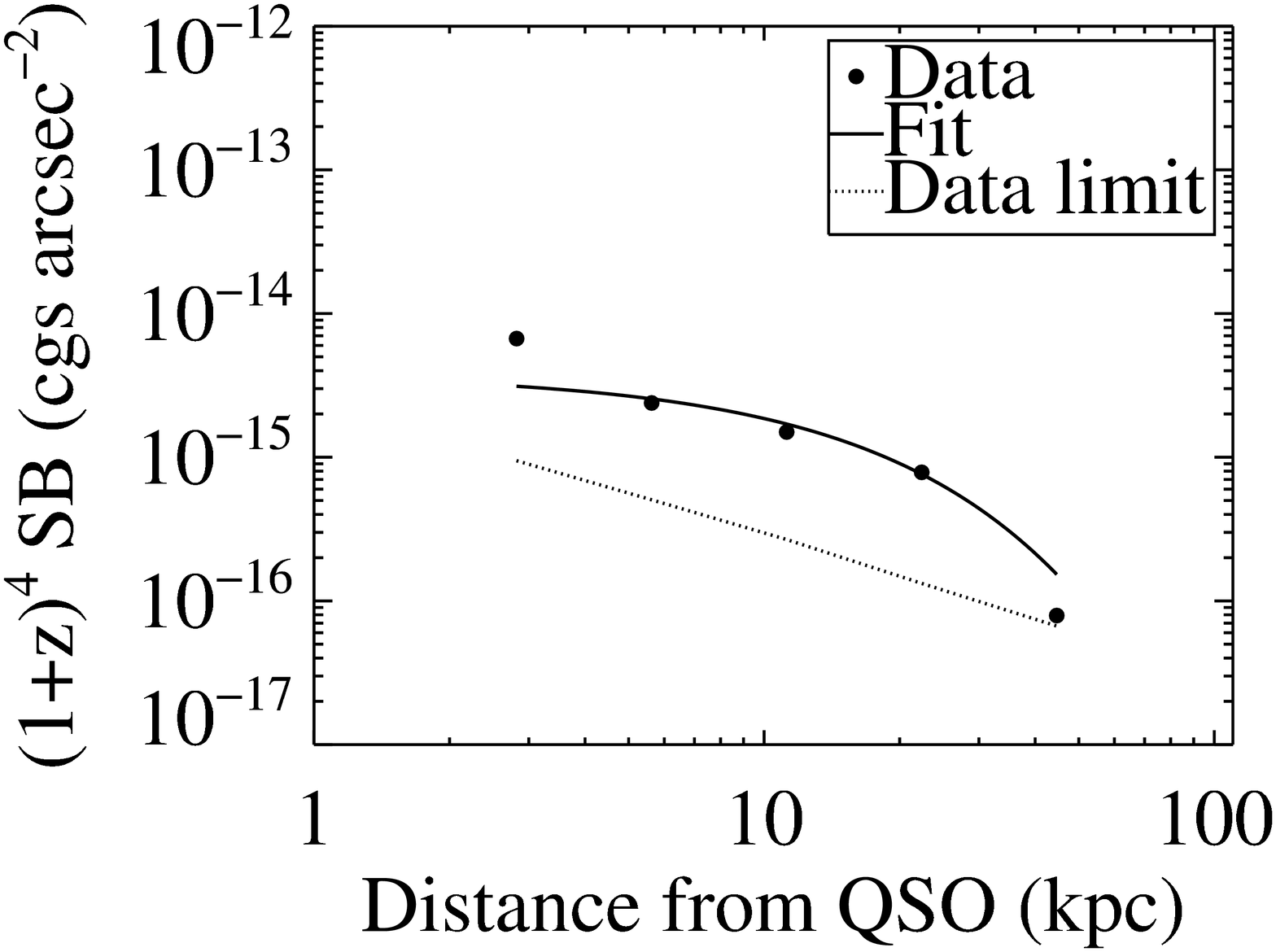}
\hspace{0.0in}
\includegraphics[width=0.27\textwidth]{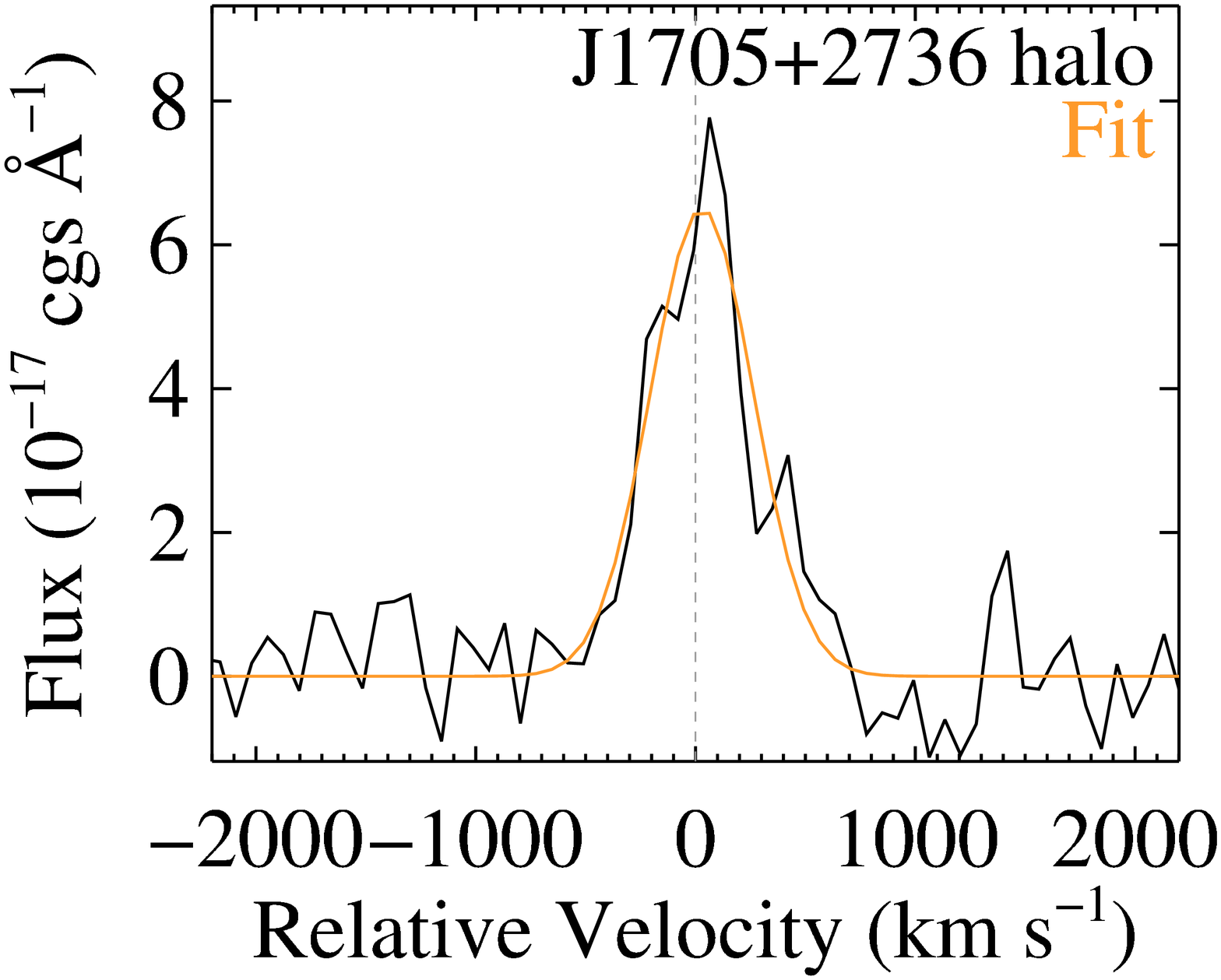}\\

\includegraphics[height=1.5in]{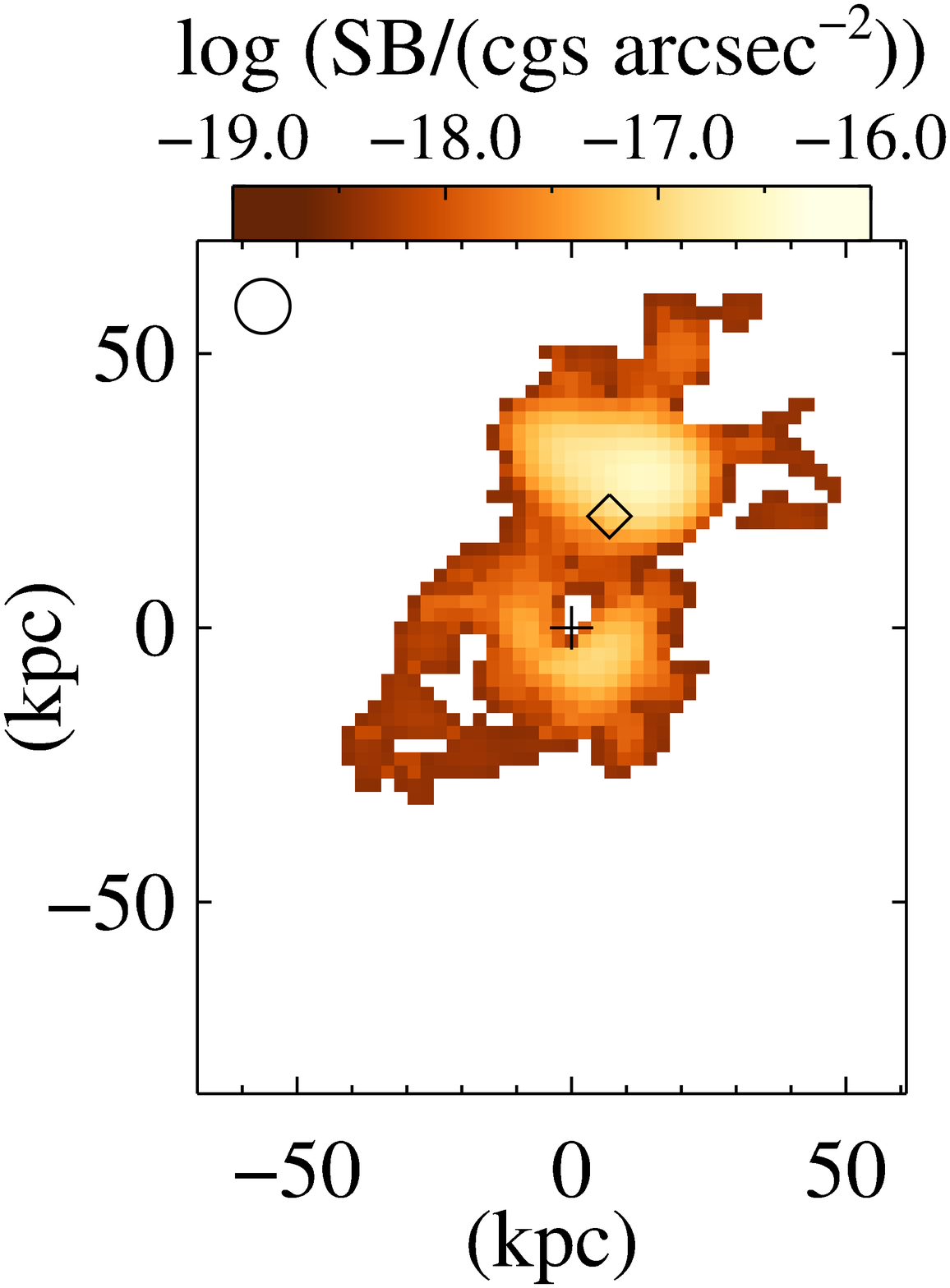}
\hspace{-0.01in}
\includegraphics[height=1.5in]{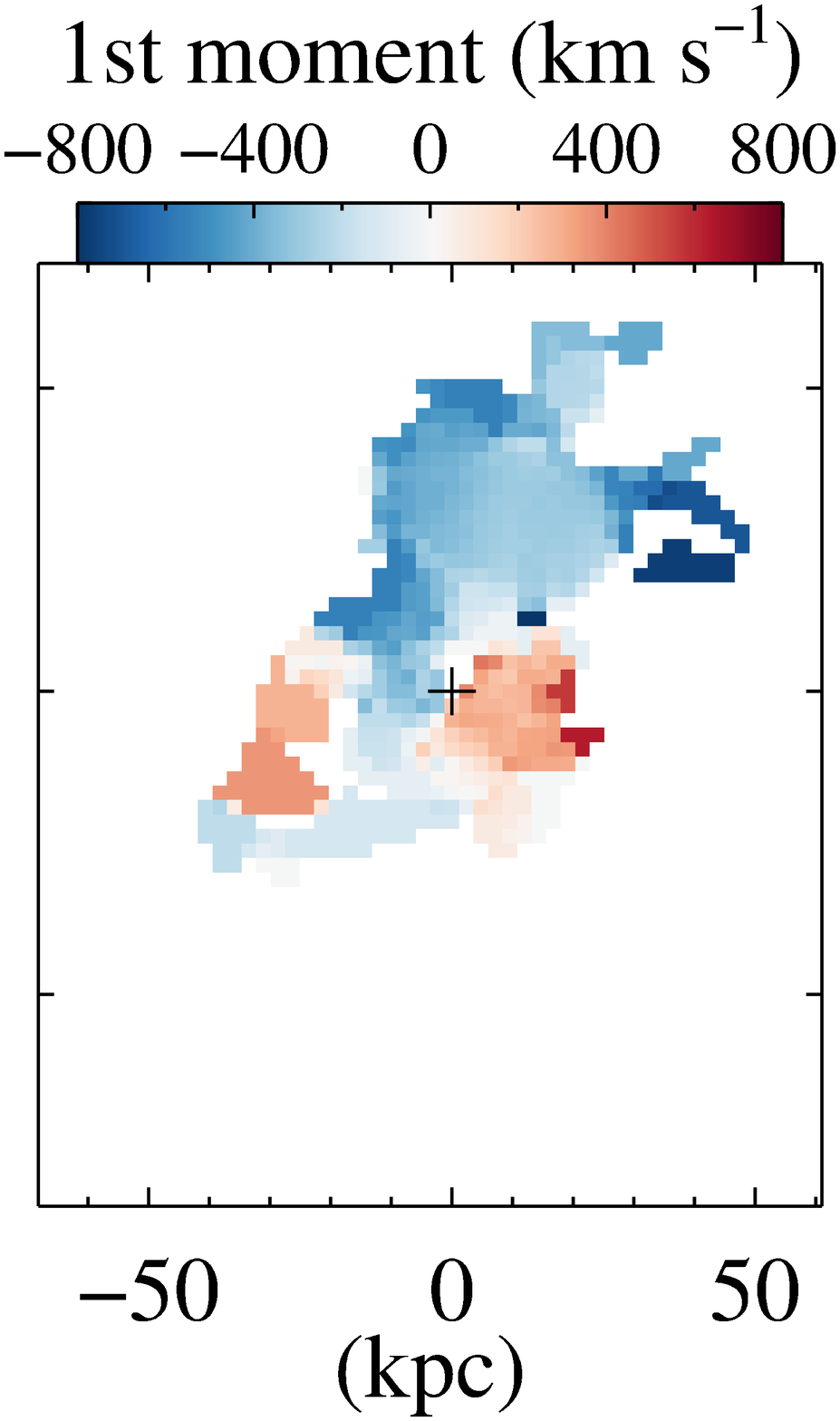}
\hspace{+0.01in}
\includegraphics[height=1.5in]{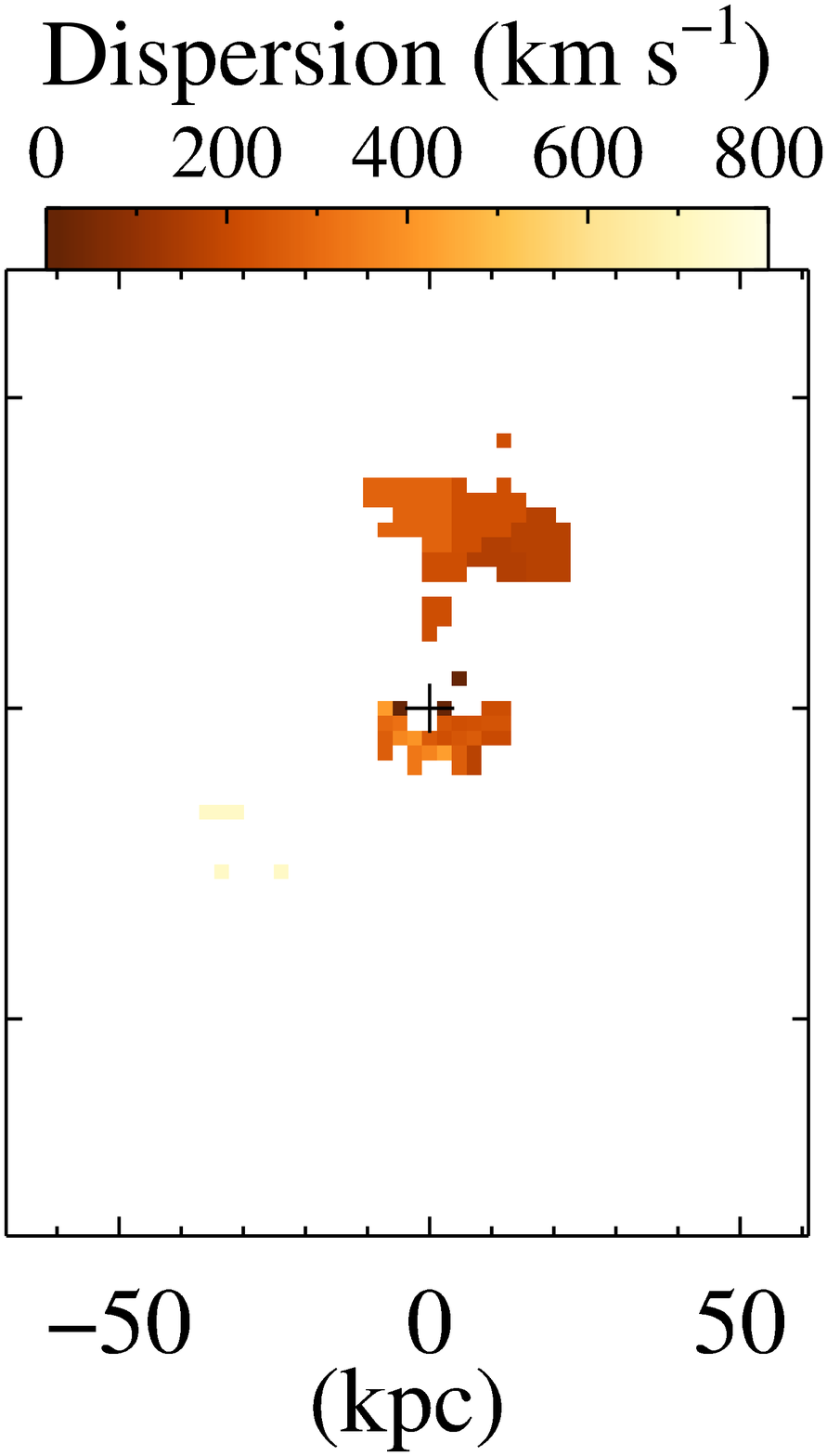}
\hspace{0.0in}
\includegraphics[width=0.27\textwidth]{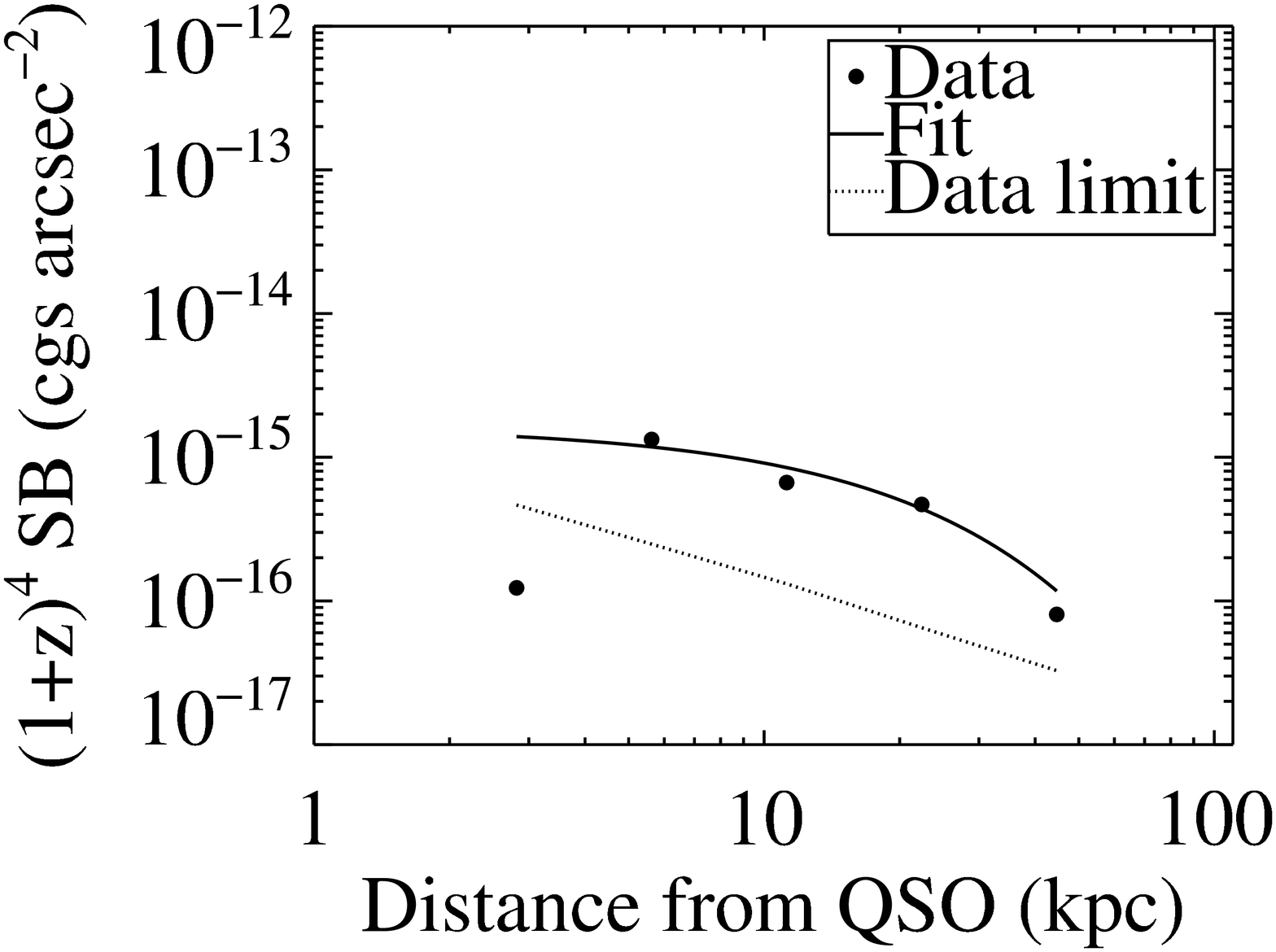}
\hspace{0.0in}
\includegraphics[width=0.27\textwidth]{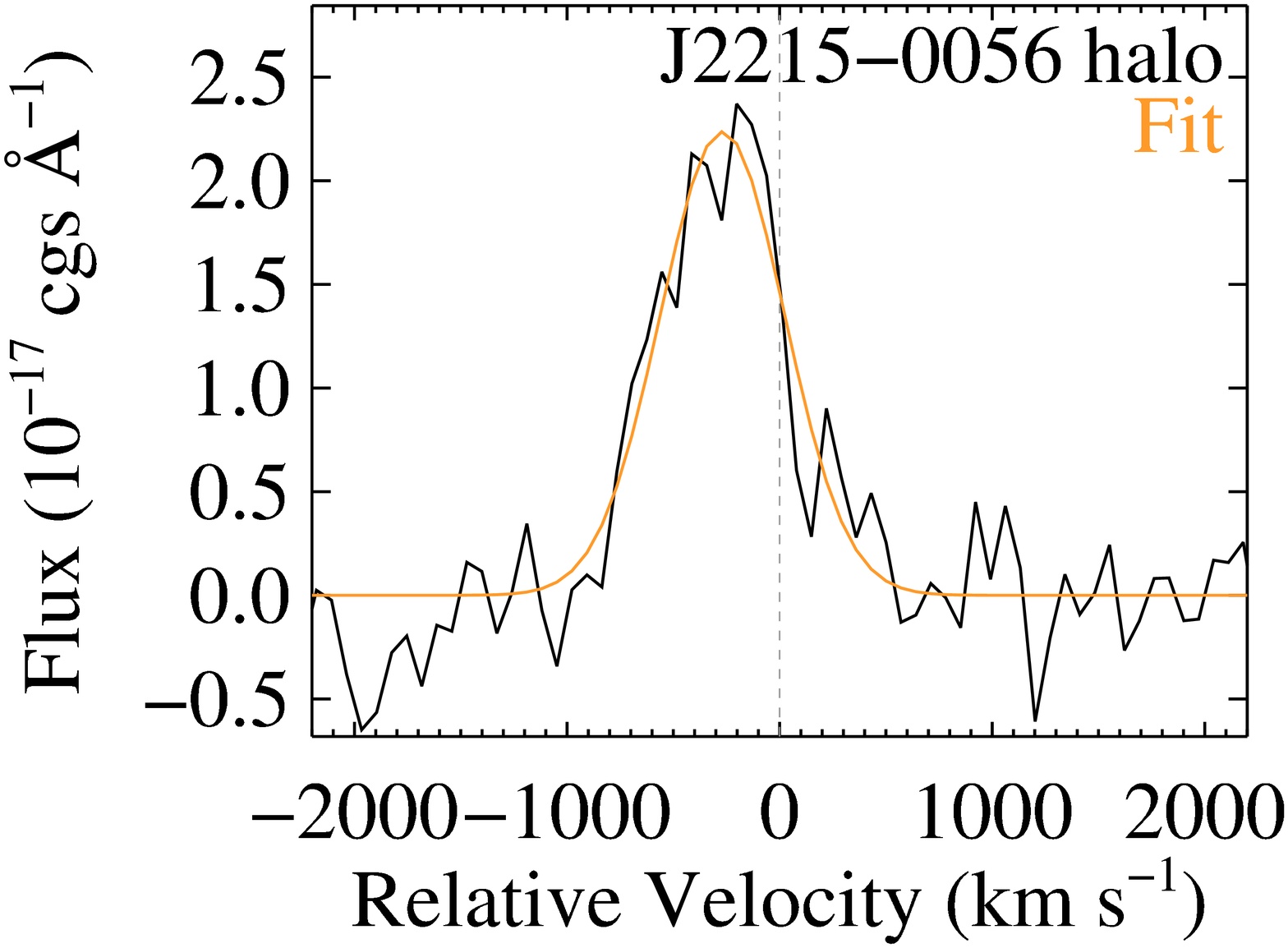} \\
\raggedright \contcaption{. J1652+1728's integrated halo centroid is offset from zero velocity because we define zero from the outer halo. J2215$-$0056 has a cloud in the north of the quasar which appears to be associated with the quasar as part of the same large scale structure, and blueshifts the spatially-integrated halo spectra. The associated object is omitted in measuring the ERQ's total Ly$\alpha$ halo emission. }
\end{figure*}

\newpage

\begin{figure*}
\includegraphics[height=1.5in]{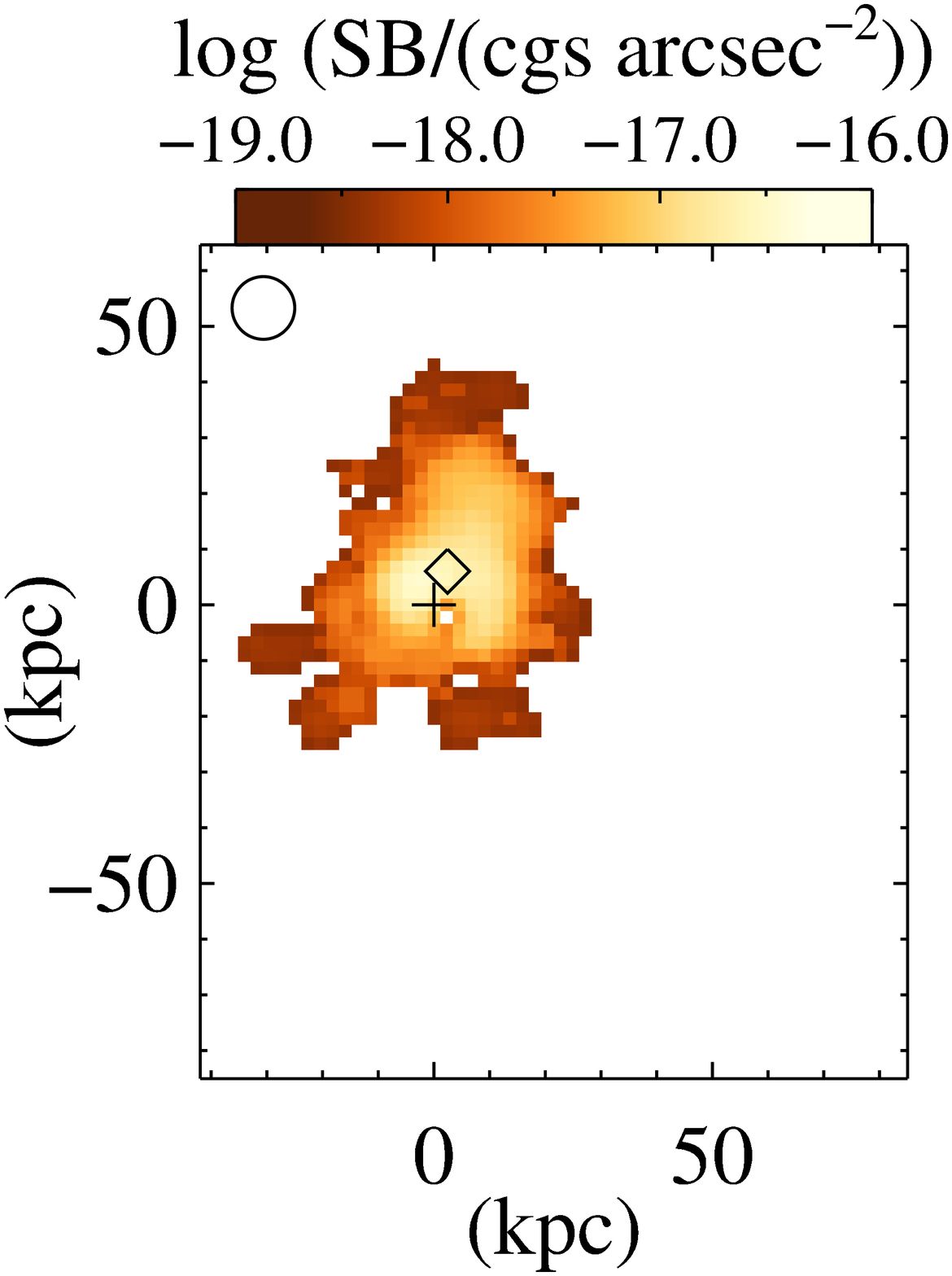}
\hspace{-0.01in}
\includegraphics[height=1.5in]{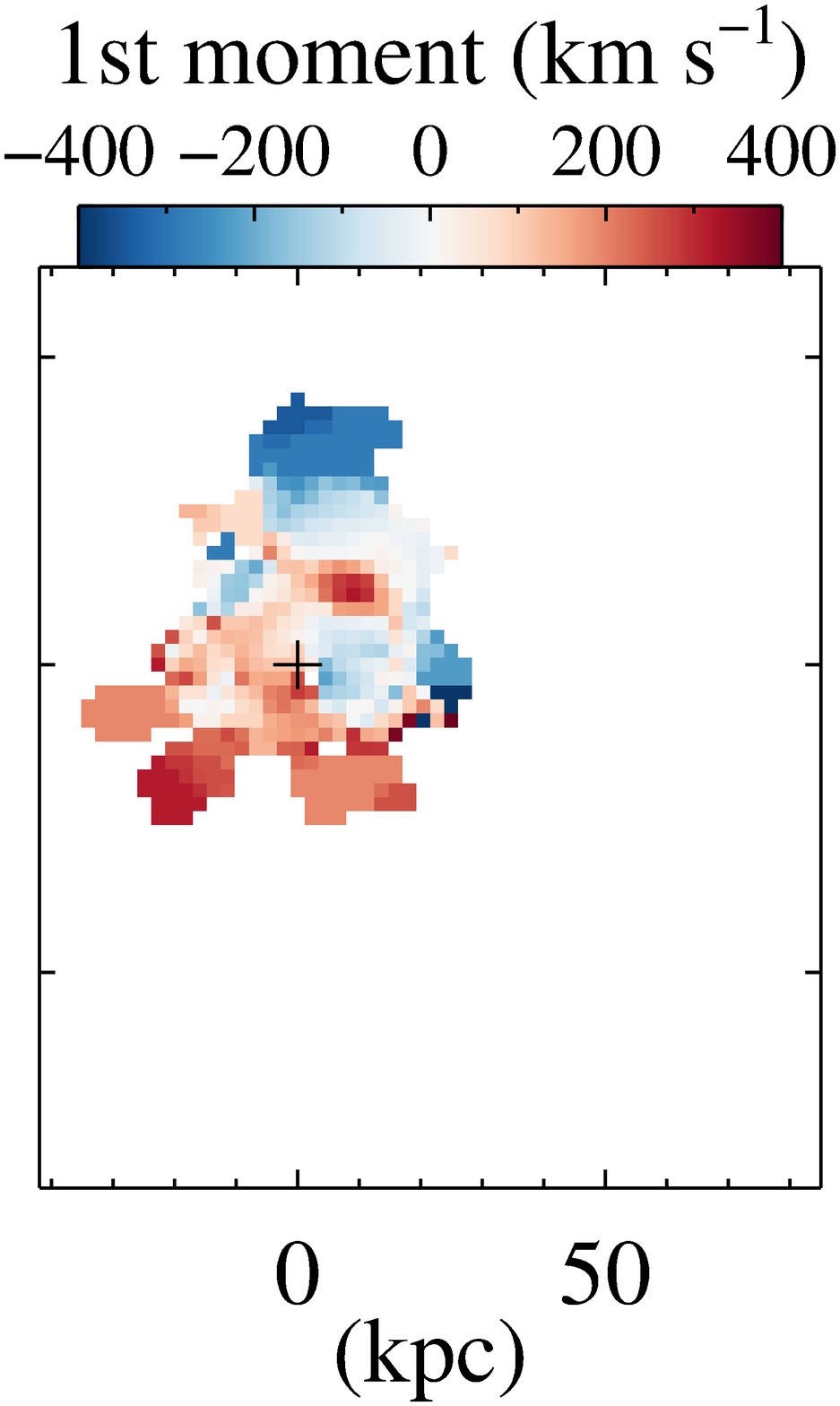}
\hspace{+0.01in}
\includegraphics[height=1.5in]{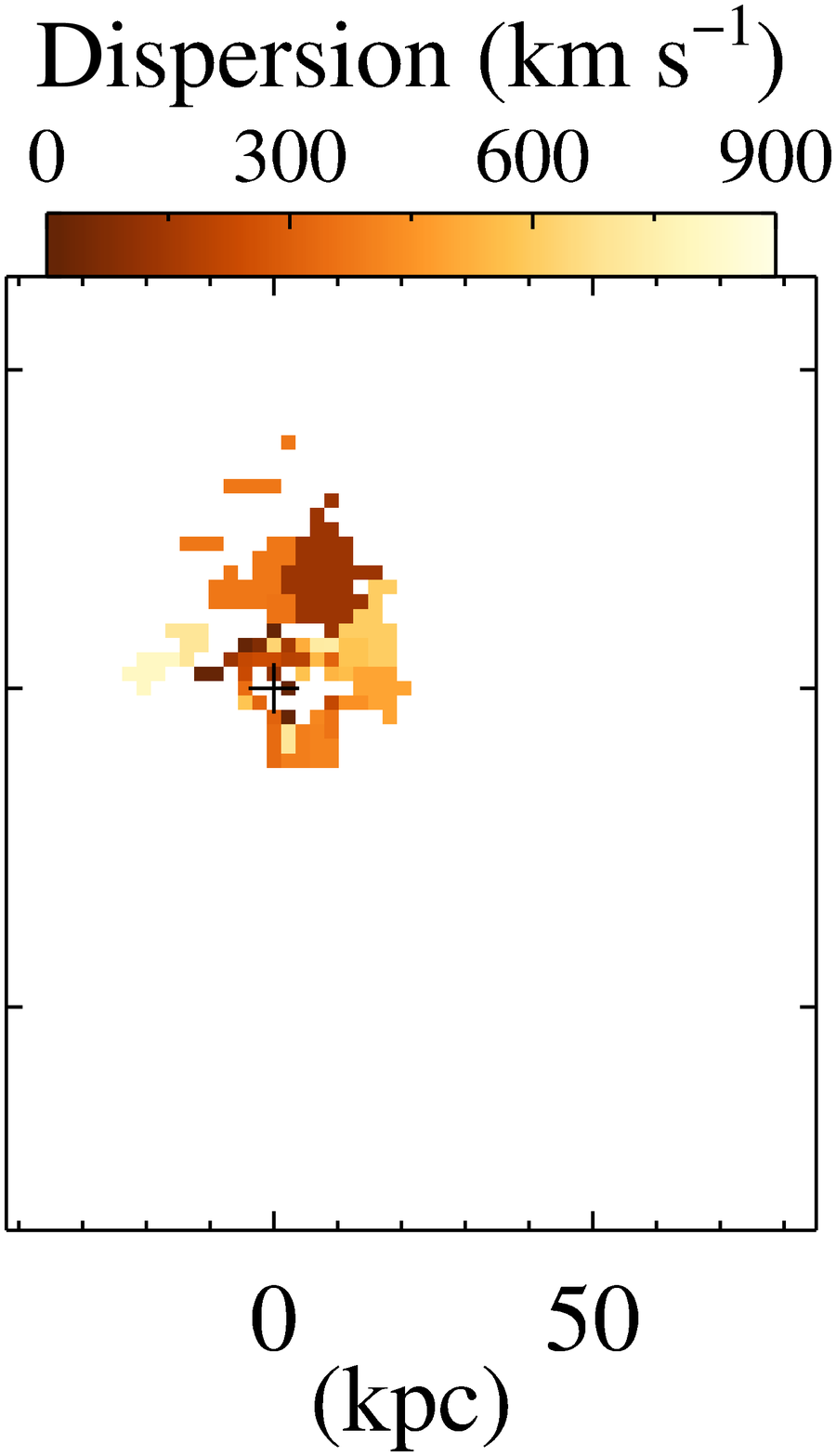}
\hspace{0.0in}
\includegraphics[width=0.27\textwidth]{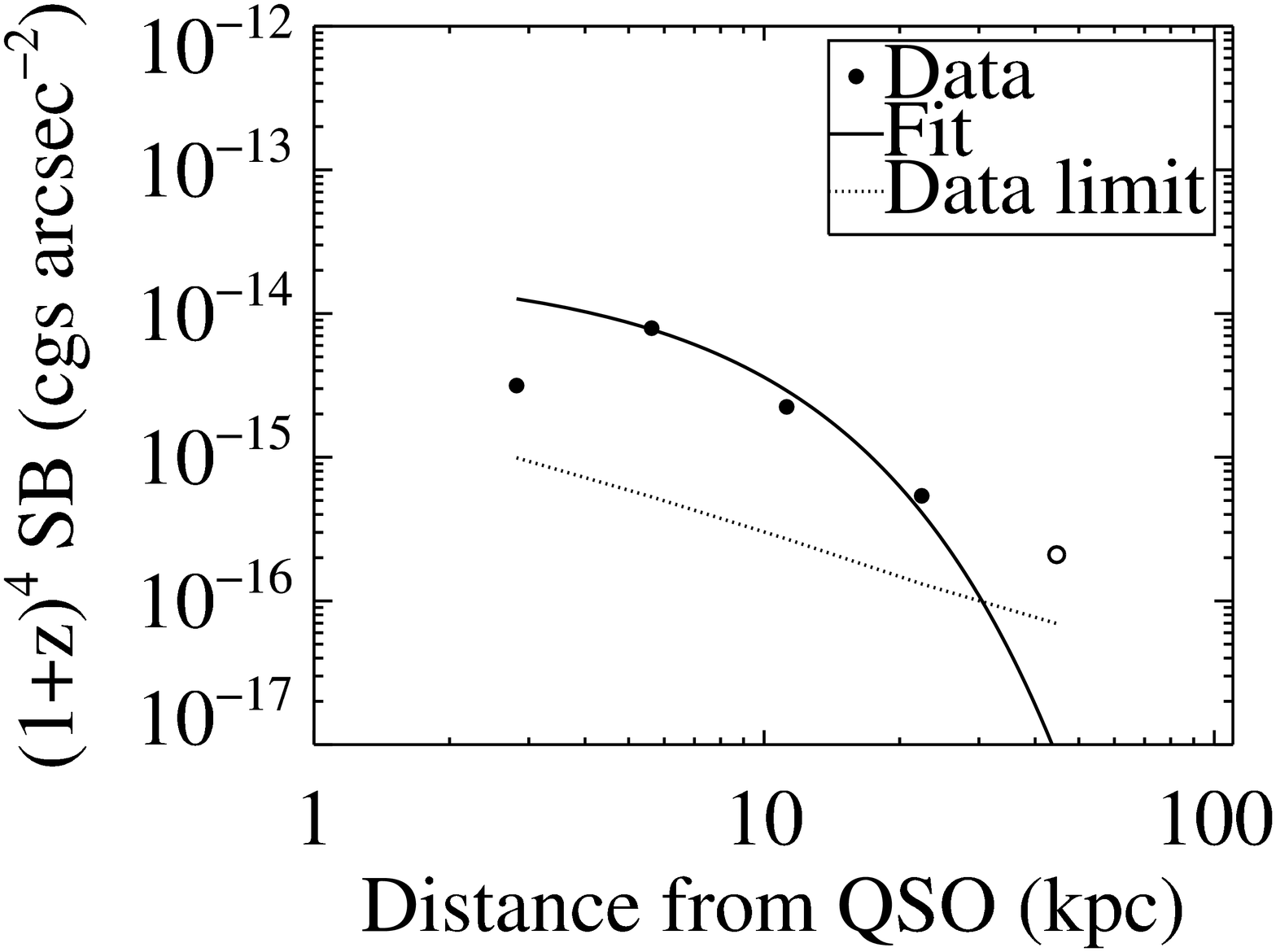}
\hspace{0.0in}
\includegraphics[width=0.27\textwidth]{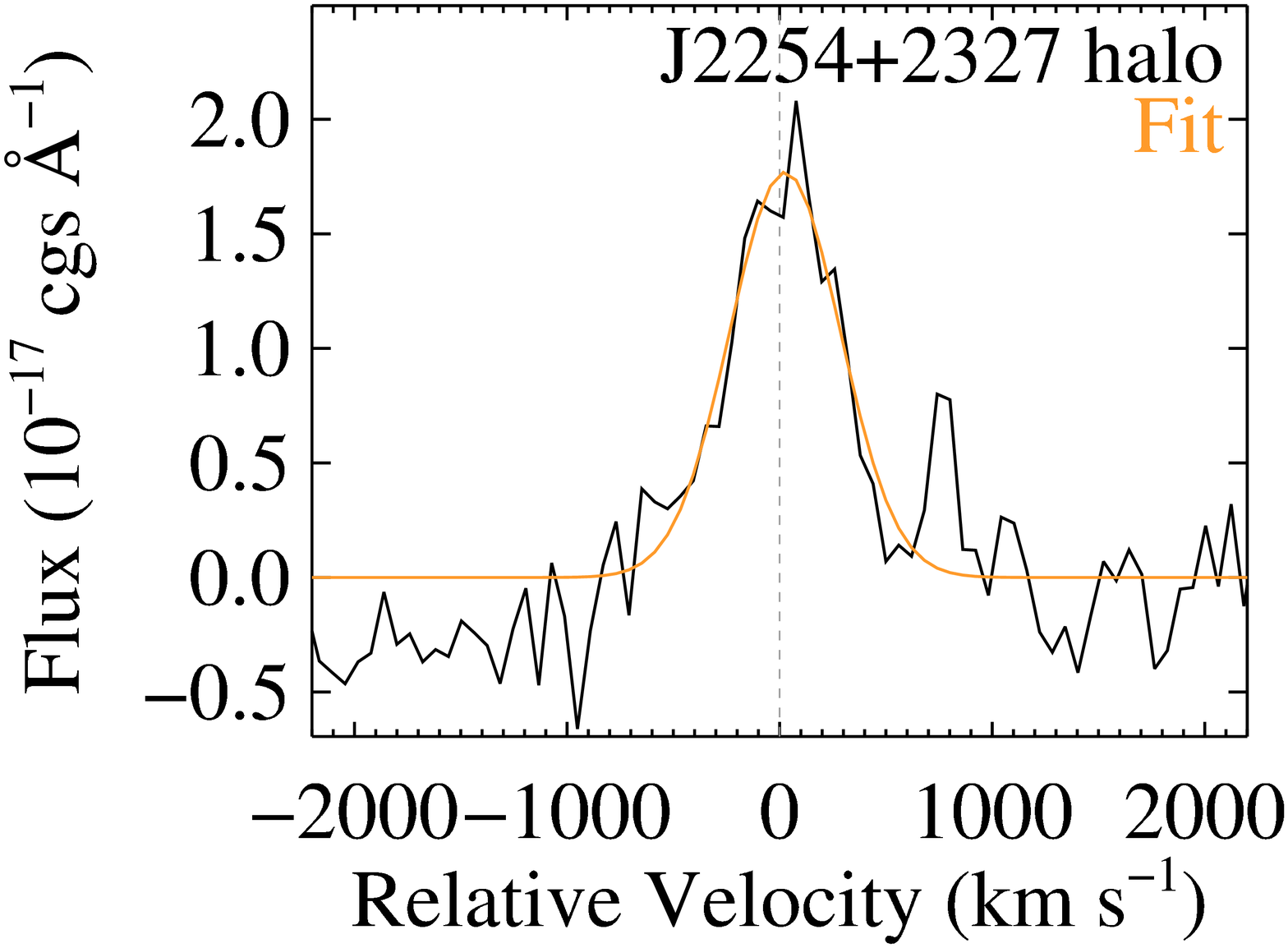} \\

\includegraphics[height=1.5in]{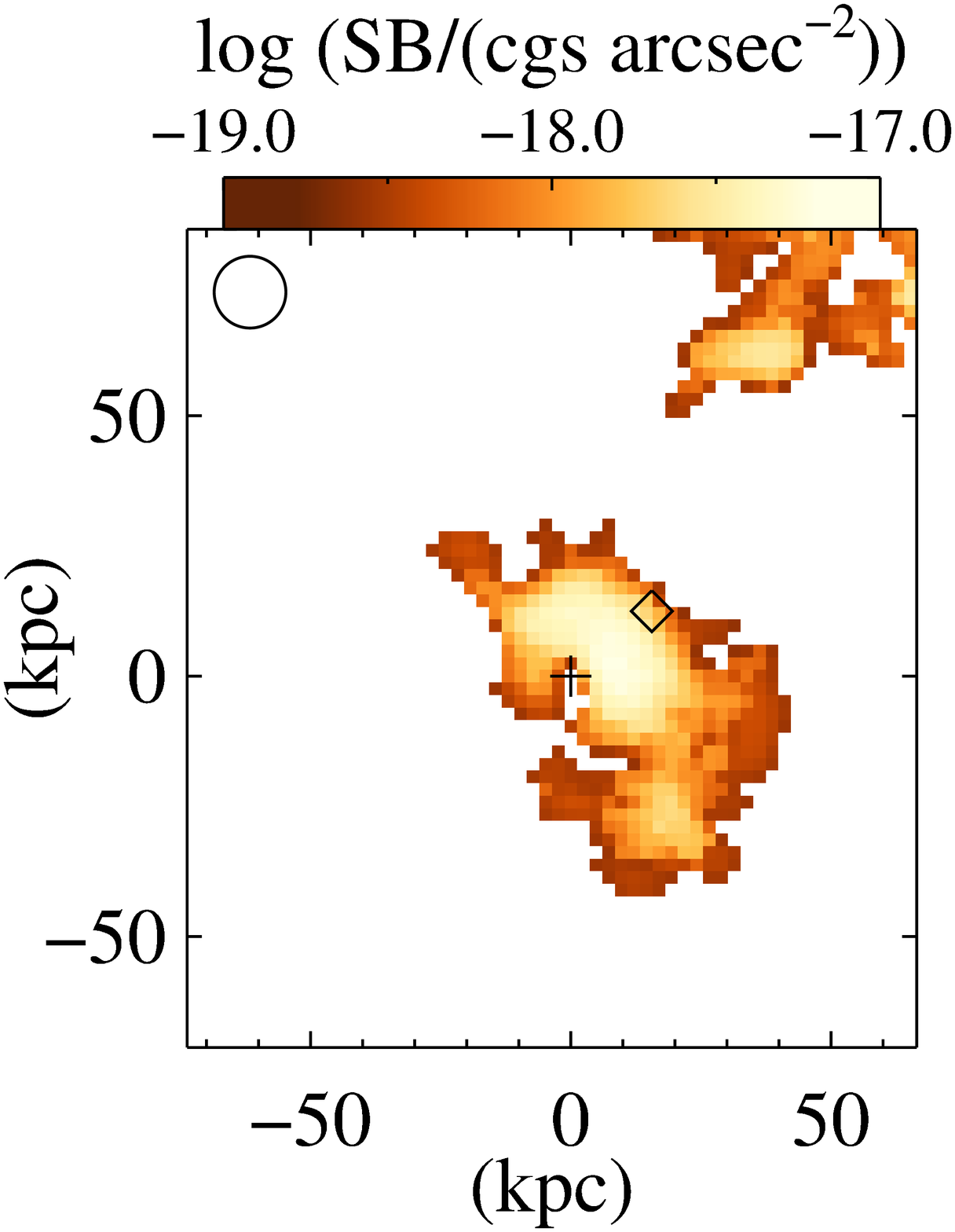}
\hspace{-0.01in}
\includegraphics[height=1.5in]{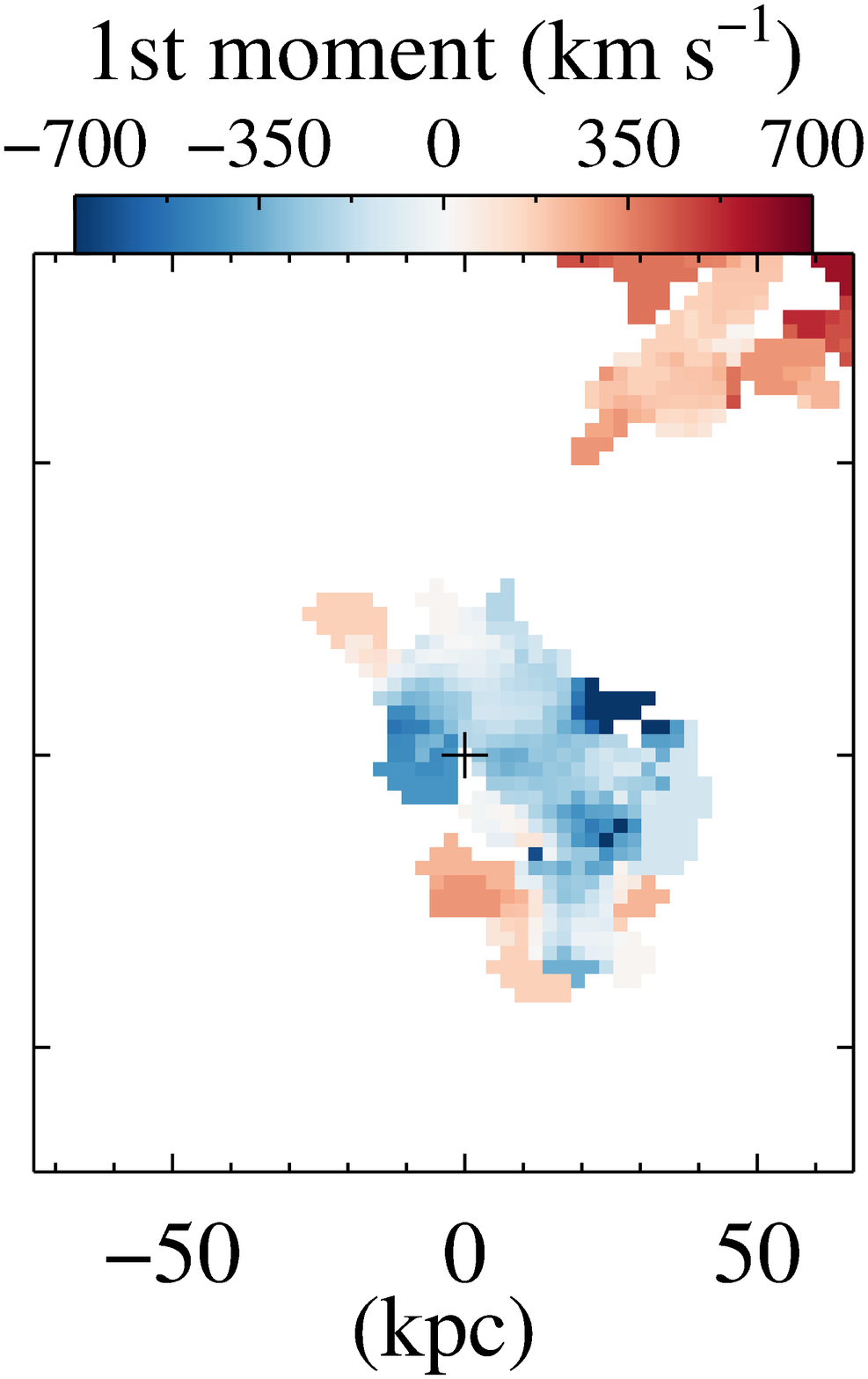}
\hspace{+0.01in}
\includegraphics[height=1.5in]{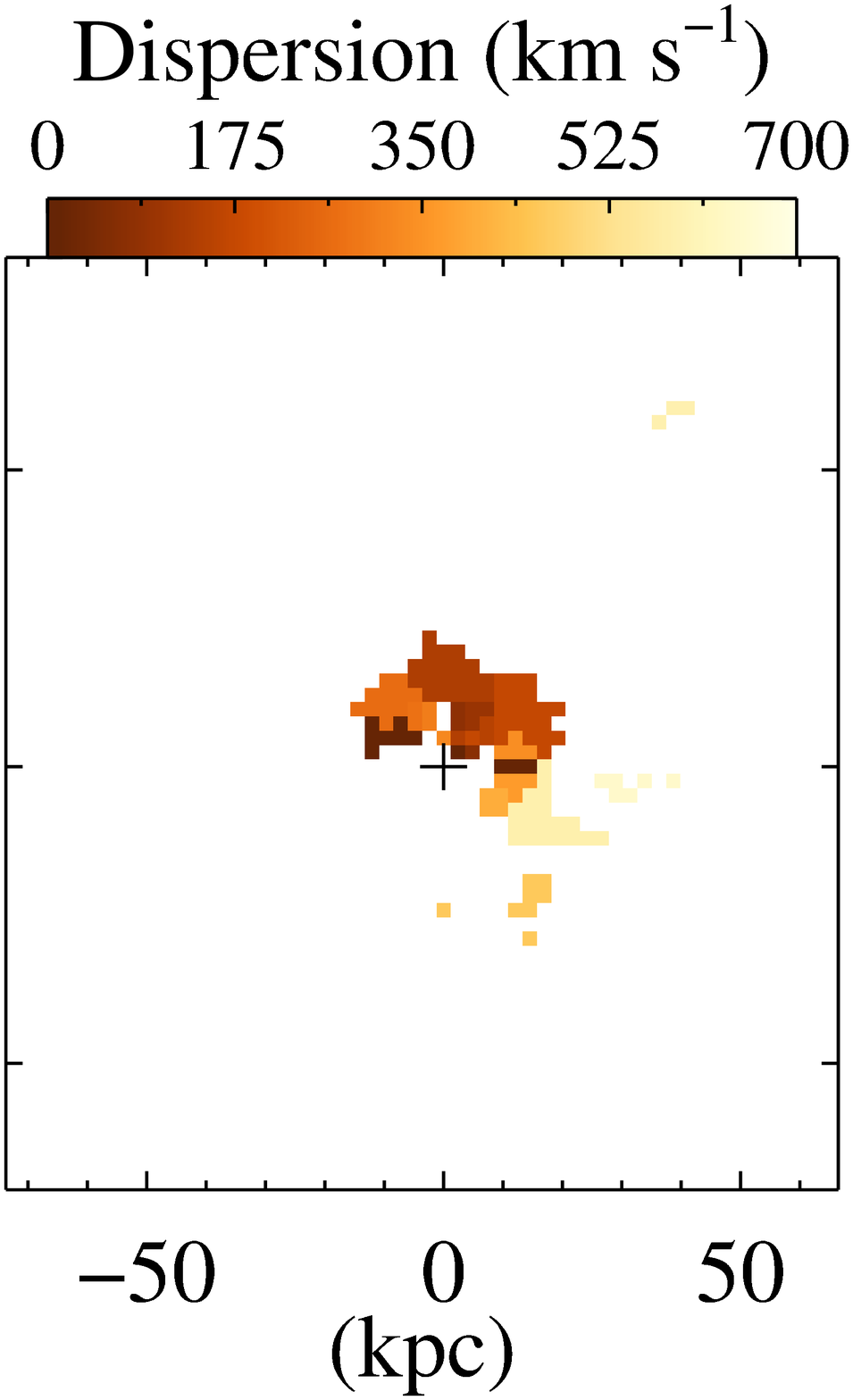}
\hspace{0.0in}
\includegraphics[width=0.27\textwidth]{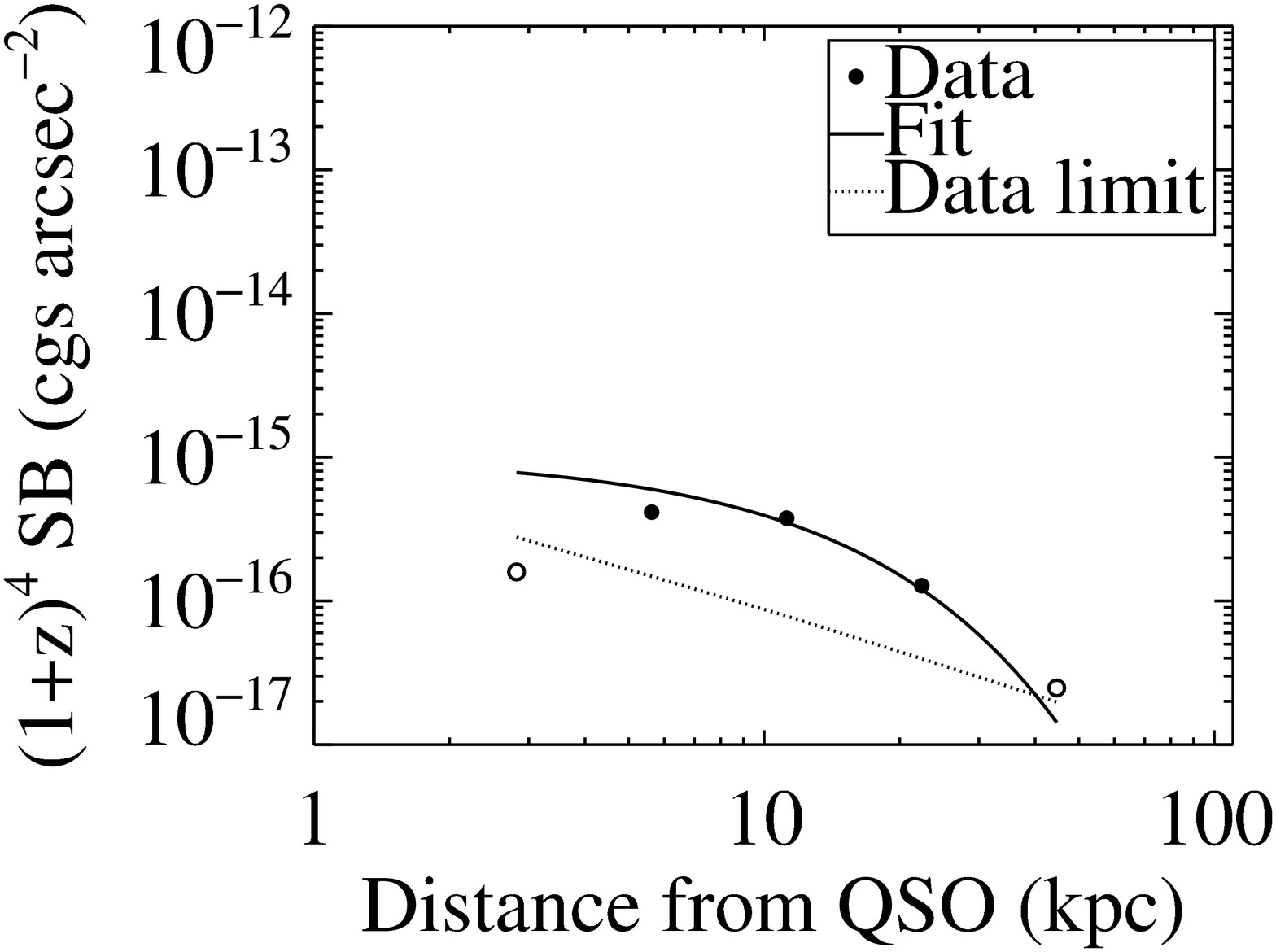}
\hspace{0.0in}
\includegraphics[width=0.27\textwidth]{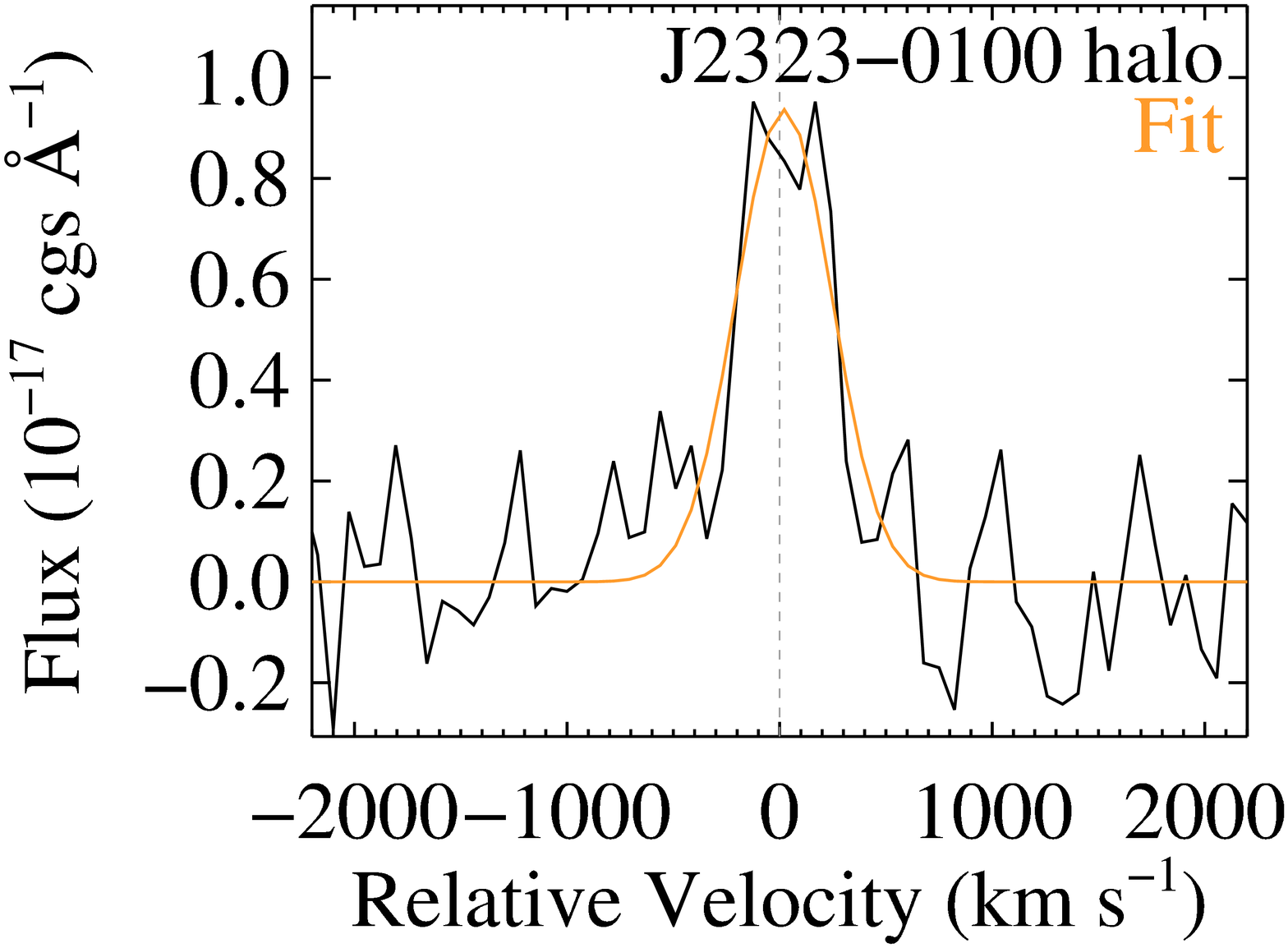}

\raggedright \contcaption{.}
\end{figure*}

\subsection{Kinematics}
\label{sec:sec_vel_maps}

Here we describe our kinematic measurements of the halos in our sample. Because of the multi-dimensional measurements of integral field spectroscopy, we can present the kinematics in both the integrated halo emission and spatially-resolved spectra. 

Figure~\ref{fig:fig_maps}'s fifth column shows the spatially integrated spectra and a single Gaussian fit to the Ly$\alpha$ halo emission. We define zero velocity by the emission centroid of the spatially-integrated halo, without clipping the narrow-emission spike from the spectral template. For the case of J2215$-$0056, we integrate the halo including the extended object to the north. Across the sample our integrated-halo emission spectra generally have similar narrow shapes, with a median spatially-integrated spectrum velocity dispersion of 293~km~s$^{-1}$. J1145+5742 has narrow absorption that is blueshifted from the centroid in its Ly$\alpha$ halo spectrum. But in spite of the absorption, the fit centroid is consistent with the Ly$\alpha$ halo emission peak. For J1652+1728, the Gaussian fit captures the line width for approximating dispersion, but the narrow Ly$\alpha$ halo emission in the central region is blueshifted relative the outer halo, blueshifting the total spatially-integrated spectrum. This centrally concentrated blueshifting, which is seen in J1652+1728's 1st moment velocity map, is also seen in the Ly$\alpha$ emission profile from the central arcsec aperture in Figure~\ref{fig:fig_psf_template} \& \ref{fig:fig_annular_spec}. Blueshifted central emission causes uncertainties in what to clip as halo emission  when subtracting the quasar. Table~\ref{tab:tab_dispersion}'s fifth column shows the resulting velocity dispersion from a Gaussian fit to the integrated halo emission.

\begin{figure*}
\hspace{-0.15in}
\includegraphics[height=2.5in]{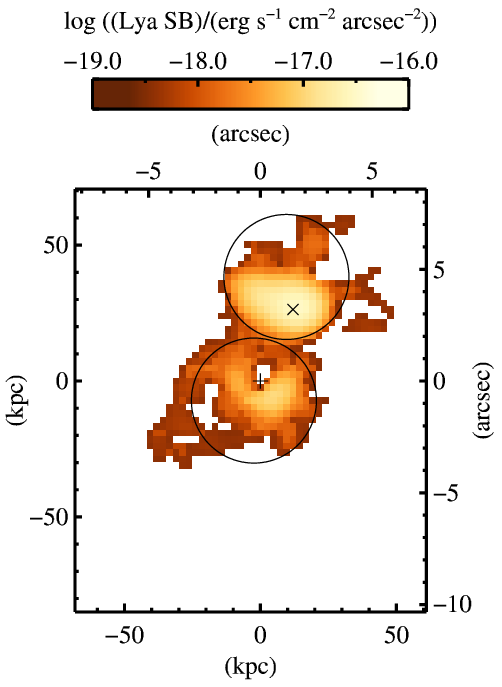}
\includegraphics[height=2.25in]{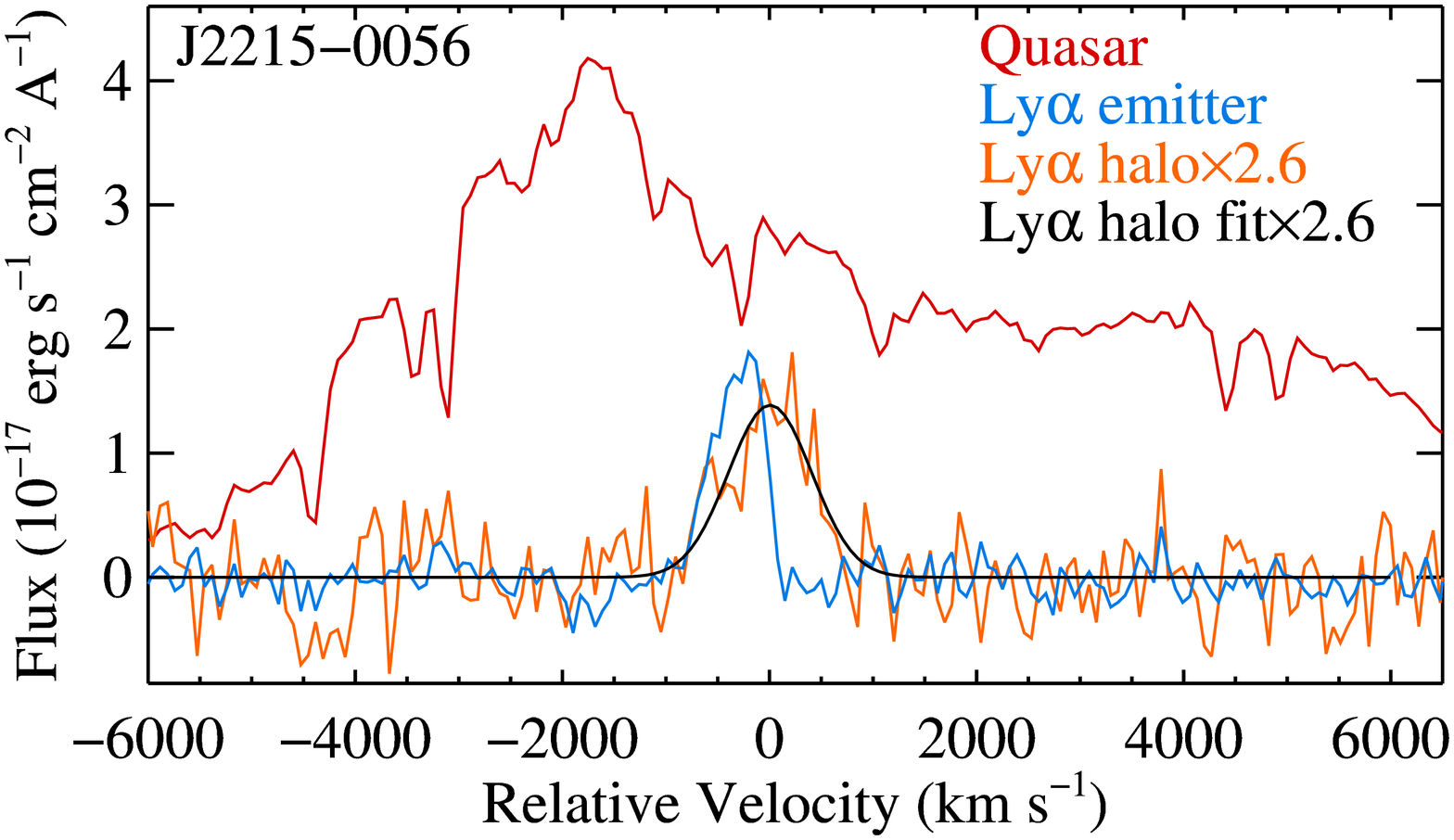}
\caption{Aperture spectra of J2215$-$0056's halo and the associated Ly$\alpha$ emitter (LAE). The left panel displays the same SB map from Fig.~\ref{fig:fig_maps}, over-plotted with the apertures used to generate emission spectra. The plus symbol is the location of the quasar, and the cross symbol is the location of peak brightness for the LAE. The aperture used to extract the halo emission includes the quasar position, but is offset to avoid the LAE. The second aperture above the quasar is used to generate the LAE spectrum. The right panel is the quasar spectra for J2215$-$0056, the extracted aperture spectra from the left panel, and a singe Gaussian fit to the halo emission. In black is J2215$-$0056's quasar spectrum, generated using a 1 arcsecond aperture from the non-PSF subtracted data, blue is the LAE emission, orange is the aperture spectrum of Ly$\alpha$ halo emission, and black is a Gaussian fit to the halo emission. Ly$\alpha$ halo emission is scaled with a constant for visualization and comparison to the other spectra. There is sharp and narrow absorption in the quasar spectrum that visibly aligns with the LAE spectrum centroid. Our extracted halo emission may also display absorption at the same spectral location. There is no strong emission detected near the boundary of these emitters, nor is there a smooth and continuous transition from the halo emission profile to the LAE profile, to suggest there is a luminous arm directly connecting the objects.}
\label{fig:fig_J2215}
\end{figure*}

Figure~\ref{fig:fig_J2215} displays evidence of an extended Ly$\alpha$ emitter (LAE) whose emission centroid coincides with absorption in the quasar spectrum. The centroid of the LAE is $\sim$30~kpc projected distance from the halo peak, and spectroscopically <~$-500$~km~s$^{-1}$ from the redshift measured by the Ly$\alpha$ halo. At higher redshifts (3~<~$z$~<~4.5) LAEs with similar velocity offset and clustering are suggested to be star-forming galaxies orbiting in the quasar halo potential \citep{Fossati+21}. It is uncertain if the extended emitter is powered by quasar radiation.

We describe kinematics at each spatial position around the quasar in 2D maps. Figure~\ref{fig:fig_maps}'s second column shows velocity centroid maps, which are the 1st-moments in velocity space of the flux distribution at each spatial position. Our velocity maps use the same detection region and PSF from the SB maps. We measured relatively low velocity shifts of the halos, at hundreds of km~s$^{-1}$, and do not exhibit energetic outflows of thousands of km~s$^{-1}$. 

Table \ref{tab:tab_dispersion} lists the spatial velocity-centroid standard deviation for each ERQ, other computed quantities of velocity dispersion, and their median. The first column is the standard deviation of the 1st velocity-moment centroid, with a sample median of 288~km~s$^{-1}$. Second is the median velocity dispersion for each object, from the Voronoi-binned velocity dispersion maps in Figure~\ref{fig:fig_maps}, and their respective standard deviation in the third column. Our sample median of the spatial velocity dispersion and its standard deviation are 374 and 114~km~s$^{-1}$ respectively. Finally we show the dispersion of the Gaussian-fit to the spatially integrated Ly$\alpha$ halo emission, with a median of 293~km~s$^{-1}$. Discussion and analysis of these quantities is in Section \ref{sec:sec_ERQ_properties}.

Our data do not have sufficient signal-to-noise ratios to measure halo velocity dispersion using second velocity moments (see L22 for more discussion). We instead use Gaussian fits to a Voronoi-binned map (see \citet{Rupke+19}). Figure~\ref{fig:fig_maps}'s third column shows the final Ly$\alpha$ Gaussian velocity dispersion maps. These maps are generally smaller than the other maps of SB and velocity shift, but as an independent detection method they confirm the general morphology of the extended halo. In the case of J2215$-$0056, the northern emitting region has unusually uniform dispersion across its area, unlike any other extended emission cloud we detect. It is not clear if this Ly$\alpha$ emitter is a distinct object. For J2215$-$0056 we determine the systemic redshift by extracting the extended halo excluding the extended object. The remaining ERQ Ly$\alpha$ halos show relatively low dispersion across their maps, in the hundreds of km~s$^{-1}$.

Some ERQs show a velocity gradient from one edge of the halo to the other of $\sim$1,000~km~s$^{-1}$ (e.g., J1232+0912 and J1451+0132), and the coherent transition from redshift to blueshift helps confirm the measured velocities are valid. A rotating disc of $\sim$100~kpc size is not expected to have formed by $z \sim 2$ \citep[e.g.,][]{DeFelippis+20,Huscher+21}. One ERQ, J0220+0137, has a kinematically distinct cloud at the eastern boundary of the FOV, $\sim$65~kpc from the quasar, and shifted in velocity space about $-$1,500~km~s$^{-1}$ from the Ly$\alpha$ halo redshift. The kinematic separation makes it unlikely to be of the quasar halo diffuse gas, and it is omitted in measuring the Ly$\alpha$ halo emission. ERQs J1451+2338 and J1652+1728 have a noticeable circular patch at the center of their 1st velocity-moment map, and are likely the residual from imperfect quasar subtraction caused by complexities in their Ly$\alpha$ profile.

\subsection{Blueshifted Absorption}
\label{sec:sec_blue_absorption}
 
\begin{figure*}
\hspace{-0.15in}
\includegraphics[height=1.78in]{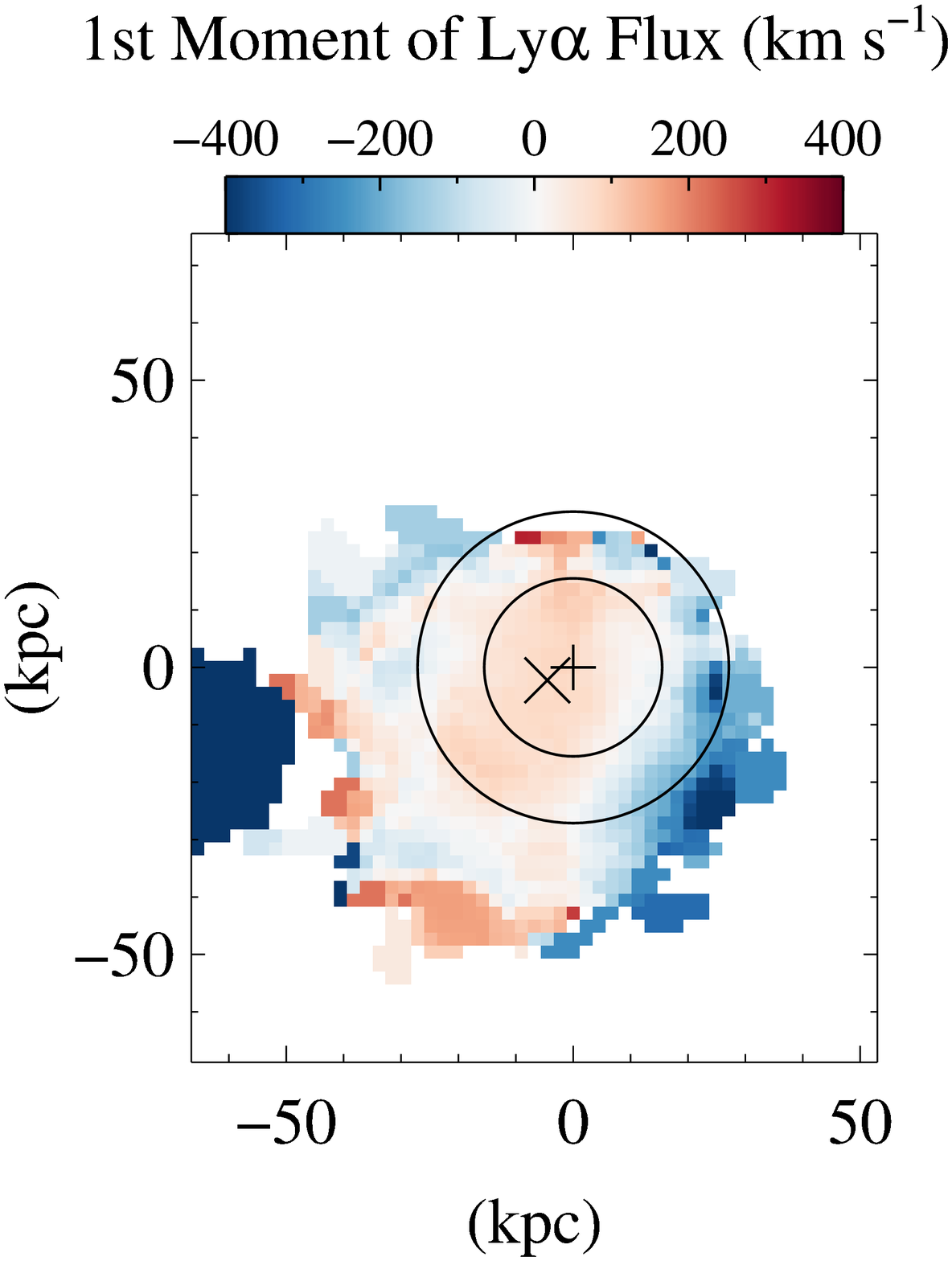}
\includegraphics[height=1.78in]{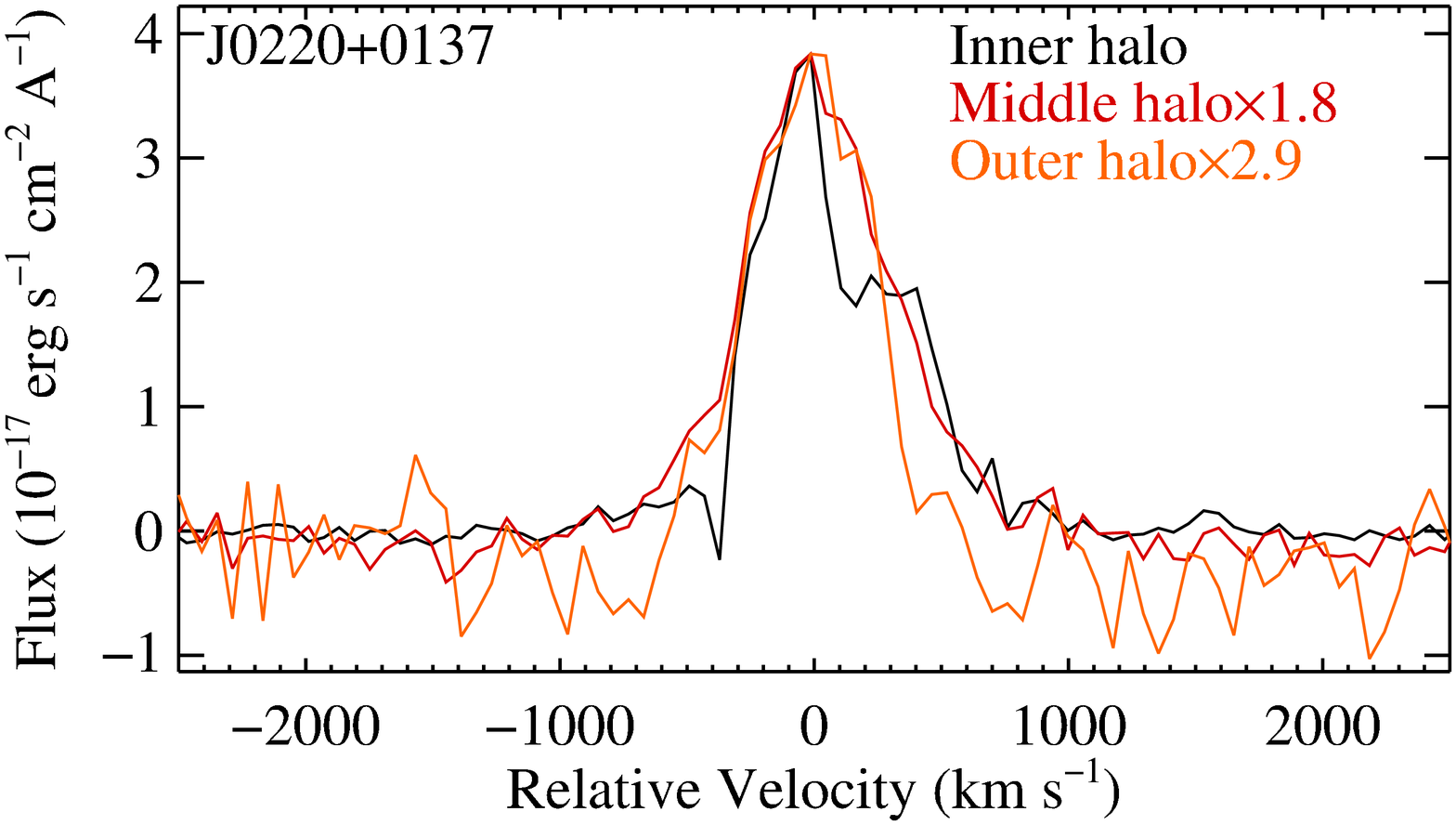} \\
\includegraphics[height=1.78in]{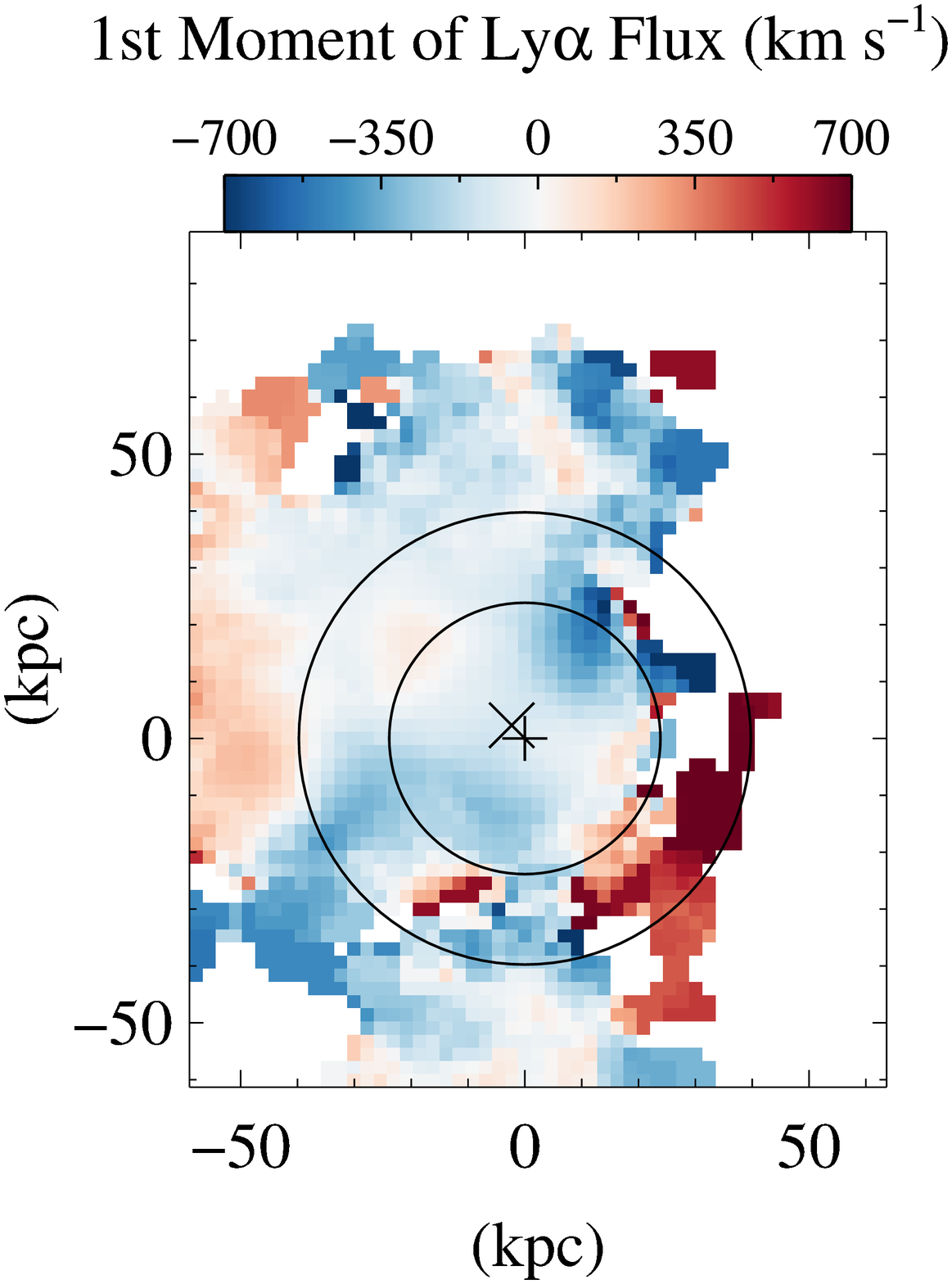}
\includegraphics[height=1.78in]{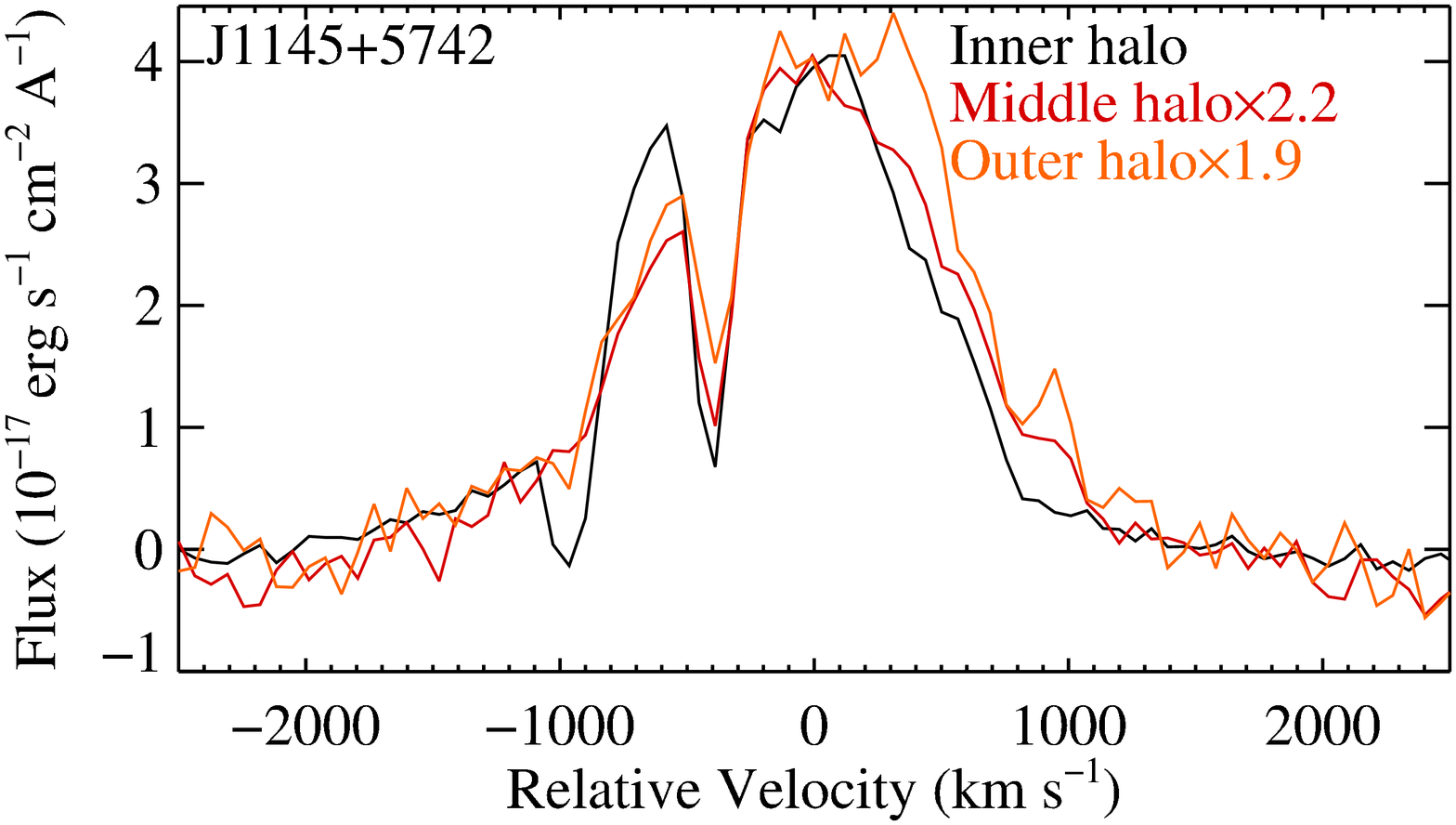} \\
\includegraphics[height=1.78in]{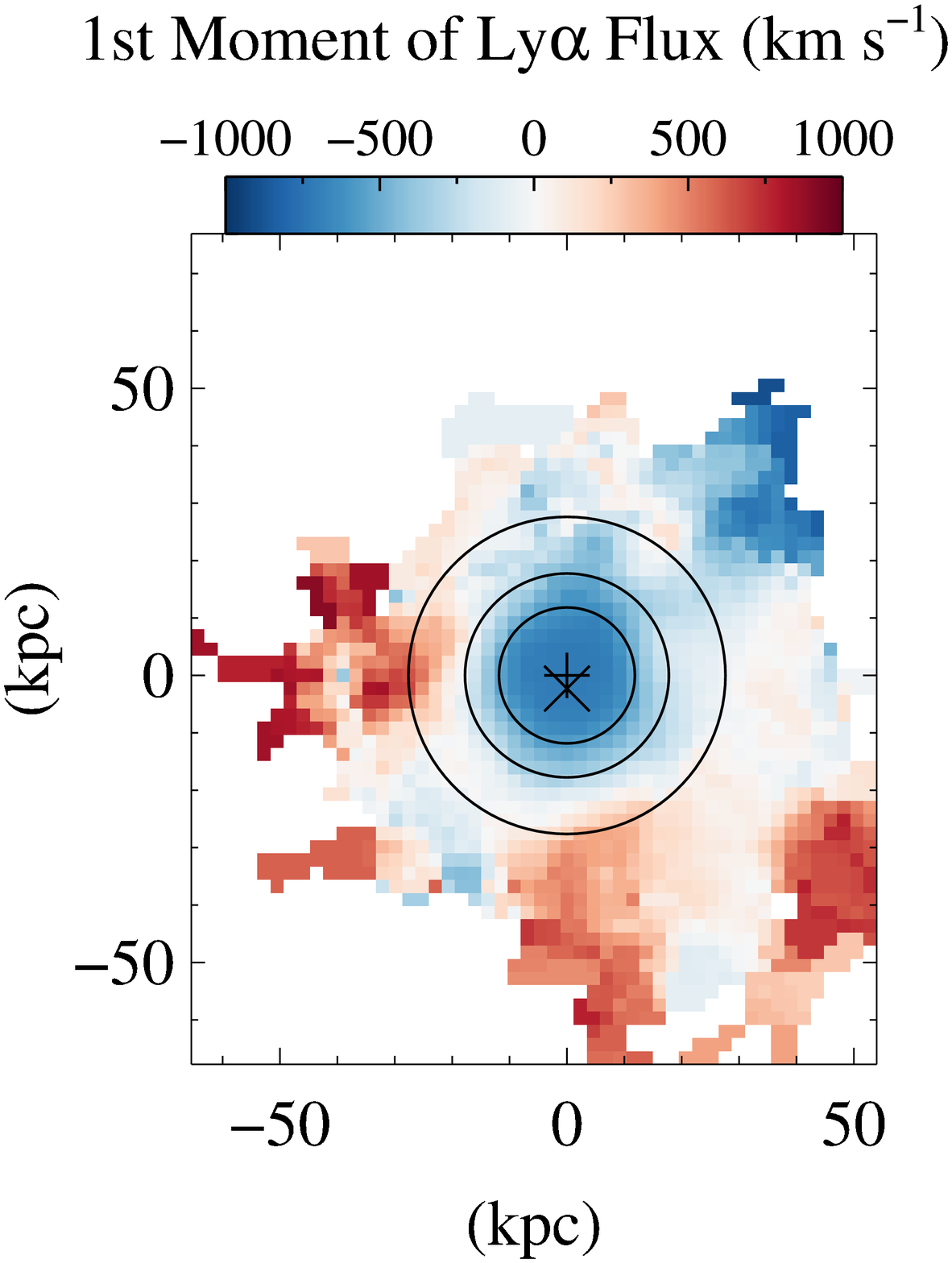}
\includegraphics[height=1.78in]{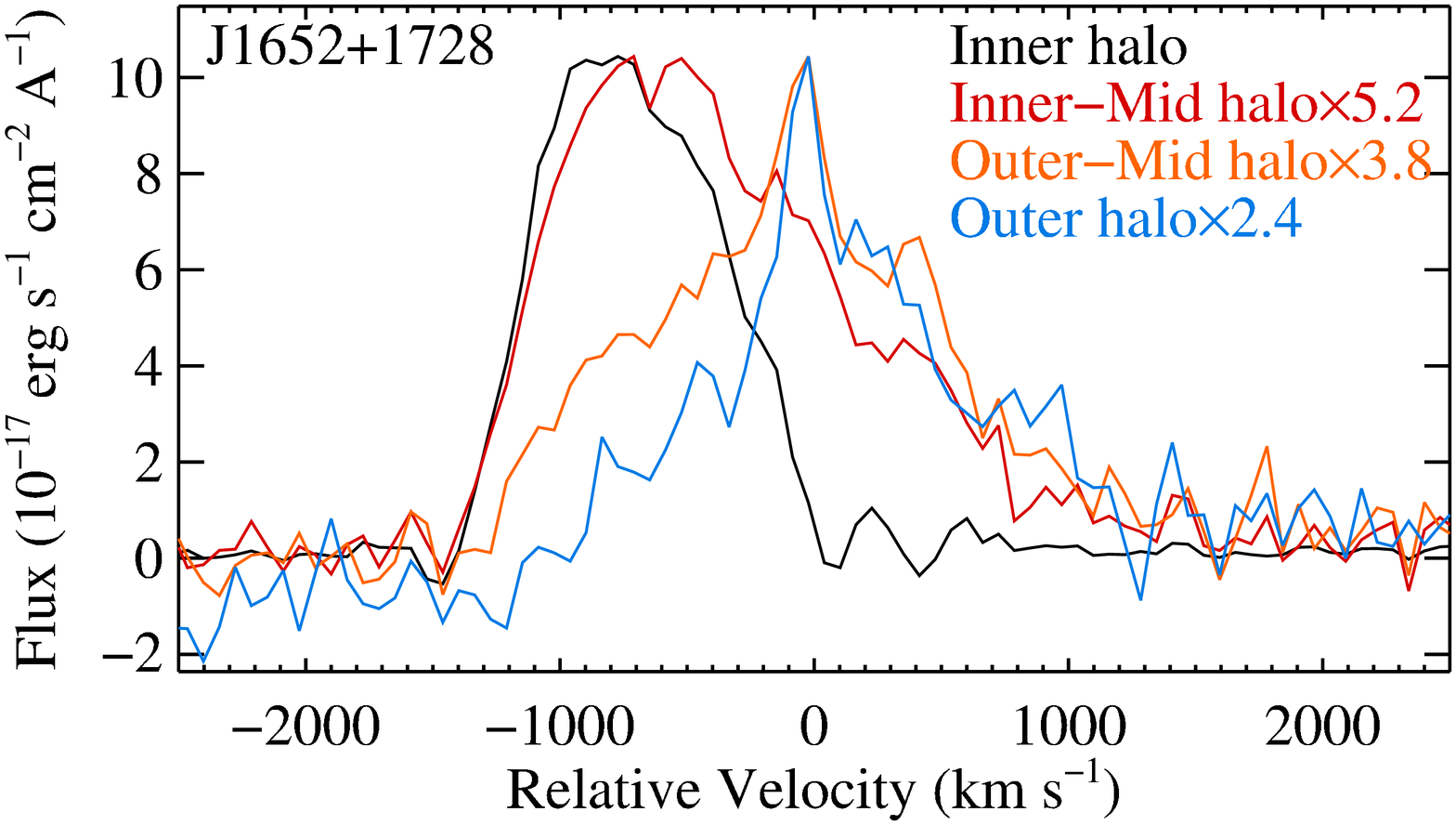} \\
\includegraphics[height=1.78in]{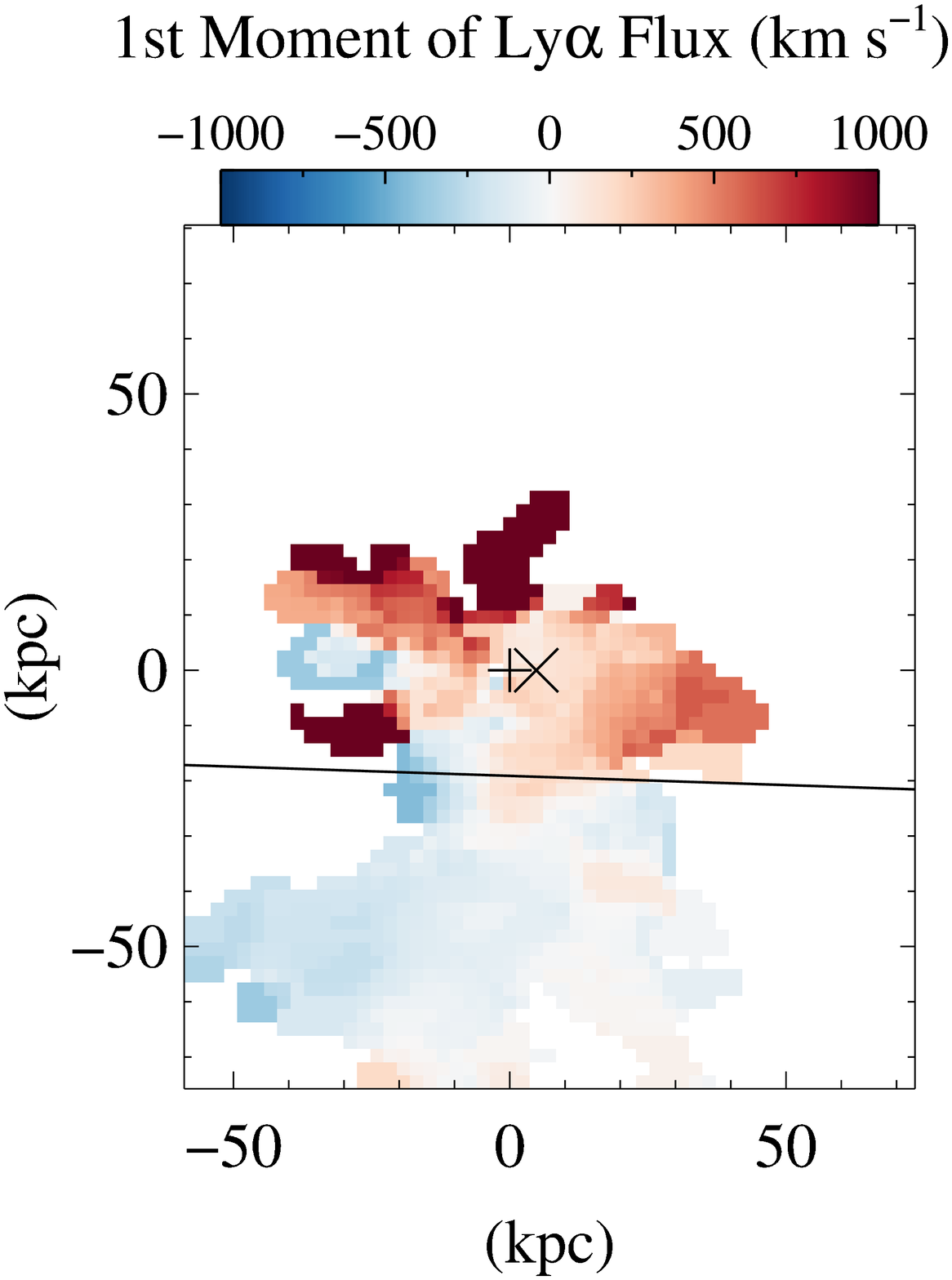}
\includegraphics[height=1.78in]{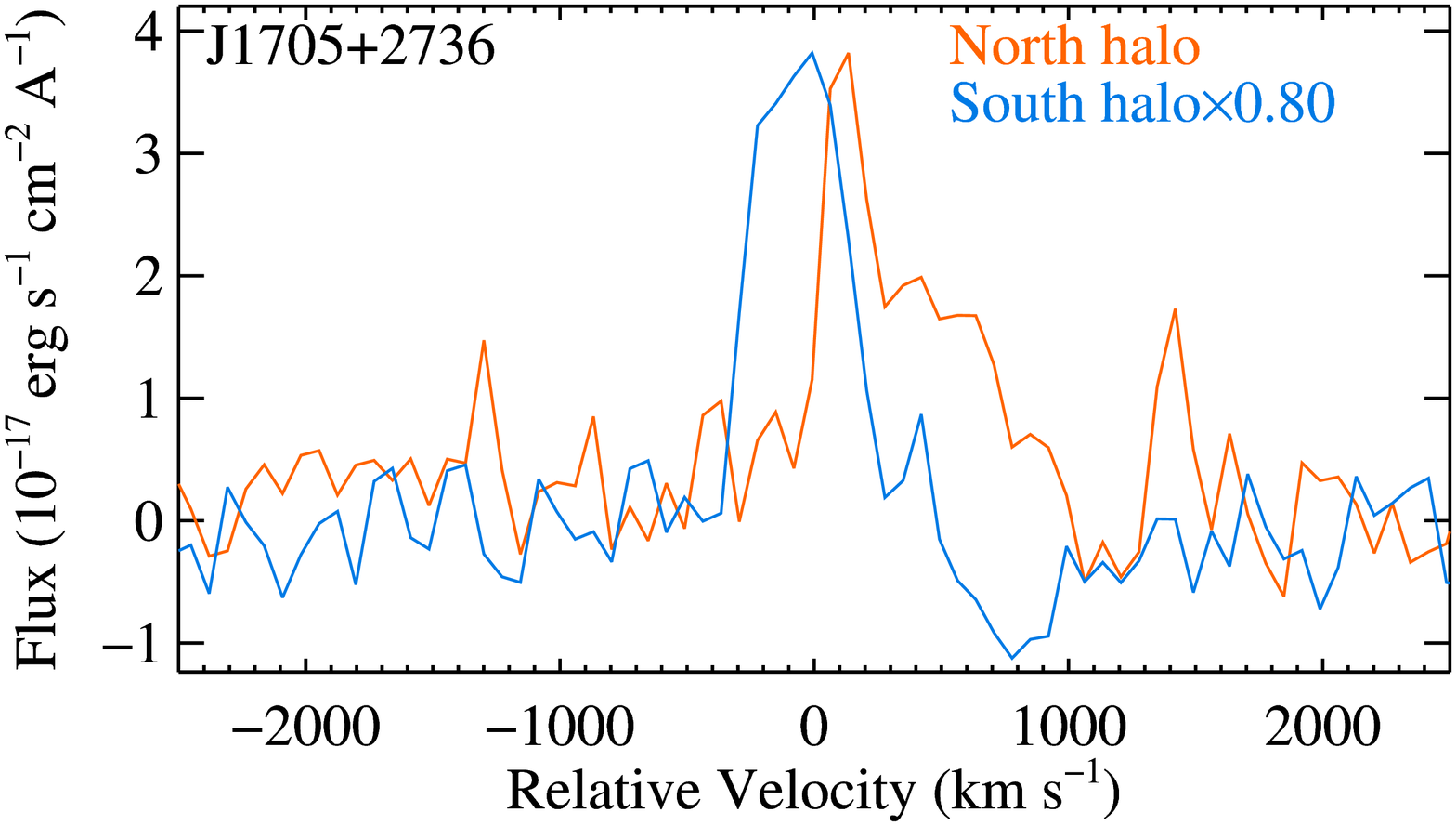}
\caption{Aperture spectra of Ly$\alpha$ halo emission for J0220+0137, J1145+5742, J1652+1728, and J1705+2736. The left column displays the same 1st moment velocity map from Fig.~\ref{fig:fig_maps}, over-plotted with the apertures used to generate Ly$\alpha$ emission spectra. The plus symbol is the location of the quasar, and the cross symbol is the peak of the Ly$\alpha$ halo emission. For J0220+0137 \& J1145+5742, circular aperture boundaries are drawn to define three regions, one innermost region, a middle transition region, and an outer halo region, which includes all emission outside the circular apertures. J1652+1728 has an additional subdivision to show a quiescent circular region of the middle halo ($\sim$20~kpc), which lacks clumpy inflows/outflows. For J1705+2736, the halo is bisected into a northern and southern region for analysis. The right panel is the corresponding quasar's Ly$\alpha$ emission spectra from each halo region, and each spectra is scaled with a constant for visualization and comparison to the other regions' emission profile. J0220+0132 is an example of a complex profile that is revealed after it's divided into apertures, instead of the smoothed emission features in 's integrated halo spectra. A flat emission spectrum between 100-400~km~s$^{-1}$ is seen in the innermost halo ($\leq \sim$20 kpc), perhaps by a combination of multi-component gas kinematics and/or absorption, and is otherwise lost in the total integrated halo spectrum shown in Fig.~\ref{fig:fig_maps}. J1145+5742 displays strong absorption at $-$400 km s$^{-1}$ across the $\sim$120 kpc span of the halo, and weaker absorption at $-$950 km s$^{-1}$ across $\leq \sim$50 kpc. Ly$\alpha$'s emission profile shape remains consistent across the $\sim$120 kpc span of the halo. J1652+1728's Ly$\alpha$ emission profile displays a rapid transition beyond $\sim$20 kpc from the quasar, and becomes sharper peaked and symmetrically broader at its base. This symmetric broadening is likely from the patches of red and blueshifted clumps in the outer regions of the velocity map. The inner halo region's peak emission is blueshifted to $\sim$900 km s$^{-1}$, and displays absorption at the the redshift of the peak outer-halo emission. J1705+2736 shows moderate blueshifting in the southern lobe of the Ly$\alpha$ halo, and becomes more narrow. The northern halo emission profile may be asymmetrically broadened on the red side by redshifted clumps at the outer boundary of the halo.}
\label{fig:fig_annular_spec}
\end{figure*}

Figure~\ref{fig:fig_annular_spec} shows aperture and annular spectra of selected ERQs, to further understand the spatial and spectral distribution of Ly$\alpha$ emission and absorption features. One ERQ shows definite blueshifted absorption, J1145+5742, similar to what was found in J0006+1215, by L22. J1145+5742 has strong absorption features in its integrated spectra, and consistently shows deep and narrow absorption in apertures across the FOV, out to $\sim$50~kpc, blueshifted at $-$400~km~s$^{-1}$. In its inner halo spectrum, only within $\sim$25~kpc, there is blueshifted absorption at about $-$950~km~s$^{-1}$. J1451+0132 has blueshifted absorption in its emission profile of the innermost and spatially unresolved inner halo, at $\sim$800~km~s$^{-1}$ and $\sim$1250~km~s$^{-1}$, but not in the outer halo profile. These blueshifted absorption features resemble those of spatially resolved Ly$\alpha$ halo spectra that require outflows \citep[e.g.,][]{Li+21b}. 

J1652+2736 has more unique profile features. It has primarily blueshifted central halo emission from apertures within 15$-$20~kpc of the quasar (about $-$900~km~s$^{-1}$), and has gradually redshifted and more symmetric Ly$\alpha$ halo emission at larger annular radii. Its aperture spectra reveal that the asymmetric emission profile in the integrated Ly$\alpha$ halo is not from blueshifted absorption, but from the central halo emission being blueshifted, relative to the extended halo. 

Our data show evidence of multi-component emission, and likely also affected by absorption. Overall, we do not find evidence for strong absorption in the Ly$\alpha$ halo emission profiles. The intrinsic emission profile of Ly$\alpha$ is not known for these quasars in order to model the level of absorption \citep[e.g.,][]{Wang+21}, and analysis of the environment which may cause absorption is beyond the scope of this work.

\subsection{Comparisons to Blue Quasars}
\label{sec:sec_compare}

We have a full sample of eleven ERQ Ly$\alpha$ halos. One of the main goals of our study is to compare the Ly$\alpha$ halos around ERQs to Type-I blue quasars roughly matched to the ERQs in redshift and luminosity ($\ge 10^{47}$ ergs s$^{-1}$, at cosmic noon). Our focus is on Type-I quasars, but we also include Type-IIs for some of the comparisons, as described below. Our median ERQ luminosity of $L_{\text{bol}} = 5\times 10^{47}$erg s$^{-1}$ and $z = 2.61$ is only slightly more luminous than other samples we compare. We use the same method described in Section \ref{sec:sec_ERQ_properties} to recalculate all of the blue quasar sample luminosities for consistency. Combining the four blue quasar samples, they have a range of medians $L_{\text{bol}} \approx 1.0 - 4.0 \times 10^{47}$erg s$^{-1}$ and $z \approx 2.3 - 3.2$, totaling 108 quasars (\cite{Cai+19,ArrigoniBattaia+19,Borisova+16b,Mackenzie+21}).



We recalculate all of the blue quasar sample luminosities in the same way as ERQs, using W3 photometry to estimate bolometric luminosities (see Section \ref{sec:sec_sample_properties}).

Our comparison samples are the following. We take the quasar sample from \citet{Cai+19}, consisting of 16 blue quasars of median $z = 2.3$, and $L_{\text{bol}} = 10^{47.2}$erg s$^{-1}$. We take the 61 blue quasars from \citet{ArrigoniBattaia+19}, with median $z = 3.2$ and $L_{\text{bol}} = 10^{47.4}$erg s$^{-1}$. \cite{Borisova+16b} sample consists of 19 luminous blue quasars of median $z = 3.2$ and median bolometric luminosity $10^{47.6}$~erg~s$^{-1}$. \citet{Mackenzie+21} have 12 blue quasars of median $z = 3.2$, and are on the fainter side of these samples, with median $L_{\text{bol}} = 10^{47.0}$erg~s$^{-1}$. \cite{denBrok+20} sample consists of four Type-II quasars of median $z = 3.4$ and median $L_{\text{bol}} = 10^{46.7}$erg~s$^{-1}$. We verify that the intrinsic bolometric luminosities estimated based on mid-IR luminosities are similar to those estimated with X-ray luminosities, and the correction factors modeled in \citet{Shen+20}. A final Type-II quasar from \citet{Sanderson+21} is at $z = 3.2$, which is a mid-IR-only source not detected in the optical.

\begin{figure*}
\includegraphics[width=\columnwidth]{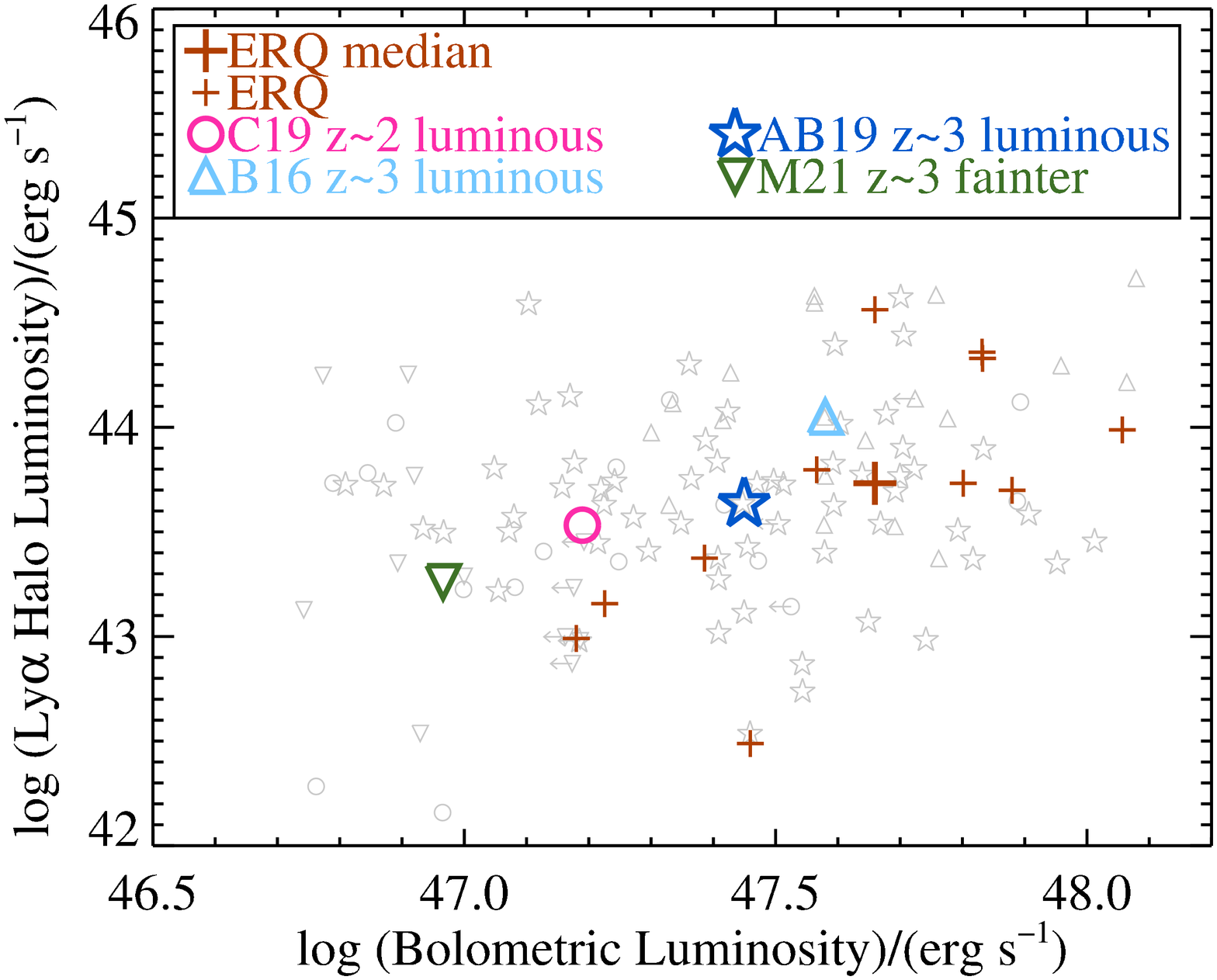}
\caption{Computed logarithmic bolometric luminosity vs logarithmic Ly\(\alpha\) halo luminosity of various literature samples and our ERQs. Upper and lower limits are shown with arrows. Brown pluses are the individual ERQs, and the large brown plus is the median of our sample. Blue quasar samples have median values shown as the large pink circle, dark blue star, light blue triangle, and the dark green downward triangle symbol. Individual quasars from other samples shown as small gray symbols matching their median symbol. There is significant overlap of individual quasars in our sample and luminous blue quasar samples. The median of these samples shows a positive luminosity trend for bolometric luminosity and Ly\(\alpha\) halo luminosity across all samples.}
\label{fig:fig_LhalLbol}
\end{figure*}

Figure~\ref{fig:fig_LhalLbol} shows the relationship between the halo luminosity and the bolometric luminosity of the quasar. In order to make these comparisons, we verified that masking or not masking the inner 1-arcsecond yields similar halo luminosities, within about 10 per cent of the total halo luminosity. We also assessed changes in computed halo luminosity with or without Ly$\alpha$ spike clipping. Four of six ERQs with a spike showed an increase in computed halo luminosity of 20$-$40 percent. Halo luminosities of J1451+0132 \& J1652+1728 more than doubled, each increasing by factors of $\sim$2.25. When subtracting the emission of the quasar, the shape of the spectral template that is subtracted is important for determining the residual halo emission. Halo emission is not as distinct in typical blue quasar spectra, and does not  change the spectral template shape (ref. Sections \ref{sec:reduction_label} \& \ref{sec:sec_narrow_lya}, and Figure~\ref{fig:fig_psf_template}). Most of ERQ Ly$\alpha$ halo emission is concentrated in the inner halo regions near the quasar PSF, so modifications to the spike are more important for ERQ Ly$\alpha$ halo luminosity.

Our ERQ median is $L_{\text{halo}} = 5.40\times 10^{43}$ergs s$^{-1}$, where the luminous blue quasar populations have medians in the range $L_{\text{halo}} \approx 5-10\times 10^{43}$ergs s$^{-1}$. L22 noted that the Ly$\alpha$ halo luminosity around the reddest quasar in our sample, J0006+1215, is roughly three times lower than expected from the matched blue quasar samples. In our larger sample, the second reddest ERQ, J2323$-$0100, stands out for having the lowest halo luminosity, roughly 10 times lower than expected from the blue quasars (see Table~\ref{tab:table_SB_maps} and Fig.~\ref{fig:fig_LhalLbol}). However, the full range of ERQ halo luminosities in our study falls within the range of those for blue quasars. We also find that the ERQs follow a weak trend for larger halo luminosity around quasars with larger bolometric luminosity that was noted previously in blue quasars by \citet{Mackenzie+21}.

\begin{figure*}
\includegraphics[width=\columnwidth]{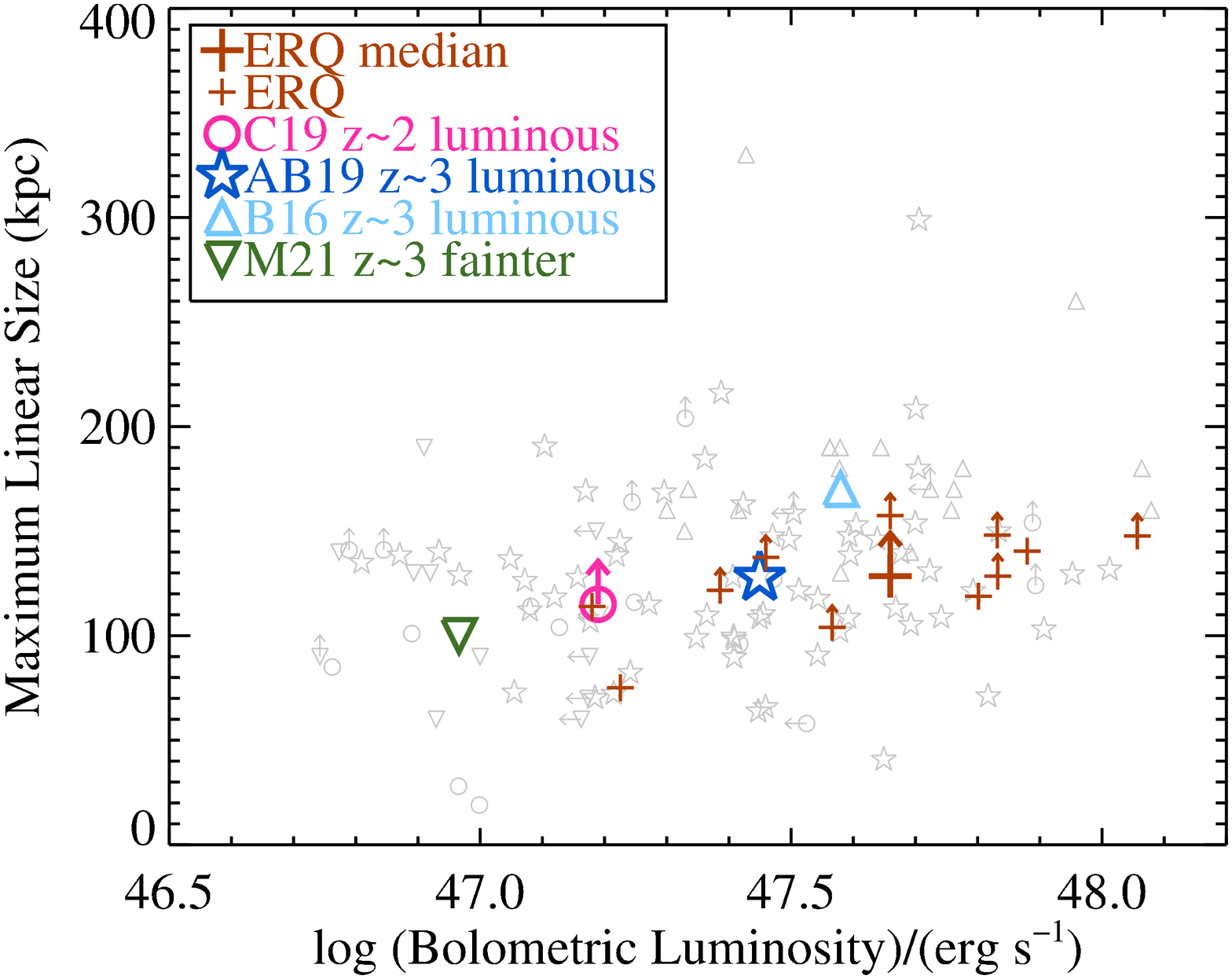}
\caption{Computed logarithmic bolometric luminosity vs maximum linear size of the halo, using identical color and symbol legend as Figure~\ref{fig:fig_LhalLbol}. Lower limits are shown for medians computed from samples that contain halos extending to the edge of the field of view. There is overlap of individual quasars with other samples, and a weak positive luminosity trend can be seen across the medians of the samples.}
\label{fig:fig_linszLbol}
\end{figure*}

Figure~\ref{fig:fig_linszLbol} shows the maximum linear size vs bolometric luminosity. Many of the ERQs are shown with only lower limits on their maximum halo size, because they exceed the field of view of KCWI (see also Table~\ref{tab:table_SB_maps} and Section \ref{sec:SB_maps}). We compare samples from \citet{Borisova+16b}, \citet{ArrigoniBattaia+19}, \citet{Cai+19},  and \citet{Mackenzie+21}. For \citet{ArrigoniBattaia+19} we estimate a maximum linear size for each quasar from the sum of maximal radial extent for the halo plus $\sqrt{(\text{covered area})/\pi}$. It was noted in \citet{ArrigoniBattaia+19} that the size of blue quasar nebulae did not increase significantly with longer exposure time, confirming our size measurements are mostly insensitive to integration time. We report the maximum linear size as lower limits when the halo SB map extends to the edge of the FOV. Similarly to the luminosity trend, ERQs populate the high-luminosity region of a positive trend with the maximum linear size of Ly$\alpha$ halos and bolometric luminosities.

\begin{figure*}
\includegraphics[width=\columnwidth]{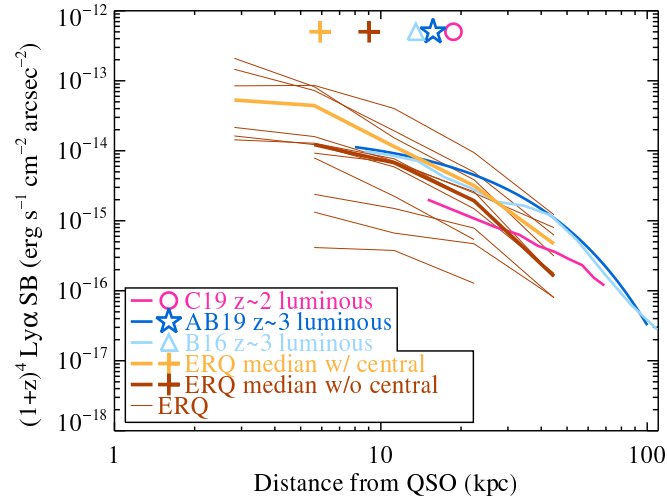}
     \caption{Circularly averaged surface brightness radial profiles of our sample of ERQs in comparison with other samples of blue quasars at comparable redshift and luminosity. The bold brown line is the ERQ sample median at each surface brightness bin for which there is detected halo emission. Thin brown lines are the individual ERQs. Other colors correspond to the other samples of blue quasars, pink, blue, and light blue for \citet{Cai+19}, \citet{ArrigoniBattaia+19}, and \citet{Borisova+16a}, respectively. Across the top of the figure~shows the exponential scale lengths, color coded to match each sample. Notice that the scale lengths of ERQs are shorter than blue quasar samples, and thus have profiles that fall off more rapidly at large distances. Also notice that ERQs with higher surface brightness tend to have detectable Ly\(\alpha\)-spike halo emission down to the central projected radius from the quasar. Negative values of surface brightness from Fig.~\ref{fig:fig_maps} are not shown in this figure. All surface brightnesses are corrected for cosmological dimming.}
\label{fig:fig_rad_prof}
\end{figure*}

Figure~\ref{fig:fig_rad_prof} shows the surface brightness radial profile of ERQs pseudo-narrowband images, the median radial profiles of ERQs and several blue quasar samples, and the median exponential scale length of each quasar sample. Each point on the radial curve shows the average SB of an annulus centered on the quasar. Table~\ref{tab:table_rad_profile} presents the data for our sample. An important feature we take advantage of is that obscuration behaves as a coronagraph, and allows probing of the inner-most regions of the halo. Six ERQs have had their quasar emission subtracted without the Ly$\alpha$ spike, and we measure their SB at smaller projected distances from the quasar than the blue quasar samples. For comparison, we take the median SB radial profile from \cite{Borisova+16b}, \citet{ArrigoniBattaia+19}, and \citet{Cai+19}. The \cite{Borisova+16b} radial profile is obtained from \citet{Marino+19}. All samples have comparable size of the point-spread function corresponding to a FWHM of 1.4 arcsec or 12 kpc. For our ERQ sample we present two median radial profiles. One median is for ERQs which can probe to zero projected distance, and the other is omitting the central 1-arcsecond region ($\sim$4~kpc radius), for comparison to other quasar studies that do not have the Ly$\alpha$ spike.

Along the top of Figure~\ref{fig:fig_rad_prof} we also present the exponential scale length fit to the median radial profiles. We fit an exponential function defined in Section \ref{sec:SB_maps} to the median SB radial profile, and record fit values in Table~\ref{tab:table_rad_profile}. Exponential scale length probes the brighter inner halo region, in contrast to the maximum linear size which depends on more diffuse low-SB emission. Our full sample's median radial profile, omitting the innermost bin (0$-$4~kpc) for comparison with blue quasar samples, has a scale lenth $r_h = 9.0$~kpc. Median profile scale lengths for other blue quasar samples are 13.5, 15.7, and 18.7~kpc for \citet{Borisova+16b}, \citet{ArrigoniBattaia+19}, and \citet{Cai+19}, respectively. For ERQs with a Ly$\alpha$ spike we can probe deeper to the inner halo regions, at projected distance 0$-$4~kpc, because of modified PSF subtraction (ref. Sections \ref{sec:reduction_label} \& \ref{sec:sec_narrow_lya}). Considering only the ERQs with a Ly$\alpha$ spike, we can compute a scale length of their median profile $r_h = 5.9$~kpc. ERQs are generally more luminous than blue quasars, and there should be a natural tendency for more luminous quasars to have larger halos, as everything scales with higher luminosity. Despite this intuitive tendency, ERQ Ly$\alpha$ halos are characteristically more compact in their inner regions than those of luminous blue quasars.

\begin{figure*}
\includegraphics[width=\columnwidth]{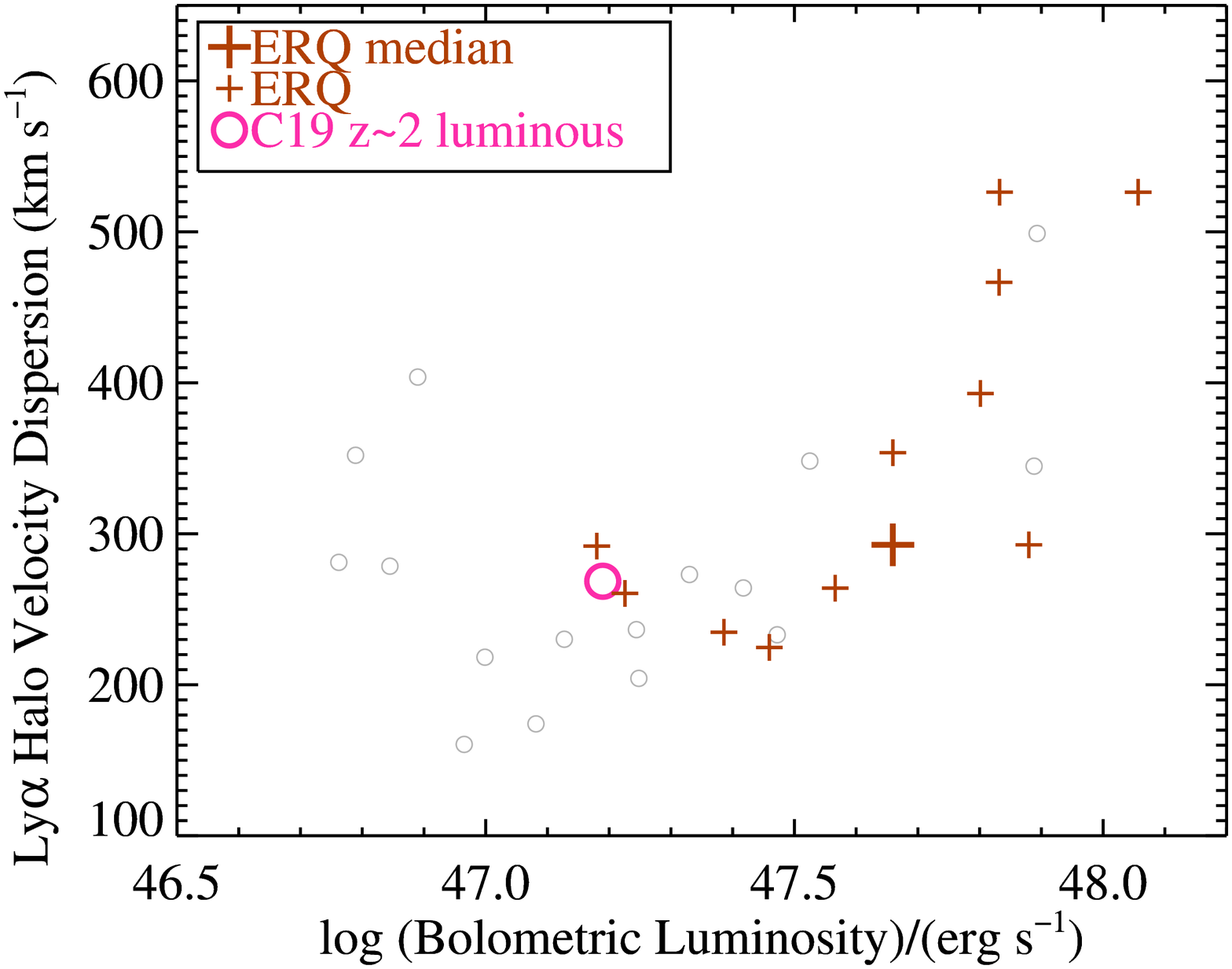} 
\caption{Comparison of logarithmic bolometric luminosity vs spatially integrated Ly\(\alpha\) halo velocity dispersion for samples which it was measured, using identical legend as Figure~\ref{fig:fig_LhalLbol}. Halo velocity dispersion for this comparison is taken from a Gaussian fit to the spatially integrated halo spectrum, shown as the last column in Figure~\ref{fig:fig_maps}. \citet{Cai+19} is the only blue quasar survey from the blue quasar surveys we have considered in this paper also computed integrated halo velocity dispersion. There is a positive luminosity trend among individual quasars, but the trend is uncertain for the two sample medians available for this analysis.}
\label{fig:fig_dispLbol}
\end{figure*}

Figure~\ref{fig:fig_dispLbol} compares ERQ Ly$\alpha$ halo spatially-integrated velocity dispersion vs bolometric luminosity with those from blue quasars which have measured spatially integrated velocity dispersion in the literature. We did not have enough signal to compute 2nd-moment velocity dispersions for the halo maps. Our ERQ velocity dispersions for direct comparison are computed from the Gaussian fits to the total integrated halo spectrum, shown in the last column of Figure~\ref{fig:fig_maps}, and are tabulated in the last column of Table~\ref{tab:tab_dispersion}. Other samples compute a 2nd-moment velocity dispersion, but \citet{Cai+19} also reports a velocity dispersion from spatially integrated velocity dispersion for the integrated halo spectrum, which we can directly compare. Line fitting of the integrated spectrum is a more robust dispersion measurement than moment maps (see discussion in  \citet{OSullivan+20}). Our ERQ sample's median integrated Ly$\alpha$ halo dispersion is $293$ km~s$^{-1}$. \citet{Cai+19} measured their sample median of $269$ km~s$^{-1}$. We see the halo velocity dispersion of ERQs follow the distribution of the luminous blue quasars of \citet{Cai+19}. Both distributions generally follow a trend for larger dispersion in more luminous quasars.

\begin{figure*}
\includegraphics[width=\columnwidth]{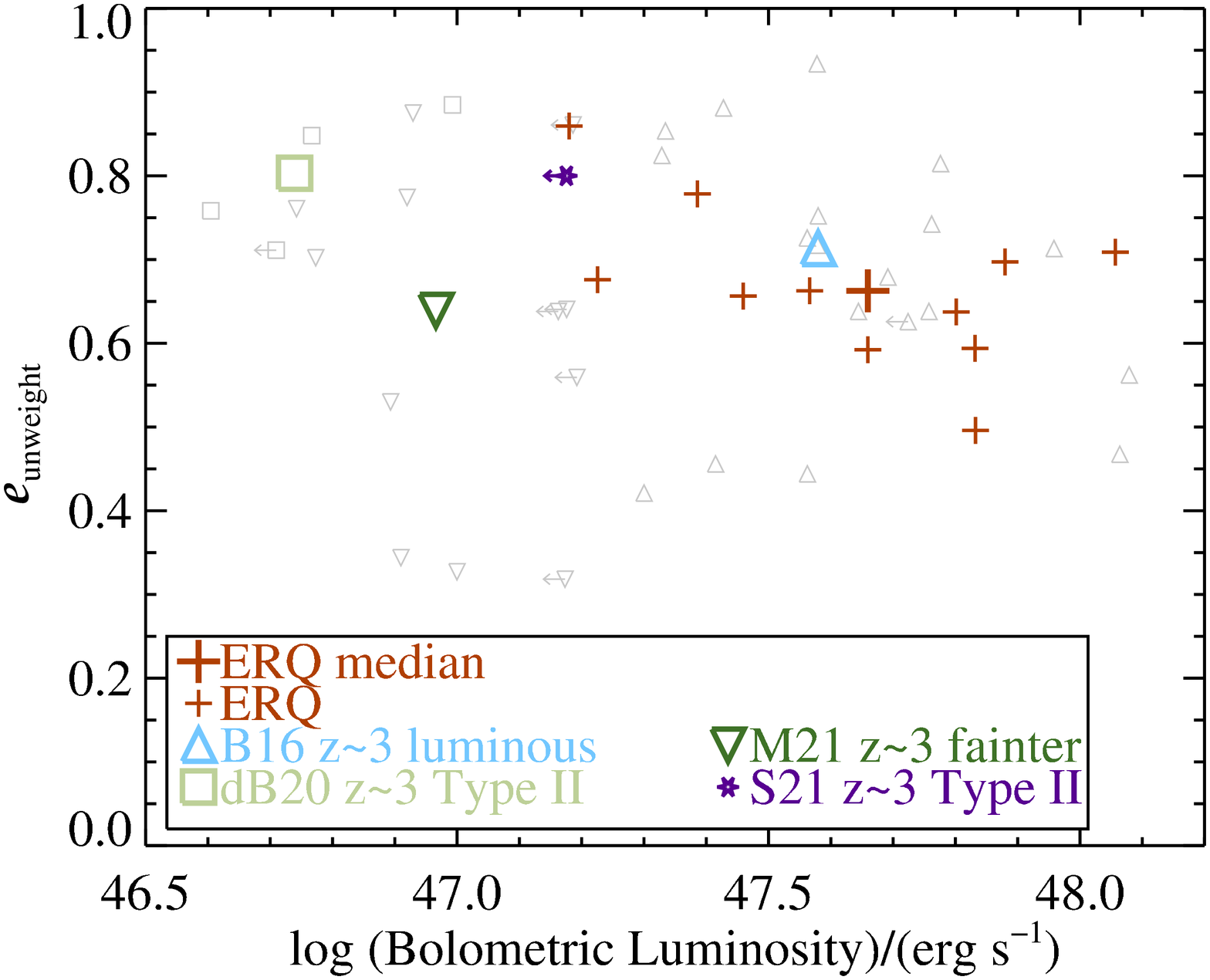}
\includegraphics[width=\columnwidth]{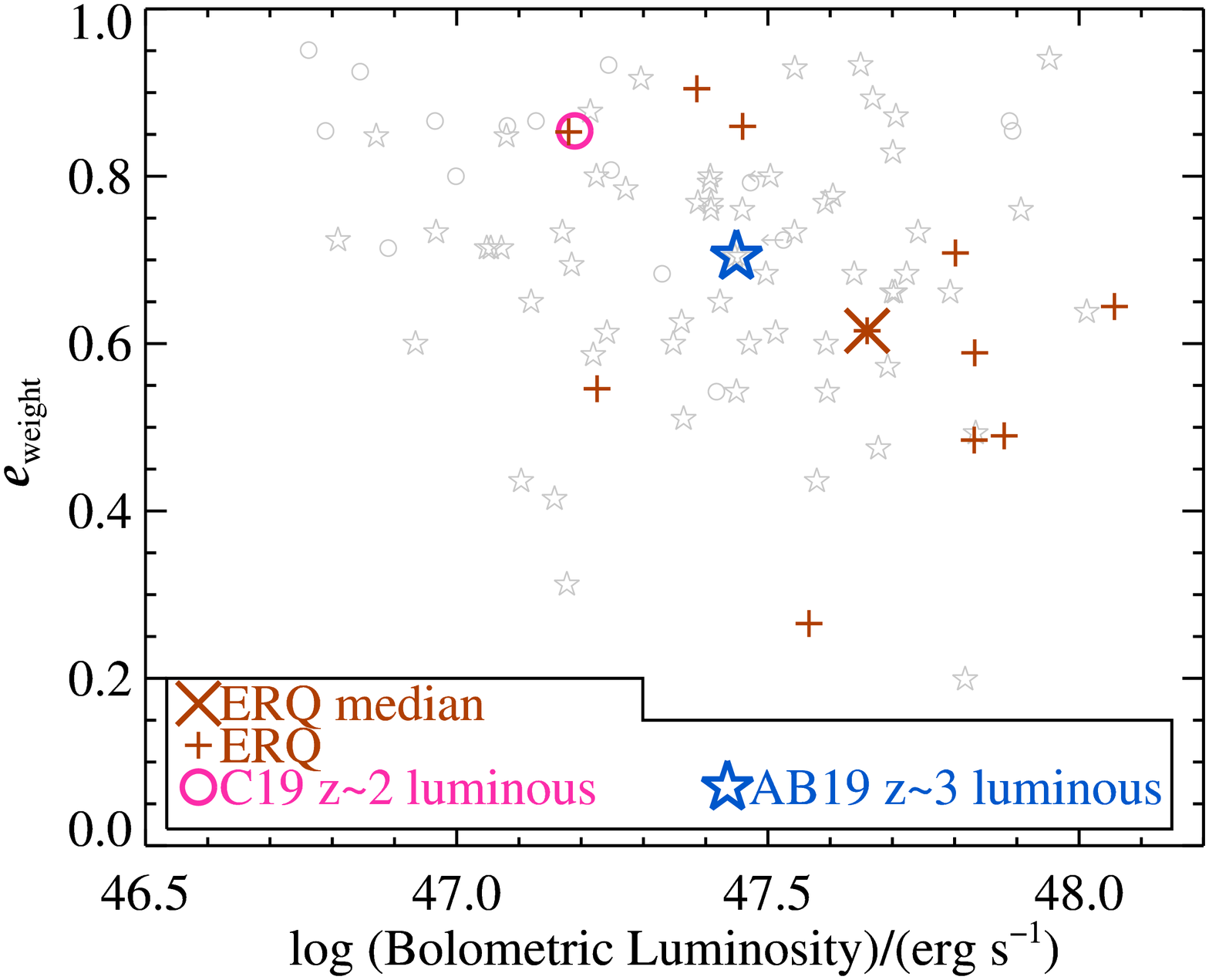}
\caption{Logarithmic bolometric luminosity vs Ly\(\alpha\) halo eccentricity, flux weighted and unweighted, for samples which have the respective measurements to compare. In addition to the identical symbols as Figure~\ref{fig:fig_LhalLbol}, Type-II quasars from \citet{denBrok+20} are shown in box symbols with their median shown as the large light green box, and an upper limit of an individual Type-II in \citet{Sanderson+21} is shown as a purple star. Eccentricity of zero is circular, and values near one are more elliptical. Plotted ERQ eccentricity values are computed with the central 1 arcsecond masked to compare with the other samples that use standard PSF subtraction, and Table~\ref{tab:table_SB_maps} shows the eccentricity values calculated without masking the central 1 arcsecond. Flux-unweighted eccentricity is a measure of the outer halo shape, whereas the flux-weighted eccentricity is more strongly influenced by the brighter inner halo regions. In the flux unweighted plot there is some separation between the populations of Type-I and Type-II quasar samples, which is more clear in the median symbols. }
\label{fig:fig_eccentLbol}
\end{figure*}

Finally, Figure~\ref{fig:fig_eccentLbol} presents morphology of halos vs bolometric luminosity, with two different eccentricity parameters. Different surveys used different parameters to characterize morphology for their population. We present ERQs with both of these parameters computed to compare with as many as possible. We take $e_\text{unweight}$ values from, \cite{Borisova+16b}, \cite{denBrok+20}, \citet{Mackenzie+21}, and \citet{Sanderson+21}. We take the computed $e_\text{weight}$ from \citet{ArrigoniBattaia+19} and \citet{Cai+19}. Their $e_\text{weight}$ is calculated with a 1-arcsecond region centered on the quasar masked to avoid residuals from point spread function subtraction. All samples we compare also mask the central 1-arcsecond region for their computed $e_\text{weight}$ or $e_\text{unweight}$, except for \cite{denBrok+20} and \cite{Sanderson+21}. The two samples with no masking are so obscured they did not need point spread function subtraction to measure the halo.

For our data in Figure~\ref{fig:fig_eccentLbol}, we also mask the inner 1-arcsecond of the quasar for plotted measurements, for fair comparison with other surveys which did not probe the inner halo as we have. Table~\ref{tab:table_SB_maps} presents our best possible eccentricity measurements, without the inner 1-arcsecond masked. Without masking, our ERQ sample has median weighted and unweighted  eccentricities of $e_\text{weight} = 0.65$ and $e_\text{unweight} = 0.66$. We give these values context by comparing them to other typical quasar values available in the literature. 

Flux unweighted eccentricity is insensitive to strong central emission, and thus can describe morphology at large scales. Blue quasar samples have $e_\text{unweight} \approx 0.7$, and are comparable to the less luminous population seen in \citet{Mackenzie+21}. Our ERQs tend to have morphologies like blue Type-I quasar samples.

There is significant scatter in $e_\text{weight}$ measurements among each sample, but medians of these populations show ERQs have more circular symmetry to other blue quasar Ly$\alpha$ halo surveys. Blue quasar samples have $e_\text{weight} \approx 0.7 - 0.9$. Luminous quasar medians are larger than the ERQ median and most individual ERQs, tabulated in Table~\ref{tab:table_SB_maps}.

We include Type-II quasars in this morphology comparison, because other studies find that Type-IIs tend to show more asymmetric halos, consistent with more edge-on views of the quasar asymmetric illumination patterns in the normal Type-I/II dichotomy \citep{denBrok+20}. Samples of Type-II quasars have an unweighted eccentricity as high as $e_\text{unweight} \approx 0.8$ \citep{denBrok+20,Sorini+21}. The ERQ in our sample with narrowest C~IV emission line FWHM, J1705+2736, is similar to the other Type-IIs, and has the largest unweighted eccentricity in our sample with $e_\text{unweight} = 0.78$. J1705+2736 also has the largest flux weighted eccentricity of our sample, and greater than of most of the Type-I blue quasars, with $e_\text{weight} = 0.91$.

\section{Discussion}
\label{sec:section_discussion}

ERQs could represent a brief transition phase in galactic and quasar evolution, characterized by an obscured galactic/quasar environment and potentially extreme feedback via outflows (see Section \ref{sec:sec_intro}). Our main goal is to determine if ERQs differ from normal blue quasars in significant ways that might identify them as a distinct, and potentially more youthful, galactic/quasar population. We obtained Keck KCWI data for 11 ERQs to measure their basic Ly\(\alpha\) halo properties. In this section we will discuss our results for the full sample.

\subsection{Quasar Systemic Redshift \& Future Work}
\label{sec:sec_dis_redshift}

Systemic redshifts are uniquely difficult to determine for ERQs due to frequent blueshifted/fast-outflowing emission-lines that are typically used as redshift indicators, such as [O~III]~$\lambda$5007 and broad emission lines (\citealt{Hamann+17}; \citealt{Perrotta+19}; Gillette et al. 2023b in prep.). Our study confirms the result in L22 that the narrow Ly$\alpha$ spike present in aperture spectra of some ERQs are the same redshift and emission profile as the inner halo emission, and therefore this spike also forms in the inner halo and is a good indicator of ERQ systemic redshifts. L22 used the Ly$\alpha$ spike redshift in the reddest ERQ, J0006+1215, to show the centroid of the broad C~IV~$\lambda$1549 emission line is blueshifted by 2240 km~s$^{-1}$. \cite{Hamann+17} provides other examples of large broad emission-line blueshifts in ERQs based on the Ly$\alpha$ spike, claiming that large blueshifts are more common in ERQs than luminous blue quasars. We present a more complete discussion of blueshifts and their implications for ERQ outflows in a forthcoming paper, Gillette et al. (2023b in prep.).

\subsection{Circularly Symmetric and Compact Halos}
\label{sec:sec_compact}

Overall we find that ERQs and blue quasars both exhibit a wide range of Ly$\alpha$ halo properties and there is considerable overlap between these samples. The main differences identified by our study are in the halo morphologies. In particular, the ERQs tend to have 1) more compact and centrally concentrated SB profiles, and 2) more circularly symmetric central regions. Fig.~\ref{fig:fig_rad_prof} shows other samples which have circularly averaged SB radial profiles measured have median exponential scale lengths between 13.5 and 18.7~kpc, and the scale length of the median profile of our full sample is 9.0~kpc. When considering only ERQs that exhibit a Ly$\alpha$ spike, the median profile's exponential scale length is only 6.0~kpc. Fig.~\ref{fig:fig_eccentLbol} shows the flux-weighted elliptical eccentricities of our sample are much smaller, probing the inner halo, and meaning more circularly symmetric. 

Reasons for ERQs having more compact and more circular morphology are unclear. They are possibly related to a younger evolution stage than blue quasars, for example, if the outflows from ERQs (e.g., \citealt{Perrotta+19}; Gillette et al. 2023b in prep.) have not had time yet to distort and disperse the gas in their inner halo. Compact and circular morphology also supports the argument that ERQs do have Type-I orientation in spite of their obscuration. 

\subsection{Type-I versus Type-II Quasars}
\label{sec:sec_type_I_vs_II}

We can gain insights into the dust distribution around ERQs from the halo morphology being more similar to Type-I blue quasar samples than Type-II samples. Previous studies have shown that Type-IIs tend to have more asymmetric Ly$\alpha$ halos, consistent with Type-IIs having a more edge-on view of the asymmetric (bipolar) radiation pattern of quasars in the standard Type-I/II dichotomy \citep{Antonucci93,UrryPadovani95,Netzer15,denBrok+20}. 

Gas distribution asymmetry is evident in the eccentricity parameters plotted in Fig.~\ref{fig:fig_eccentLbol}, although the sample sizes are small and there is considerable overlap in values between samples. ERQs tend to have more symmetric halo morphologies than Type-I blue quasars, making them even more different from the Type-IIs.

One ERQ in our sample, J1705+2736, has considerable asymmetry in its outer Ly$\alpha$ halo morphology, and has an asymmetric and luminous arm that extends >70 kpc from the quasar. J1705+2736 has substantially narrower FWHM(C~IV) than the other ERQs, similar to other Type-II quasars, although FWHM(C~IV) alone is known not to be a good discriminator for Type-I versus Type-II classification \citep{Greene+14, Hamann+17}. Fig. \ref{fig:fig_annular_spec} shows J1705+2736's extended emission, that is uniformly blueshifted from the central halo by about $-$100~km~s$^{-1}$, and out to $>$70~kpc in the southern direction. 

Thus it appears that the large line-of-sight extinctions toward ERQs cannot be attributed to viewing effects analogous to Type-IIs in the simplified picture of Type-IIs versus Type-I quasars. Our conclusion that ERQs are more similar to Type-I's is supported by the finding in L22 that the He~II/Ly$\alpha$ line ratio in the ERQ J0006+1215 (the only ERQ for which those lines are measured) is more similar to Type-I blue quasars than Type-IIs (see section 4.5 in L22 for more discussion).

\subsection{No Evidence of Halo Feedback}
\label{sec:sec_no_feedback}

Figure~\ref{fig:fig_dispLbol} reveals ERQs appear to follow the same trend as blue quasars for larger velocity dispersions around quasars with larger bolometric luminosity. Overall the halos are kinematically quiet, with integrated velocity dispersion in the range 225$-$526~km~s$^{-1}$, similar to matched blue quasars. 

We compare these velocity dispersions with emission-line broadening expected from the orbital motions of gas in the host galaxy's dark matter halo. Our sample is on the high end of quasar luminosities, and further studies have found that obscured quasars and hyperluminous quasars may on average reside in more massive halos, typically $10^{13}h^{-1} M_{\odot}$ \citep{DiPompeo+17,Geach+19}. We assume a dark matter halo profile from \citet{NavarroFrenkWhite97}, and halo concentration parameter of 4, from the COLOSSUS software and references therein.\footnote{https://bdiemer.bitbucket.io/colossus/halo\_concentration.html} We then use the halotools software to calculate the circular velocity for an upper limit for the projected 1D velocity.\footnote{https://halotools.readthedocs.io/en/latest/api/halotools.empirical\_models. \\NFWProfile.html} Halo gas with dispersion above $379$~km~s$^{-1}$ could be considered fast moving, and corresponds to 1D circular velocity of 536 km~s$^{-1}$. Table~\ref{tab:tab_dispersion} presents our integrated Ly$\alpha$ halo dispersion. Velocity dispersions of our sample have a median dispersion of 293~km~s$^{-1}$, are within a few hundred km~s$^{-1}$ of each other, and none are above the fast moving circular velocity estimate. Quiet kinematics do not rule out the possibility of youth in ERQs, because they may have had less time for energetic outflows to extend outward to generate feedback in the inner halo.

Across the sample we generally do not find Ly$\alpha$ halo emission-line broadening above the expected dark matter halo velocities down to $\sim$0.7~arcsec or $\sim$6~kpc radius. Episodic lifetimes of quasars are typically $10^5-10^7$ yrs \citep[e.g.,][]{Khrykin+21}. With most Ly$\alpha$ halo outflow speeds $\sim$400~km~s$^{-1}$ they can at most only travel to $\sim$4~kpc. Therefore any Ly$\alpha$ emission on circumgalactic scales are not outflowing due to the present quasar episode. Spatially resolved velocity dispersion beyond 20$-$30~kpc is more commonly measured in ERQs with a Ly$\alpha$ spike. These more extended halos gradually decrease velocity dispersion toward the edges, away from the quasar. Instances of extended clouds (eg. J1145+5742 and J1451+0132) also show a decrease in dispersion farther along the extended arm, but J0006+1215 shows increasing dispersion with increasing distance along its extended arm. We do not find evidence for large velocity dispersion in the Ly$\alpha$ halos that may be caused by outflows and feedback effects from the central quasar, but an absence of fast outflows in our sample has no implications on the quasar evolutionary stage.

In conclusion, the quiet kinematics, and the compact and circular symmetry discussed in Section \ref{sec:sec_compact}, are evidence against feedback being present at CGM halo scales, now or in the past.

\subsection{General Halo Properties}
\label{sec:sec_ERQ_properties}

If ERQs are in younger host galaxies, then they may have different ionizing escape fractions compared to typical blue quasars. If the galaxy is dusty, then the escape fraction could be lower. But if the dust distributions are clumpy, then they may have substantial ionization escape fractions. Luminosities of the Ly$\alpha$ halos are comparable to typical luminous blue quasars of similar redshift in Fig.~\ref{fig:fig_LhalLbol}. We also see in Fig.~\ref{fig:fig_linszLbol} that ERQ halo linear size is within the size distributions in other samples of luminous quasars. The median ERQ halo luminosity is offset roughly two times lower than expected from the blue quasar data, but this does not appear significant given the small sample size and the width of the halo luminosity distributions spanning nearly a factor of 100. Thus the appearance of narrow Ly$\alpha$ emission spikes in ERQ spectra can be attributed to extinction toward the central quasar, and not from unusually bright halo emission, relative to the central quasars.

If the Ly$\alpha$ halo emissions are powered by hydrogen-ionizing radiation from the central quasars, the similar halo luminosities between ERQs and blue quasars might indicate that similar fractions of the H-ionizing photons emitted by the quasars escape to the circumgalactic medium in ERQs. This similarity is notable because of the much larger line-of-sight extinctions observed in ERQs, estimated to be typically $\sim$3 mags or a factor of $\sim$16 in the near UV around 1500 \AA\ to 2000 \AA\ \citep{Hamann+17}. One possible explanation is that other lines-of-sight toward ERQs have lower extinctions than we observe, allowing their ionizing photons to escape in quantities similar to blue quasars. Another possibility is that the dust distributions are clumpy and inhomogeneous, which can permit larger UV photon escape fractions than expected from the extinctions along direct lines of sight to the central source, due to scattering. The general notion that dust scattering plays an important role is consistent with the large UV polarizations found in some ERQs by \citet{Alexandroff+18}; however, they attribute their results to a particular axisymmetric scattering geometry that might conflict with our finding that ERQs tend to have circularly symmetric halos, resembling Type-I blue quasars (Section \ref{sec:sec_type_I_vs_II}).

An important caveat to keep in mind for the $L_{halo}$ comparisons is that Ly$\alpha$ halos around quasars are generally believed to be at least partly matter bounded, meaning that the observed luminosities depend at least partly on the amount of halo gas available for ionisation, not simply on the flux or escape fraction of ionising photons emitted by the quasar. ERQs show a weaker than expected Ly$\alpha$ halo luminosity for their bolometric luminosity, which could be evidence of matter bounded halo emission (\citealt{DempseyZakamska18}, also Fig.~\ref{fig:fig_LhalLbol}). In an ionization bounded scenario we expect the luminosity of the halo to scale linearly with bolometric luminosity, but when a halo is matter bounded there is no dependence. However, real quasars can show a weak dependence of the halo properties on the bolometric luminosity, and any relation between Ly$\alpha$ halo luminosities and extinction toward the quasars will also be weak. 

A unique feature of our study is that extreme obscuration allows us to map the 2D halo emissions all the way down to the quasar positions, and possibly down to galactic emission, in roughly half of our ERQ sample. ERQs that exhibit the Ly$\alpha$ spike show similar inner halo characteristics as J0006+1215 in L22, but vary in their outer region symmetry and morphology beyond 20$-$30 kpc from the emission peak. ERQs without the spike are most luminous around the central region of the quasar, and tend to be less extended to outer regions beyond 30$-$40 kpc from the halo emission peak (see Fig.~\ref{fig:fig_maps}).

\subsection{Multi-component Emission}
\label{sec:sec_spatial_properties}

In several ERQs with spatially resolved measurements, a transition at $\sim$20$-$30~kpc from the quasar frequently occurs (e.g., a drop off in SB, change in blueshift, or velocity dispersion). This transition leads us to define a boundary of inner-halo and outer-halo Ly$\alpha$ emission. This boundary could be evidence of a 2-component halo structure of compact and extended gas (e.g., J1145+5742 and J1652+1728). L22 showed a halo transition phase 20$-$30~kpc from the position of ERQ J0006+1215, defining an inner and outer halo component. The inner region is more coherent, circularly symmetric, and has quiet gas kinematics, and the outer region has more asymmetric and disrupted kinematics. One possibility is that the outer halo is inflowing CGM gas, and the inner halo is dominated by outflow  (see section 4.2 in L22 for more discussion). This is further supported in multi-component JWST observations of J1652+1728, which found complex gas kinematics and outflows on kpc scales \citep{Vayner+23b}.  Figure~\ref{fig:fig_annular_spec} shows aperture spectra of ERQs with distinct inner and outer halo emission regions that show different kinematics from each other.

Many of our ERQ Ly$\alpha$ halos show a smooth gradient in their velocity shift from one side of the halo to the other, with zero velocity centered near the halo emission peak. None have extreme velocities consistent with powerful outflow. In cosmological radiation-hydrodynamic simulations from \cite{Costa+22} quasar driven outflows on circumgalactic scales move $\sim$1,000~km~s$^{-1}$, and are also less dense than most Ly$\alpha$ emitting gas.

In summary, our comparison of ERQ Ly$\alpha$ halo properties to blue quasars has revealed many similarities, and some contrasting characteristics that distinguish them from simple orientation effects. Extended morphology of halos around ERQs appear similar to those of blue quasars, but the inner halo is more circularly compact. Future work with larger samples and/or deeper maps could help resolve the evolution/youth question. Further exploration into ERQs as an evolutionary stage (eg., Perrotta et al. 2023 and Hamann et al. 2023) will investigate the host galaxies for comparison to blue quasars.

\section{Conclusions} 
\label{sec:section_conclusion}

We present a sample of 11 ERQs observed with KCWI Integral Field Spectroscopy, which have median redshift $z$~$=$~2.6, a median color of $i-W3$~$=$~5.9~mag, and median bolometric luminosity $L_{\text{bol}}$~$\approx$~5~$\times$~$10^{47}$~erg~s$^{-1}$. Except for one ERQ observed under cloudy conditions, all have detected Ly$\alpha$ halos, and have a median halo luminosity $L_{\text{halo}}$~$=$~5.83~$\times$~$10^{43}$~erg~s$^{-1}$.

The median of Ly$\alpha$ emission's maximum linear size is >128 kpc, and exponential scale length of the circularly averaged SB radial profile median is $9.0$ kpc. Morphology is generally circular around the inner halo regions, with a median flux-weighted eccentricity of $e_\text{weight} = 0.65$ and unweighted $e_\text{unweight} = 0.66$. One ERQ in our sample, J1705+2736, that has substantially narrower emission line widths, has the most asymmetric morphology in its outer halo with $e_\text{unweight} = 0.78$, and compared to the rest of the ERQ sample has the most asymmetric inner halo with $e_\text{weight} = 0.91$.

Kinematics of the halos are relatively calm, with velocity shifts of the Ly$\alpha$ emission centroid to be in the hundreds of km~s$^{-1}$, and much weaker than the shifts found in [O III] for ERQs in the thousands of km~s$^{-1}$ by \citet{Perrotta+19}. Velocity maps are coherent, with some showing a gradual gradient from red to blue velocity shifts across the halo. Dispersion of the halo emission is also quiet, with a spatial median dispersion of 374 km~s$^{-1}$, and standard deviation of 114 km~s$^{-1}$.

Our measured quantities of size, luminosity, blueshift, and dispersion show ERQs are mostly similar to those obtained in blue quasar surveys. ERQs generally follow luminosity trends that are seen across faint to luminous blue quasar samples for halo linear-size, luminosity, and velocity dispersion. However, ERQs do stand apart from similarly luminous blue quasars in that their halos are more circularly compact.

Most of the statements and inferences about ERQ populations made in L22 are supported by this work with the addition of 10 other Ly$\alpha$ halos, summarized below:

\begin{itemize}
    \item We do not see a color correlation with Ly$\alpha$ luminosity or kinematics across our sample.
    
    \item At circumgalactic scales we do not find clear evidence of feedback based on the circularly symmetric inner halos and low velocity dispersions (see Sections~\ref{sec:sec_compact} \& \ref{sec:sec_no_feedback}).

    \item Ly$\alpha$ halo velocity dispersions are mostly consistent with circular velocities of halo gas at a typical dark matter halo mass (see Section~\ref{sec:sec_no_feedback}).

    \item Our sample's illumination patterns are similar to other blue quasars based on Ly$\alpha$ halo morphologies, characterized by elliptical eccentricity parameters (see Section~\ref{sec:sec_compare}).

    \item ERQ halos are more circularly concentrated, and could mean they have had less time to extend by outflows from the innermost unresolved regions (see Section~\ref{sec:sec_compact}).

    \item Obscuration acting as a chronograph allows for measurements of narrow Ly$\alpha$ emission lines, and these lines can be used as systemic redshift estimators for constraining broad line emission blueshifts and outflows (see Section~\ref{sec:sec_dis_redshift}).
\end{itemize}

\section*{Acknowledgements}

JG, MWL, and FH acknowledge support from the USA National Science Foundation grant AST-1911066. NLZ and AV acknowledge support from the NASA Astrophysics Data Analysis Program Grant 80NSSC21K1569. The data presented herein were obtained at the W. M. Keck Observatory, which is operated as a scientific partnership among the California Institute of Technology, the University of California and the National Aeronautics and Space Administration. The Observatory was made possible by the generous financial support of the W. M. Keck Foundation. Data presented herein were partially obtained using the California Institute of Technology Remote Observing Facility. The authors wish to recognize and acknowledge the very significant cultural role and reverence that the summit of Maunakea has always had within the indigenous Hawaiian community.  We are most fortunate to have the opportunity to conduct observations from this mountain. 


\section*{Data Availability}

The data are available upon request.



\bibliographystyle{mnras}
\bibliography{ms}






\bsp	
\label{lastpage}
\end{document}